\documentclass[apj,twocolappendix,numberedappendix]{emulateapj}

\usepackage{graphicx}
\usepackage{apjfonts}
\usepackage{amsmath}
\usepackage[raggedright,FIGBOTCAP]{subfigure}
\usepackage{color}
\usepackage{hyperref}

\newcommand{\eref}[1]{(\ref{#1})}
\newcommand{\Eref}[1]{Equation~(\ref{#1})}
\newcommand{\Sref}[1]{Section~\ref{#1}}
\newcommand{\Fref}[1]{Figure~\ref{#1}}
\def\EAH{\textit{Einstein@Home}}
\def\Fermi{\textit{Fermi}}
\DeclareMathOperator{\sinc}{sinc}
\def\sig{{\tiny\textrm{sig}}}
\newcommand{\vecb}[1]{\boldsymbol{\mathrm{#1}}}
\def\Doppler{\mathrm{u}}
\def\vDoppler{\vecb{\Doppler}}
\def\av#1{{\langle #1 \rangle}}

\def\Nullvec{\vecb{0}}

\newcommand{\rmd}{\mathrm{d}}
\newcommand{\msun}{\ifmmode\mbox{M}_{\odot}\else$\mbox{M}_{\odot}$\fi}
\newcommand{\mm}{m}

\def\erfc{\mathrm{erfc}}
\def\sinc{\mathrm{sinc}}

\DeclareGraphicsExtensions{.pdf}

\shorttitle{Blind Gamma-ray Pulsar Searches} 
\shortauthors{\sc Pletsch and Clark}

\begin{document}

\title{Optimized Blind Gamma-ray Pulsar Searches at Fixed Computing Budget} 

\author{Holger~J.~Pletsch and Colin~J.~Clark}
\affil{Max-Planck-Institut f\"ur Gravitationsphysik (Albert-Einstein-Institut), D-30167 Hannover, Germany, and 
Institut f\"ur Gravitationsphysik,\\ Leibniz Universit\"at Hannover, D-30167 Hannover, Germany; 
holger.pletsch@aei.mpg.de}

\begin{abstract} 
\noindent
The sensitivity of blind gamma-ray pulsar searches in multiple years worth of photon data, as from the \textit{Fermi} LAT, is primarily limited by the finite computational resources available. Addressing this ``needle in a haystack'' problem, we here present methods for optimizing blind searches to achieve the highest sensitivity at fixed computing cost. For both coherent and semicoherent methods, we consider their statistical properties and study their search sensitivity under computational constraints. The results validate a multistage strategy, where the first stage scans the entire parameter space using an efficient semicoherent method and promising candidates are then refined through a fully coherent analysis. We also find that for the first stage of a blind search incoherent harmonic summing of powers is not worthwhile at fixed computing cost for typical gamma-ray pulsars. Further enhancing sensitivity, we present efficiency-improved interpolation techniques for the semicoherent search stage. Via realistic simulations we demonstrate that overall these optimizations can significantly lower the minimum detectable pulsed fraction by almost $50\%$ 
at the same computational expense.
\end{abstract} 


\section{Introduction}\label{s:intro}

The \Fermi{} Large Area Telescope \citep[LAT;][]{generalfermilatref} 
has an unprecedented sensitivity to detect the periodic gamma-ray emission 
from spinning neutron stars. Owing to the LAT,
the number of detected gamma-ray pulsars has vastly increased from a 
handful to about 150 \citep[for a recent review see e.g.,][]{Caraveo2013}, 
making these objects a dominant Galactic source class at GeV energies. 

So far, the largest fraction of LAT-detected gamma-ray pulsars 
has been uncovered indirectly \citep{Fermi2PC}. In this approach,
pulsar ephemerides known from previous radio observations are used to 
assign rotational phases to the gamma-ray photons, which are then tested 
for pulsations. Dedicated radio searches at positions of unidentified gamma-ray 
sources in the \Fermi-LAT Second Source Catalog \citep[2FGL;][]{FermiSecondSourceCatalog} 
have been particularly successful in discovering many new radio pulsars, and have 
provided ephemerides for subsequent gamma-ray 
phase-folding \citep[e.g.,][]{Ransom2011,Guillemot2012,Fermi2PC}.

The direct detection of new gamma-ray pulsars, which are not known beforehand from other 
wavelengths, requires  \emph{blind searches} for periodicity in the sparse 
gamma-ray photon data \citep[e.g.,][]{Chandler2001}.
With the \Fermi-LAT, for the first time such blind searches have been successful \citep{16gammapuls2009}.
Notably, many of the gamma-ray pulsars found this way have so far remained undetected at radio
wavelengths \citep{Fermi2PC}, implying that blind searches are the only way
to access this pulsar population.
Currently, hundreds of  \Fermi-LAT sources still remain unidentified, but feature pulsar-like properties 
\citep{2012ApJ...753...83A,Lee+2012} and thus likely harbor undiscovered pulsars.

The key problem in blind searches for gamma-ray pulsars is the enormous
computational demand involved, which is what limits the search sensitivity.
Since the relevant pulsar 
parameters are unknown in advance, one has to search a dense grid covering 
a multidimensional parameter space. The number of search grid points increases rapidly 
with longer observation times. For observations spanning multiple years,
``brute-force'' (most sensitive but most expensive) methods, which involve fully coherently
tracking the pulsar rotational phase over the entire observational data time span, are unfeasible.
Therefore, the \emph{efficiency} of blind-search methods is crucial, because
optimal strategies are those that provide the best search sensitivity at fixed computing cost. 
This is the main theme of this work.

The problem is generally best addressed by a multistage search scheme
\citep[e.g.,][]{Meinshausen2009}. This also applies to blind searches for gravi\-tational-wave 
pulsars, i.e. spinning neutron stars emitting periodic gravitational waves 
\citep{bccs1:1998,bc2:2000,cutler:2005,PrixShaltev2012}. The basic idea is that in a first stage, the entire 
search parameter space is scanned but employing a much lower resolution, and therefore 
at much lower computing cost, which can most efficiently discard unpromising regions. 
This reduction in parameter resolution is accomplished by semicoherent methods,
in which only time intervals of data much shorter than one year are coherently
analyzed whose results are then incoherently summed over multiple years.
In subsequent stages, only small promising regions (i.e. pulsar candidates) are followed up with higher 
resolution at higher computational expense, 
by using longer coherent integration times.

One semicoherent method appropriate for the first search stage in gamma-ray pulsar
searches is the seminal ``time differencing technique'' by \citet[][hereafter A06]{Atwood2006}.
It can basically be seen as the application of the classic Blackman--Tukey method
\citep{BlackmanTukey1958} to gamma-ray data: 
To search along the $f$-dimension (estimating the power spectrum)
A06 calculated the discrete Fourier transform (DFT) of the 
autocorrelation function between photon arrival times up to a maximum lag.
This significantly improved the efficiency over earlier methods 
\citep[e.g.,][summing power of many DFTs from subintervals]{bc2:2000,Chandler2001}, 
because the autocorrelation function can be computed at negligible cost 
thanks to the sparsity of the photon arrival times.
The success of the A06 method has been spectacularly demonstrated by
the blind-search discovery of $24$ gamma-ray pulsars \citep{16gammapuls2009,8gammapuls2010}
within the first \Fermi{} mission year.

Using further improved methods, in part originally developed for blind searches for gravitational-wave 
pulsars \citep{PletschAllen2009,Pletsch2010}, analyzing about three years of LAT data revealed $10$ 
new gamma-ray pulsars \citep{Pletsch+2012-9pulsars,Pletsch+2012-J1838}. 
Crucial methodological improvements included the use of an analytic metric on 
parameter space 
to construct the grid over both sky position and frequency derivative.
This allowed pulsars to be found that are much farther from the
LAT catalog sky position than was possible previously.
In addition, a photon weighting scheme \citep[first studied by][]{Kerr2011}  
was used for both photon selection and for the
search computations to ensure near optimal detection significance.
For enlarged computational resources we have recently moved this ongoing search effort onto 
the volunteer computing system 
\EAH{}.\footnote{\href{http://einstein.phys.uwm.edu/}{http://einstein.phys.uwm.edu/}} 
So far, this has resulted in the discovery of another $4$ young pulsars \citep{Eah4GammaPulsars}.
We here give a more detailed description of the strategies and methods
exploited in these searches, and consider related questions one be might
faced with when setting up a blind search:
Could a fully coherent blind search using a subset of data perhaps be more sensitive
than a semicoherent search using all of the data? Is harmonic summing worthwhile
under computational constraints? What is the optimal search-grid point density
to balance sensitivity versus computing effort? 
In addressing such questions, we present the technical framework to optimize 
the sensitivity of blind pulsar searches in gamma-ray data at fixed computing cost.
Moreover, we present further important methodological advances to 
improve the overall blind-search efficiency.

The paper is organized as follows. 
In \Sref{s:statisticaltests}, we describe the
statistical detection of pulsations in general. In \Sref{s:coherent},
we discuss the statistical properties of coherent blind searches
and study their computational cost scalings using the parameter-space metric.
We also investigate the efficiency of harmonic summing for 
different pulse profiles. In \Sref{s:semicoherent}, we describe
the statistical properties of a semicoherent blind-search method and
compare the respective computing demand using the semicoherent metric.
\Sref{s:implementation} presents a collection of technical
improvements for the implementation of the semicoherent search stage, 
including efficient interpolation methods 
and automated candidate follow-up procedures. 
We demonstrate the superiority from combining these advances 
through realistic simulations in \Sref{s:perfdemo}. 
Finally, conclusions follow in \Sref{s:concl}.

\section{Statistical detection of pulsations}\label{s:statisticaltests}

In blind pulsar searches the pulse profile (the periodic light curve)
and the exact parameters describing the rotational evolution 
of the neutron star are \emph{unknown} in advance.
As \citep{Bickel+2008} have pointed out, unless the pulse profile shape is
precisely known, there is no universally optimal statistical test, because
any most powerful test for one template profile will not be most powerful against
another. Any test can only be most sensitive to a finite-dimensional
class of targets. Thus, for computational feasibility of a blind search
an efficient (potentially suboptimal) template pulse profile to test against should
attain only modest reduction in detection sensitivity compared to an optimal template.
The construction of such a test can be guided by the profiles of known
gamma-ray pulsars, which we will consider below.

For isolated pulsars the search parameters describing the rotational phase
of the neutron star is at least four-dimensional, consisting of
frequency $f$, spindown rate $\dot f$, and sky position with 
right ascension~$\alpha$ and declination~$\delta$.
To the LAT-registered arrival times $t_{\tiny\rm LAT}$
 sky-position $(\alpha,\delta)$ dependent corrections (``barycentric corrections'') 
are applied in order to obtain the photon arrival times~$t$ 
at the solar system barycenter (SSB).
Then the rotational phase $\Phi(t)$  is  described by
\begin{equation}
  \Phi(t) = \phi_0 + 2\pi\,f (t-t_0) + 2\pi\,\dot f \frac{(t-t_0)^2}{2} \,,
  \label{e:phase1}
\end{equation}
where $f$ and $\dot f$ are defined at reference time~$t_0$, 
when the phase equals the constant $\phi_0$. 

Apart from the arrival time, for each of $N$ detected gamma-ray photons, 
indexed by $j$, the LAT also records the photon's reconstructed energy and direction.
From these a weight, $w_j$, can be computed measuring the probability that it
has originated from the target source \citep{Bickel+2008,Kerr2011}. 
Using these probability weights efficiently avoids testing different 
hard selection cuts on energy and direction (implying binary weights),
providing near optimal pulsation detection sensitivity 
\citep{Kerr2011,Pletsch+2012-9pulsars}. 

The observed gamma-ray pulse profile $F(\Phi)$, the flux as a function of $\Phi$,
can be written as
\begin{equation}
  F(\Phi) \propto \frac{1-p}{2\pi} + p\; F_s(\Phi) \,,
  \label{e:F1}
\end{equation}
where $p$ is the \emph{pulsed fraction} that is estimated by the number of pulsed
gamma-ray photons divided by the total number of photons.
$F_s(\Phi)$ represents the pulse profile (undisturbed by background) and
is a probability density function on $[0,2\pi]$,
which can be expressed as a Fourier series
\begin{equation}
  F_s(\Phi) = \frac{1}{2\pi} \left(1 + \sum_{n \neq 0} \alpha_n \; e^{i\, n\, \Phi} \right) \,,
\end{equation}
with the complex Fourier coefficients $\alpha_n$, defined at harmonic order~$n$ as
\begin{equation}
 \alpha_n = \int_{0}^{2\pi} F_s(\Phi) \; e^{-i\,n\,\Phi} \, d\Phi \,.
 \label{e:FourierCoeff}
\end{equation}
Hence the total flux $F(\Phi)$ can be rewritten as
\begin{equation}
  F(\Phi) \propto 1+ p\;  \sum_{n \neq 0} \alpha_n \; e^{i\, n\, \Phi} \,.
  \label{e:F2}
\end{equation}
If $F_s(\Phi)$ is an exact sinusoidal pulse profile, then from \Eref{e:FourierCoeff}
it follows that $|\alpha_1| = 1/2$ and all other coefficients vanish, $|\alpha_{n>1}|=0$.
As another example, if the pulse profile $F_s(\Phi)$ is
a Dirac delta function, i.e. the narrowest possible profile, then all coefficients
are equal, $|\alpha_n|=1$, implying equal Fourier power at all harmonic orders.

In general, the null hypothesis is given by $p=0$, meaning that all phases are 
uniformly distributed (i.e. no pulsations). From the likelihood for photon 
arrival times \citet{Bickel+2008} derived a score test statistic~$Q_M$ for $p > 0$,
\begin{equation}
  Q_M =  \frac{1}{K^2}\sum_{n = 1}^M |\alpha_n|^2 \; |A_n|^2  \,,
  \label{e:Qstat}
\end{equation}
where we defined the normalization constant~$K$ \citep[different from][]{Bickel+2008}
as
\begin{equation}
  K^2 = \frac{1}{2M} \sum_{n = 1}^M |\alpha_n|^2\,,
\end{equation}
and $A_n$ is given by
\begin{equation}
  A_n = \frac{1}{\kappa} \sum_{j=1}^N w_j \; e^{-i\,n\,\phi(t_j)} \,,
\end{equation}
with the time-dependent part of the phase \mbox{$\phi(t) = \Phi(t) - \phi_0$} and the normalization constant~$\kappa$ defined as
\begin{equation}
  \kappa^2 = \frac{1}{2} \sum_{j=1}^N w_j^2 \,.
  \label{e:kappa}
\end{equation}
Thus, we denote by $\mathcal{P}_n$ the coherent Fourier power at the $n$th harmonic,
\begin{equation}
  \mathcal{P}_n = |A_n|^2 = \frac{1}{\kappa^2} \left| \sum_{j=1}^N w_j \; e^{-i\,n\,\phi(t_j)} \right|^2 \,.
  \label{e:Pn}
\end{equation}
Appealing to the Central Limit Theorem (since $N\gg1$ in all practical cases)
the normalization choice of \Eref{e:kappa} has the convenient 
property that the coefficients $\Re(A_n)$ and $\Im(A_n)$ 
become independent Gaussian random variables 
with zero mean and unit variance under the null hypothesis.
Therefore, to good approximation each $\mathcal{P}_n$ is $\chi^2$-distributed with 
$2$~degrees of freedom, as will be discussed below. 
Thus, $Q_M$ is the weighted sum of coherent Fourier powers,
\begin{equation}
  Q_M =  \sum_{n = 1}^M \frac{|\alpha_n|^2}{K^2} \; \mathcal{P}_n  \,.
  \label{e:Qstat2}
\end{equation}
Therefore, as noted by \citet{Bickel+2008}, the test statistic $Q_M$ is invariant under phase shifts (i.e. 
independent of reference phase~$\phi_0$) and only depends on the 
amplitudes of the Fourier coefficients~$\alpha_n$, but not on their phases.
Moreover, \citet{Beran1969} showed earlier that
if the pulse profile is known a priori,
a test statistic following from $Q_M$ for binary weights is locally most powerful 
for testing uniformity of a circular distribution, assuming unknown and 
weak (small $p$) signal strength.

\section{Coherent Test Statistics }\label{s:coherent}

In what follows, we examine the sensitivity of coherent blind searches at fixed computational cost,
taking into account the statistical properties and sensitivity scalings in terms of relevant 
quantities. For simplicity, during the remainder of this section we here assume hard photon 
selection cuts, i.e., binary weights only, $w_j \in \{0,1\}$, such that $\mathcal{P}_n$ reduces to
\begin{equation}
  \mathcal{P}_n = \frac{2}{N} \left| \sum_{j=1}^N e^{-i\,n\,\phi(t_j)} \right|^2 \,.
  \label{e:Pnbweights}
\end{equation}
However, the main conclusions obtained will also have applicability 
when arbitrary (i.e., non-binary) weights are used.

\subsection{Statistical Properties}
\label{s:Statcoherent}

Under the null hypothesis $p=0$ and assuming $N\gg1$,
the coherent power $\mathcal{P}_n$ as of \Eref{e:Pnbweights} follows a central $\chi^2$-distribution 
with $2$ degrees of freedom (see Appendix~\ref{s:statcohpow}),
whose the first two moments are,
\begin{equation}
  E_0 \left[ \mathcal{P}_n \right] = 2 \,,\quad Var_0 \left[ \mathcal{P}_n \right] = 4 \,.
  \label{e:PnE0Var0}
\end{equation}
Suppose the photon data contains a pulsed signal, $p>0$, whose
pulse profile can be expressed in terms of complex Fourier coefficients, 
$\gamma_n$ as in \Eref{e:FourierCoeff}. In this case, we show in Appendix~\ref{s:statcohpow}
that for moderately strong pulsed signals the distribution of $\mathcal{P}_n$ 
can be well approximated by a noncentral $\chi^2$-distribution \citep[][]{Groth1975,Guidorzi2011}
with $2$ degrees of freedom. Thus, in the perfect-match case 
(the pulsar parameters $f$, $\dot f$, and sky position are
precisely known), the first two moments are approximately given by
\begin{subequations}
\begin{align}
   &E_p\left[\mathcal{P}_n\right] \approx 2 + 2 p^2 N \left|\gamma_n\right|^2\,, \label{e:EpPn}\\
   &Var_p\left[\mathcal{P}_n\right] \approx 4 + 8 p^2 N \left|\gamma_n\right|^2\,, \label{e:VarpPn}
\end{align}
\end{subequations}
where $pN$ photons  are assumed to be ``pulsed'' and accordingly $(1-p)N$ photons 
are  ``non-pulsed'' (i.e., background). 
Thus, the second summand in \Eref{e:EpPn} represents the noncentrality 
parameter.\footnote{A random variable X following
a non-central $\chi^2$-distribution with $2$ degrees of freedom and noncentrality parameter~$\lambda$,
has expectation value $2+\lambda$.}
We can also identify the amplitude signal-to-noise ratio (S/N) at the $n$th harmonic, 
$\theta_{\mathcal{P}_n}$, as
\begin{equation}
  \theta_{\mathcal{P}_n}^2  
    = \frac{E_p \left[ \mathcal{P}_n \right] - E_0\left[ \mathcal{P}_n \right]}{\sqrt{ Var_0 \left[ \mathcal{P}_n \right]}} 
    \approx p^2\,N \, |\gamma_n|^2\,.
   \label{e:thetan}
\end{equation}
Therefore, by comparison to \Eref{e:EpPn} 
the noncentrality parameter is just~$2\theta_{\mathcal{P}_n}^2$.

A similar calculation for $Q_M$, based on the above relations shows that if $p=0$,
\begin{equation}
  E_0 \left[ Q_M \right] = 2M \,,\quad 
  Var_0 \left[ Q_M \right] = \frac{4}{K^4}\sum_{n =1}^M |\alpha_n|^4 \,,
\end{equation}
and for $p>0$, one obtains
\begin{equation}
  E_p \left[ Q_M \right] \approx 2M + \frac{2\,p^2\;N}{K^2} \, \sum_{n=1}^M |\alpha_n|^2 |\gamma_n|^2 \,.
\end{equation}
Thus, the amplitude S/N~$\theta_{Q_M}$ for the test statistic $Q_M$ can be expressed as
\begin{equation}
  \theta_{Q_M}^2 \approx  \frac{p^2\;N \, \sum_{n = 1}^M |\alpha_n|^2 |\gamma_n|^2}{\sqrt{\sum_{n = 1}^M |\alpha_n|^4}} \,.
  \label{e:SNR-Q}
\end{equation}
A similar expression has been derived by \citet{Bickel+2008}
who used this parameter as an approximate measure of the sensitivity 
of the test statistic~$Q_M$, since the larger the S/N $\theta_{Q_M}$ the higher
the probability of detection. However, it is only an approximate sensitivity measure,
because any meaningful sensitivity comparison must be done at
fixed probability of false alarm as will be described below.
\Eref{e:SNR-Q} also shows that the S/N is maximized 
if $|\alpha_n|^2 \propto |\gamma_n|^2$, i.e., when the 
template pulse profile~$\alpha_n$ perfectly matches the $\gamma_n$, 
representing the signal pulse profile. However, as \citet{Bickel+2008} correctly note, 
practical blind searches  can only
test for a finite-dimensional class of template pulse profiles.

A particularly simple template profile for a given value of $M$ is
\begin{equation}
  |\alpha_n|=
  	\begin{cases}
	1, &  n\leq M \\
	0, &  n>M 
	\end{cases}\,.
  \label{e:template1}
\end{equation}
With this choice, $Q_M$ measures the coherent Fourier power summed over the 
first $M$ harmonics, which we therefore refer to as \emph{incoherent harmonic summing}.
The resulting statistic is also known as $Z_M^2$ \citep{Buccheri+1983},
\begin{equation}
 Z_M^2  = \sum_{n=1}^M \mathcal{P}_n \,.
\end{equation}
Maximizing $Z_M^2 $ over different values of $M$ 
as $H = \max_{1 \leqslant M \leqslant 20} \left( Z_M^2 - 4M + 4  \right)$
also recovers the widely used $H$-test by \citet{deJaeger+1989}.

The template of \Eref{e:template1} has the additional benefit that
the statistical distribution of $Z_M^2$ is known analytically. 
Therefore, we use this to obtain realistic sensitivity scalings
for such coherent test statistics. Since $\mathcal{P}_n$ is
$\chi_2^2$-distributed\footnote{We use the notation $\chi^2_{k}$ to indicate a
$\chi^2$-distribution with $k$ degrees of freedom.}, it follows that $Z_M^2$ is distributed 
as $\chi_{2M}^2$. Thus, one obtains
\begin{equation}
  E_0\left[ Z_M^2 \right] = 2M \,,\quad 
  Var_0 \left[ Z_M^2 \right] = 4M \,,
\end{equation}
and
\begin{equation}
  E_p\left[ Z_M^2 \right] \approx 2M + 2\theta_M^2\sqrt{M} \,.
\end{equation}
Correspondingly, the S/N $\theta_M$ is written as
\begin{equation}
  \theta_M^2 =  \frac{1}{\sqrt{M}} \sum_{n=1}^M \theta_{\mathcal{P}_n}^2 
   \;=\;  \frac{p^2\;N}{\sqrt{M}}\, \sum_{n=1}^M |\gamma_n|^2 \,.
   \label{e:SNRM}
\end{equation}

In the Neyman--Pearson sense, 
we define \emph{search sensitivity} from the \emph{lowest threshold pulsed fraction}
required to achieve a certain detection probability~$P_{\rm DET}^{\ast}$
for a given number of photons~$N$ and at given false alarm probability~$P_{\rm FA}^{\ast}$.
For $Z_M^2$ the false alarm 
probability is computed as
\begin{equation}
  P_{\rm FA}(Z_{M,\rm{th}}^2) = \int_{Z_{M,\rm{th}}^2}^{\infty} \; \chi^2_{2M}(Z_{M}^2;0)\; d Z_{M}^2 \,,
  \label{e:pfa}
\end{equation}
where $\chi^2_{k}(X;\lambda)$ denotes the probability density function for the $\chi^2_k$-distributed variable $X$ 
with noncentrality parameter $\lambda$.
The probability of detection for a noncentrality parameter of $2\theta_M^2\sqrt{M}$ is 
\begin{equation}
  P_{\rm DET}(Z_{M,\rm{th}}^2,2\theta_M^2\sqrt{M}) = \int_{Z_{M,\rm{th}}^2}^{\infty} \; \chi^2_{2M}(Z_{M}^2; 2\theta_M^2\sqrt{M})\; d Z_{M}^2 \,.
  \label{e:pdet}
\end{equation}
The minimum detectable pulsed-fraction threshold for summing coherent power from $M$ harmonics,
$p_{{\rm coh}, M}$, is obtained by first inverting \Eref{e:pfa} to get the
threshold test-statistic value~$Z_{M,\rm{th}}^2(P_{\rm FA}^{\ast})$, 
which in a second step is substituted in \Eref{e:pdet}
to numerically find the required threshold S/N:
\begin{equation}
	\theta_M^{\ast} = \theta_M(P_{\rm FA}^{\ast},P_{\rm DET}^{\ast}) \,.
\end{equation}	
Finally, \Eref{e:SNRM} can be used to convert the threshold S/N $\theta_M^{\ast}$ 
into~$p_{{\rm coh}, M}$, which defines the coherent search sensitivity as
\begin{equation}
  p_{{\rm coh},M}^{-1} 
  = \frac{\sqrt{N} }{M^{1/4}\; \theta_M^{\ast}} \,  
  \left[ \sum_{n=1}^M |\gamma_n|^2 \right]^{1/2}\,.
  \label{e:searchsens}
\end{equation}
Assuming the overall photon count rate, $\mu = N/T_{{\rm coh},1}$, is constant 
throughout the entire coherent integration time, $T_{{\rm coh},1}$ then the 
search sensitivity 
increases with the well-known square-root scaling of~$T_{{\rm coh},1}$,
\begin{equation}
  p_{{\rm coh},M}^{-1} 
  = \frac{\sqrt{\mu\, T_{{\rm coh},1}} }{M^{1/4}\; \theta_M^{\ast}} \,  
  \left[ \sum_{n=1}^M |\gamma_n|^2 \right]^{1/2}\,.
  \label{e:searchsens2}
\end{equation}
Thus, we have obtained an expression for the search sensitivity,
separating the two effects of photon count rate (or integration time)
and pulse profile shape. Regarding the latter effect,
\Eref{e:searchsens2} reveals that the sensitivity
only improves with including higher harmonics (i.e. increasing~$M$) 
if the pulse profile shape is such that $\bigl(\sum_{n=1}^M |\gamma_n|^2\bigr)^{1/2}$ 
increases more quickly than the ``statistical penalty'' 
factor~$M^{1/4}\,\theta_M^{\ast}$.
While this is true for the narrowest possible pulse profile (a
Dirac delta function), we show below that the same does not hold in general
for typical gamma-ray pulsar profiles.

\subsection{Effects of Pulse Profile on Sensitivity}

From \Eref{e:searchsens2} in the previous section, we have seen how
the sensitivity for pulsation detection depends on the shape of the
pulse profile, represented by the Fourier coefficients $\gamma_n$.
Therefore, it is instructive to examine the change in sensitivity as
a function of the number of harmonics~$M$ for some exemplary profiles.
Thus, we consider the following ratio,
\begin{equation}
  \frac{p_{{\rm coh},M}^{-1}}{p_{{\rm coh},1}^{-1}} 
  = \frac{\theta_1^{\ast}}{M^{1/4}\; \theta_M^{\ast}} \;\frac{1}{|\gamma_1|}\;
  \left[ \sum_{n=1}^M |\gamma_n|^2 \right]^{1/2}\,,
  \label{e:searchsens3}
\end{equation}
which compares in the statistical sense the search sensitivity of 
including $M$ harmonics, compared to using the fundamental only 
(in absence of any computational constraints).

\begin{figure}
	\centering
		\includegraphics[width=0.99\columnwidth]{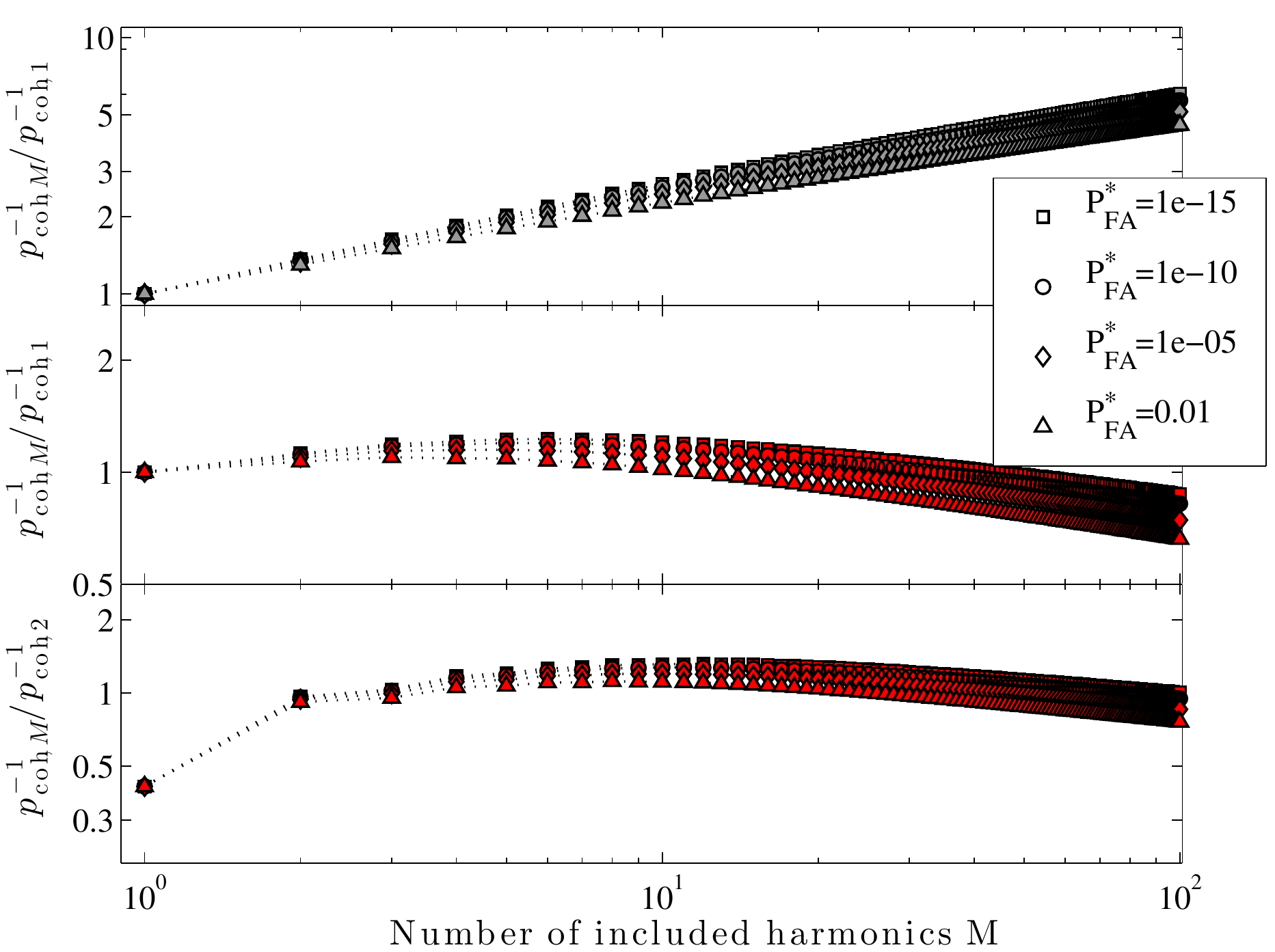}
	\caption{
Sensitivity as a function of the number of harmonics $M$ included 
in absence of computational constraints and for three 
different pulse-profile models.
In each panel, we fixed the detection probability 
$P_{\rm DET}^{\ast} = 90\%$ and the four curves correspond to
different values of false alarm probability~$P_{\rm FA}^{\ast}$ as shown by the
legend.
The upper panel is for a Dirac delta function pulse profile 
(implying equal Fourier power at all harmonics).
The middle panel is for a typical pulse profile,  
obtained from the known gamma-ray pulsars  by averaging those profiles that are
mostly
single-peaked (i.e. the $\gamma_n$ values shown in the 
bottom left panel in \Fref{f:2pc-harm}).
The bottom panel is also for a realistic pulse profile,  
obtained from the known gamma-ray pulsars  by averaging those profiles that are
mostly
two-peaked (i.e. the $\gamma_n$ values shown in the 
bottom right panel in \Fref{f:2pc-harm}). Since for these profiles
the Fourier power $|\gamma_2|^2$ is highest at the second harmonic ($n=2$),
in this plot the vertical axis shows the sensitivity compared to a blind search
which would report the highest
detection significance at the second harmonic (i.e. ``misidentify'' the
fundamental).
}
\label{f:coh-sens-scale}
\end{figure}

In the ideal case, where all harmonics have equal power~$|\gamma_n|^2=1$,
the pulse profile is a Dirac delta function as described above. 
In this case, $\bigl(\sum_{n=1}^M |\gamma_n|^2\bigr)^{1/2} = M^{1/2}$, 
and the sensitivity is a monotonically increasing function of $M$ 
at fixed detection probability, $P_{\rm DET}^{\ast}$, and
fixed false alarm probability, $P_{\rm FA}^{\ast}$.
To illustrate this, consider the following example,
assuming that $P_{\rm FA}^{\ast}=1\%$ and $P_{\rm DET}^{\ast}=90\%$. 
Then, to good approximation, the corresponding S/N threshold $\theta_M^{\ast}$ 
can be described by
\begin{equation}
   \theta_M^{\ast} \approx  \left(3.715 + \frac{4.987}{\sqrt{M}}\right)^{1/2} \,.
\end{equation}
Hence, with increasing~$M$, the threshold S/N $\theta_M^{\ast}$ decreases
and becomes constant in the limit of large~$M$, in which case the
statistical penalty factor ($M^{1/4}\,\theta_M^{\ast}$) becomes $\propto M^{1/4}$. 
Since this scaling is slower than the pulse profile factor 
$\bigl(\sum_{n=1}^M |\gamma_n|^2\bigr)^{1/2} = M^{1/2}$ in this case,
the sensitivity is monotonically increasing with $M$.
This is also shown in \Fref{f:coh-sens-scale}, using
the exact values for $\theta_M^{\ast}$ that we calculated numerically.

%---------------------------------------------------------------------------------------------------
%---------------------------------------------------------------------------------------------------

%---------------------------------------------------------------------------------------------------
\begin{figure}
	\centering	
		\subfigure
		{\includegraphics[width=\columnwidth]{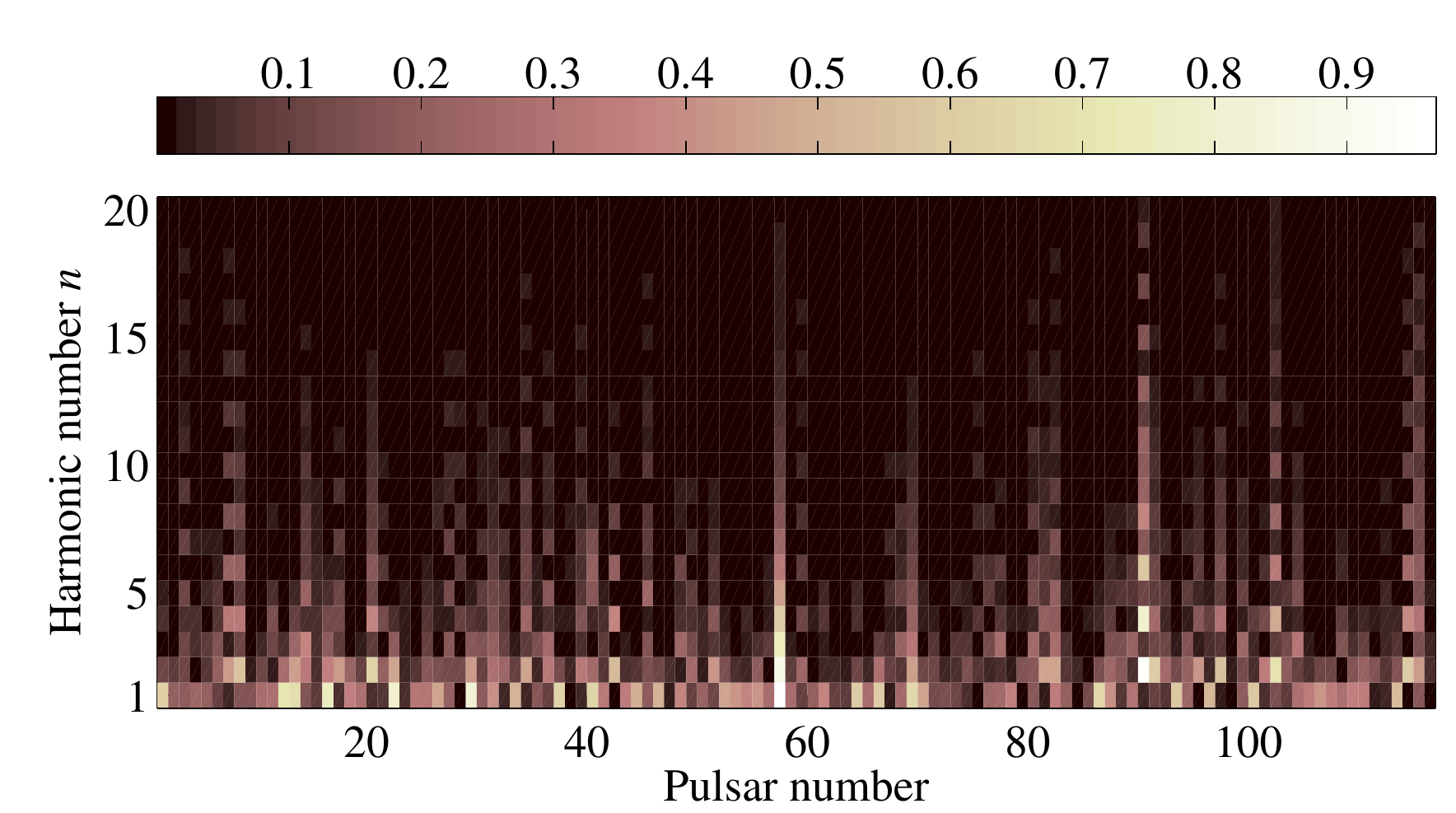}}\\
		\subfigure
		{\includegraphics[width=0.48\columnwidth]{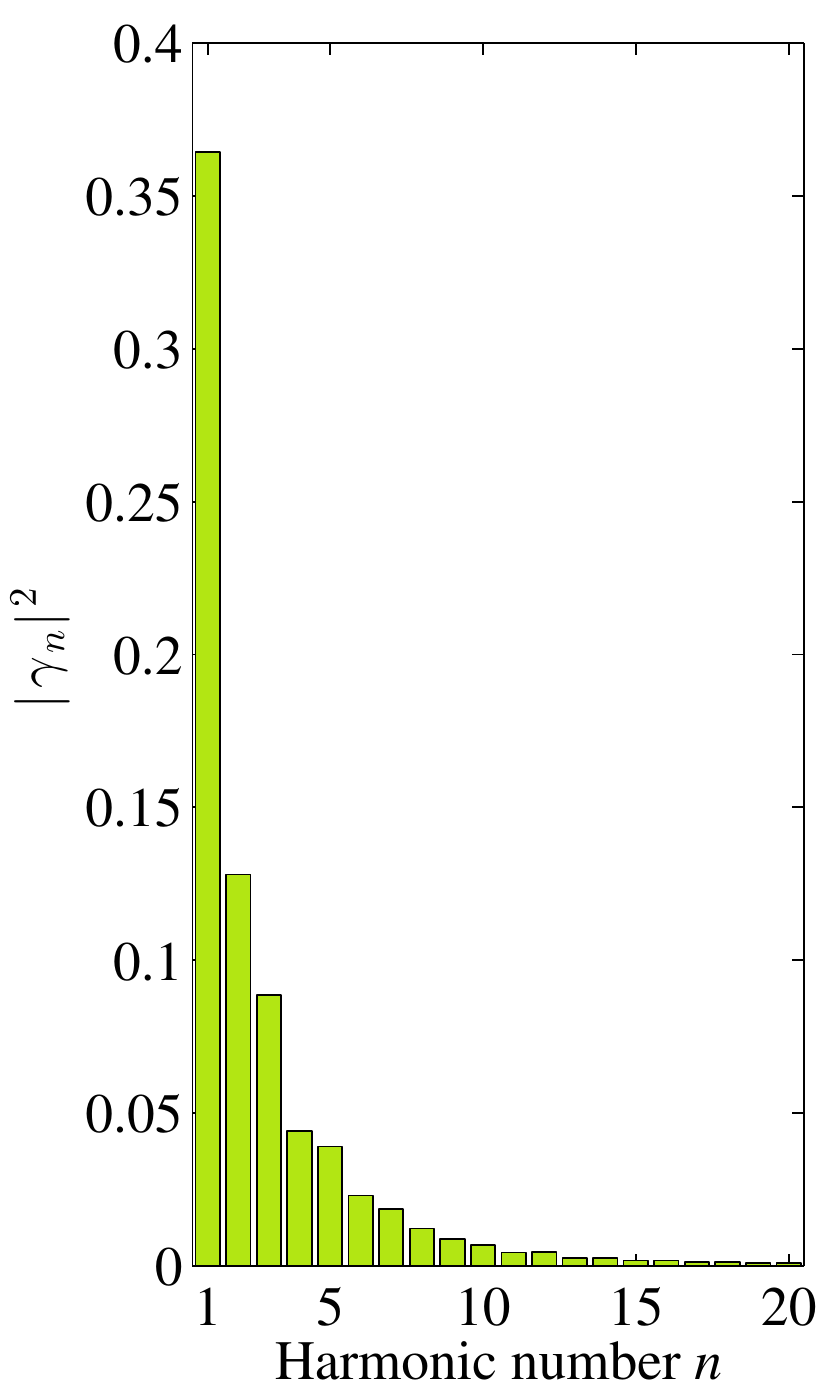}}\;\;
		\subfigure
		{\includegraphics[width=0.48\columnwidth]{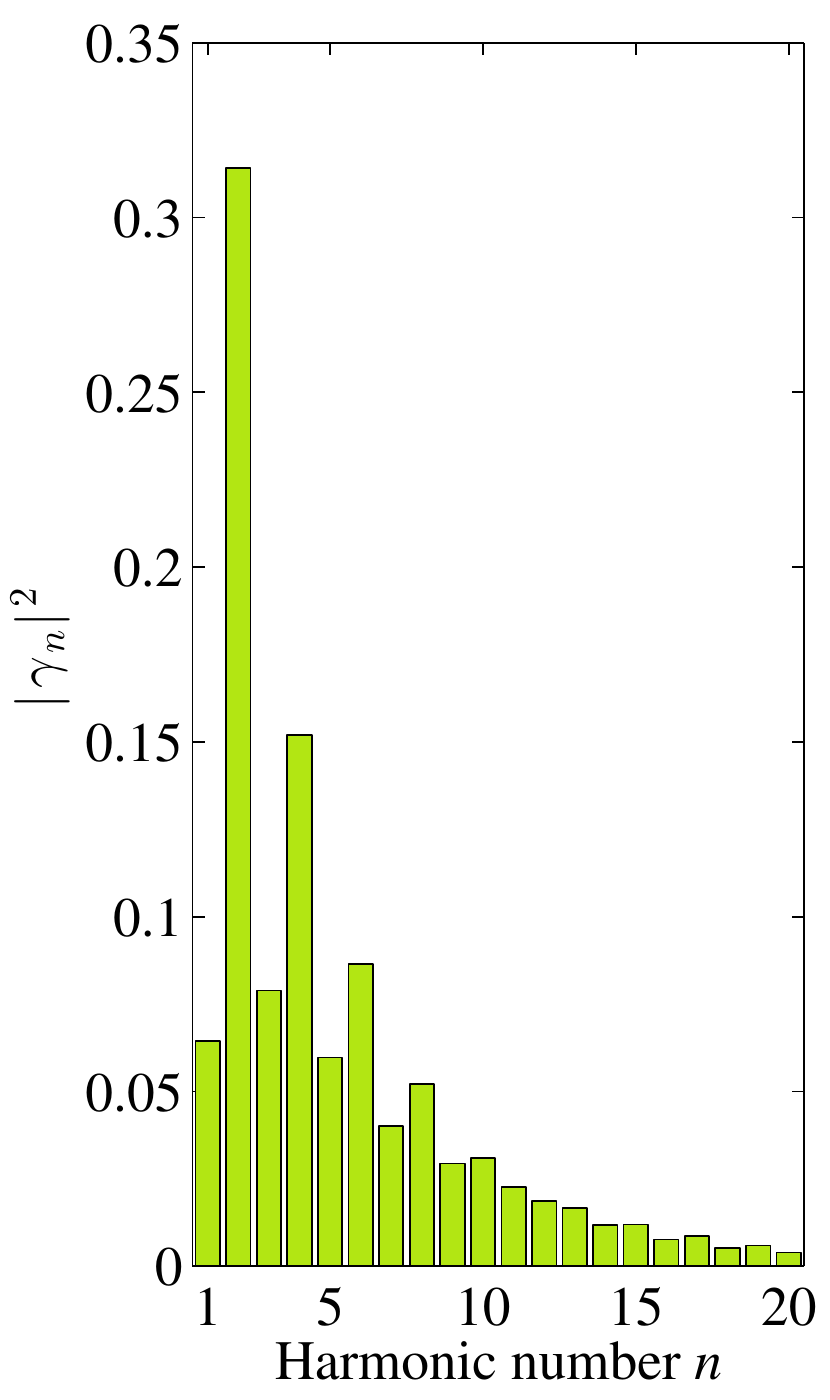}}
\caption{
Harmonic analysis of pulse profiles of the $117$ gamma-ray pulsars
in the second \Fermi{} LAT pulsar catalog~\citep{Fermi2PC}.
\emph{Top panel:} Fourier power~$|\gamma_n|^2$ (color-coded) at 
the $n$th harmonic (vertical axis) for each of the 117 pulsars
(horizontal axis).
\emph{Bottom left panel:} Fourier power~$|\gamma_n|^2$ at the $n$th
harmonic averaged over the 76 out of the 117 pulsars, whose power at the fundamental is highest
(mostly single-peaked profiles).
\emph{Bottom right panel:}  Fourier power~$|\gamma_n|^2$ at the $n$th
harmonic averaged over 41 out of the 117 pulsars, whose power at the second harmonic is highest
(mostly two-peaked profiles).
\label{f:2pc-harm}
}
\end{figure}
%---------------------------------------------------------------------------------------------------

To obtain a more realistic signal pulse-profile model, we considered those
of the known gamma-ray pulsars. We carried out a harmonic analysis of
the pulse profile shapes of the $117$ known gamma-ray pulsars 
listed in the second \Fermi{} LAT pulsar catalog~\citep{Fermi2PC}
and computed their Fourier coefficients,~$\gamma_n$.
These are shown in \Fref{f:2pc-harm} (top panel) and illustrate that for most
of the known gamma-ray pulsars the largest fraction of Fourier power is 
typically in a single harmonic that is either the first (mostly single-peaked profiles) 
or the second (mostly two-peaked profiles). Therefore, before computing 
an average profile (by averaging the $|\gamma_n|$), it makes sense to 
divide the pulsars into these two groups (based on whether or not $|\gamma_1| >  |\gamma_2|$).
These results, separately for each group, are displayed 
in the two bottom panels of \Fref{f:2pc-harm}.

We use the resulting two sets of coefficients $\gamma_n$ to calculate
the sensitivity scaling with $M$ from \Eref{e:searchsens2}
as also shown in \Fref{f:coh-sens-scale}.
Notice that for the typical pulse profiles, in contrast to the Dirac delta pulse-profile, when summing 
more than a certain number of harmonics, the sensitivity starts to decrease
(at fixed $P_{\rm DET}^{\ast}$ and $P_{\rm FA}^{\ast}$).
This is because the Fourier powers~$|\gamma_n|^2$ at the higher
harmonics become vanishingly small and thus effectively 
only contribute ``noise'' when summed (i.e.
the statistical penalty factor cannot be overcome anymore).

These results also illustrate the success of the $H$-test for targeted pulsation 
searches in gamma-ray data with known pulsar ephemerides, because this test maximizes 
the Fourier power sums over the first $20$ harmonics. Maximizing only over fewer harmonics 
could likely already be sufficient (or even be more sensitive due to the reduced trials factor) 
in most cases, as suggested by \Fref{f:coh-sens-scale}.
Besides, further improvements over the $H$-test could also be achieved
by employing one or more template profiles~$\alpha_n$ that are more 
representative of the typical gamma-ray profile (than the delta function) 
to compute the $Q_M$ test statistic. Using the average profile from the known pulsars 
from above for this seems the simplest first step. While also conducting
a principal component analysis appears worthwhile, we defer
a detailed study of this to future work.

So far, we have not considered the computational costs involved,
which is only justifiable for computationally inexpensive targeted searches.
In contrast, blind searches are limited by computational power. Therefore, 
in the following section, we will revisit the efficiency of harmonic summing 
under the constraint of a fixed computational cost.

\subsection{Grid-point Counting for Coherent Search}\label{s:cohmetric}

In blind searches, the pulsar's rotational
and positional parameters are unknown a priori. Therefore, one has to construct
a grid in the multidimensional search parameter space that is explicitly searched, 
i.e., the test statistic is to be computed at each grid point. 
Therefore the question arises: 
What is the most efficient scheme for constructing the search grid?
If grid points are placed too far apart potential pulsar signals might be missed. 
On the other hand, it is highly inefficient to place grid points too closely
together, because of redundancy resulting from 
strongly correlated nearby grid points. The problem of constructing efficient search 
grids has been intensively studied in the context of gravitational-wave searches
\citep[see, e.g.,][]{bccs1:1998,bc2:2000,prix:2007ks,PletschAllen2009,Pletsch2010}
and we employ some of these concepts here.

The key element is a distance \emph{metric} on the search space \citep{Sathy1:1996,owen:1996me}.
The metric provides an analytic geometric tool measuring the expected fractional loss 
in squared S/N for any given pulsar-signal location at a nearby grid point. 

Let the vector $\vDoppler_\sig$ collect the actual pulsar signal parameters.
In a blind search for isolated pulsars, this vector is at least 
four-dimensional, \mbox{$\vDoppler_\sig =(f_\sig,\dot f_\sig,\alpha_\sig,\delta_\sig)$}.
For simplicity, we begin by considering the metric at the fundamental harmonic
($n=1$). As will be shown below, it is subsequently straightforward to generalize the results
to higher harmonic orders. 
Following \Eref{e:thetan}, let $\theta_{\mathcal{P}_1}(\vDoppler_\sig)$ denote the S/N for the perfect-match 
case, i.e., at the signal parameter-space location.
In a blind search the signal parameters generally will not coincide with a 
grid point~$\vDoppler$, but will typically have some offset,
\begin{equation}
   \Delta\vDoppler = \vDoppler - \vDoppler_\sig \,.
\end{equation}
These offsets lead to a (time-dependent) residual phase 
\mbox{$\phi(t;\vDoppler) - \phi(t;\vDoppler_\sig)$}
and therefore a fractional loss in squared S/N results, which is commonly
referred to as \emph{mismatch},
\begin{equation}
  m(\Delta \vDoppler) = 1 - \frac{ \theta^2_{\mathcal{P}_1}(\vDoppler) 
  \;\;\;\;}{ \theta^2_{\mathcal{P}_1}(\vDoppler_\sig)} 
     = 1-\frac{ \theta^2_{\mathcal{P}_1}(\vDoppler_\sig + \Delta\vDoppler)}{ \theta^2_{\mathcal{P}_1}(\vDoppler_\sig)} \,.     
   \label{e:mmcoh}
\end{equation}
The metric is obtained from a Taylor expansion of the mismatch
to second order in the offsets $\Delta\vDoppler$ at the signal location~$\vDoppler_\sig$,
\begin{equation}
   \mm(\Delta \vDoppler) \approx \sum_{k,\ell} G_{k\ell}\,\Delta\Doppler^k 
   \,\Delta\Doppler^\ell 
   + \mathcal{O}(\Delta\Doppler^3)\,,
   \label{e:mm-metric}
\end{equation}
This equation defines a positive definite metric tensor~$G$ with components $G_{k\ell}$,
where $k$ and $\ell$ label the tensor indices. 
In Appendix~\ref{s:appcohmet}, we derive explicit expressions for the 
coherent metric for a simplified phase model that is appropriate for the purpose 
of grid construction. We also find that the resulting  metric tensor~$G$ is diagonal,
which greatly simplifies the grid construction.
The results of this derivation will therefore be used in what follows.

As noted by \citet{PrixShaltev2012}, the probability distribution of signal 
mismatches in a given search grid constructed with a certain maximal 
mismatch~$m$ depends on the structure and dimensionality of the search 
parameter space. The corresponding average mismatch in each dimension, 
$\xi\,m$, 
will generally be smaller by a characteristic geometric factor $\xi\in(0,1)$,
depending on the actual search-grid construction.
For example, for hyper-cubical lattices, $\xi$ is known to be $\xi=1/3$. 
In order to construct a hyper-cubical grid in 
which the maximum mismatch due to an offset in each parameter is $m$, then the 
grid point spacing in each parameter should be,
\begin{equation}
 \Delta \Doppler^k = 2 \sqrt{\frac{m}{G_{kk}}} \,.
 \label{e:parspacing}
\end{equation}
Denote by $\mathcal{U}$ the four-dimensional parameter space, spanned 
by $\vDoppler$, which is to be searched. Thus, when searching for pulsars with 
spin frequencies in the range $[0,f_{\rm max}]$, with spin-down rates in the
range $[\dot{f}_{\rm max},0]$, and whose sky location is confined by the LAT 
to a region of area $A_{\rm sky}$, the proper volume~$U$ can be written as
\begin{equation}
\mathcal{U} = f_{\rm max} \left|\dot{f}_{\rm max}\right| A_{\rm sky}\,.
\label{e:parspace} 
\end{equation} 
In principle, the metric coefficients (and hence also the grid point spacings) 
can 
vary throughout the parameter space. Indeed, for the metrics considered in this 
work, the grid point spacing in the sky dimensions depends on the spin 
frequency of the pulsar. In order to avoid having to construct a separate sky 
grid for each
search frequency value, we adopt the conservative approach of using the highest 
frequency searched~$f_{\rm max}$ for the sky grid construction. 
The metric (and hence also the grid point spacing) becomes uniform throughout 
$\mathcal{U}$. 
The total number of search-grid points~$\mathcal{N}_{{\rm coh},1}$
for a coherent blind search over~$\mathcal{U}$ is therefore simply the product 
of the 
number of grid points in each dimension.
\begin{equation}
 \mathcal{N}_{{\rm coh},1} = \mathcal{U} \, \prod_k \frac{1}{\Delta \Doppler^k} = 
 \frac{1}{16} \, \mathcal{U} \, m^{-2} \sqrt{\det{G}}  \,,
\end{equation}
as $G$ is found to be diagonal.
In Appendix~\ref{s:appcohmet} we derive that
\begin{equation}
 \sqrt{\det G} = \frac{\pi^4}{\sqrt{135}} \; T_{{\rm coh},1}^3 \, f^2 \, r_E^2 
 \, \Psi(T_{{\rm coh},1}) \,,
 \label{e:detGcohText}
\end{equation}
where we defined,\footnote{We use the definition $\sinc(x)=\sin(\pi x)/(\pi x)$ throughout this manuscript.}
\begin{align}
 \Psi^2(T_{{\rm coh},1}) 
  = &\;\;\left[1 +  \sinc\left(\Omega_E\,T_{{\rm coh},1}/\pi\right) 
   - 2\,  \sinc^2\left(\Omega_E\,T_{{\rm coh},1}/2\pi\right)\right] \nonumber\\
     &\;\times \left[1 - \sinc\left(\Omega_E\,T_{{\rm coh},1}/\pi\right)\right] 
     \,,
\end{align}
and where we have denoted the Earth's orbital angular frequency as $\Omega_E = 2\pi/1{\rm yr}$, and the light travel-time from the Earth to the SSB as $r_{E}=1\textrm{AU}/c\sim 500$s.

\begin{figure}[t]
\centering
\includegraphics[width=1.0\linewidth]{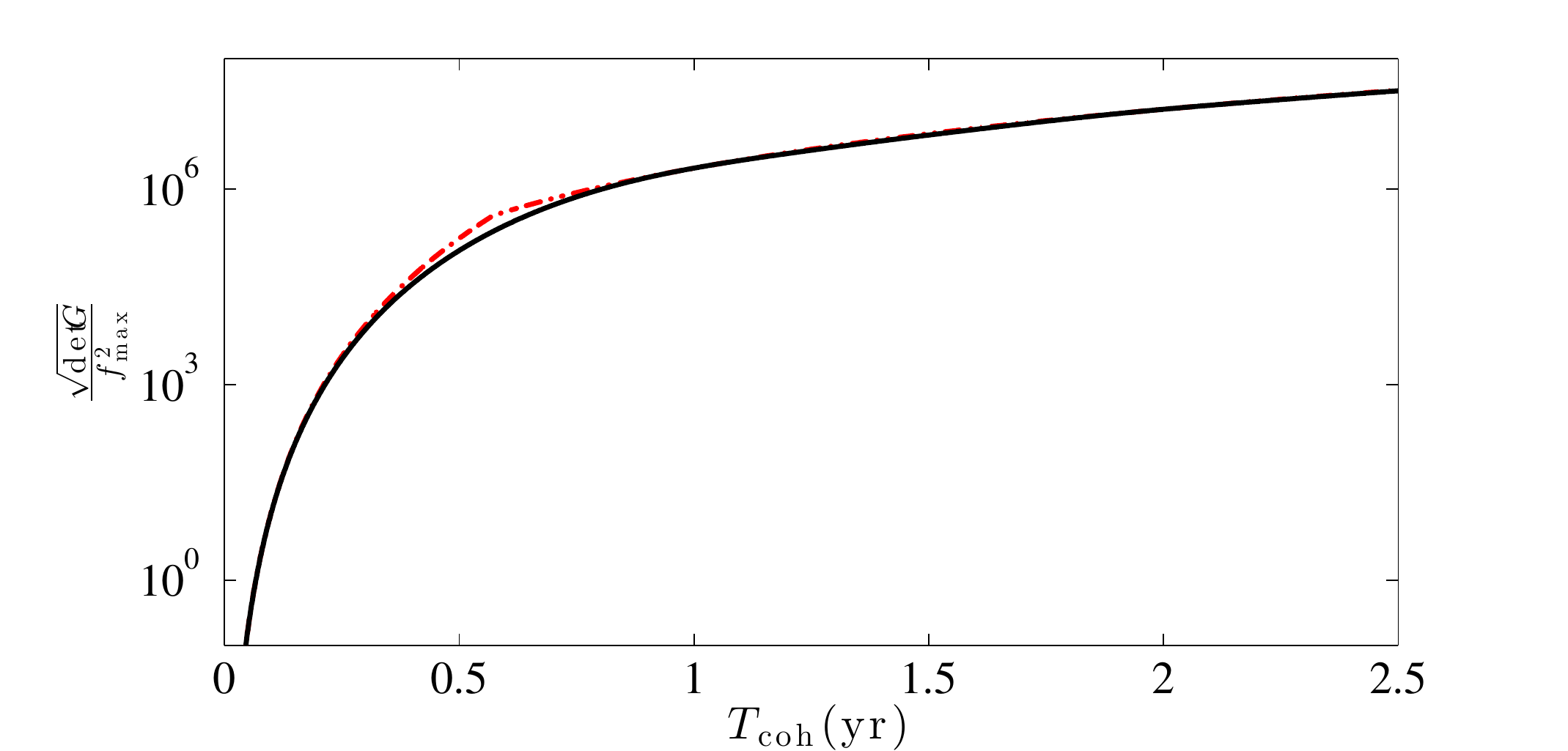}
\caption{
Scaling of the determinant of the coherent metric $G$ as function of the coherent 
integration time $T_{{\rm coh},1}$ (black solid curve). The red dot-dashed 
curve shows 
the model of the coherent metric determinant from the approximation 
of \Eref{e:CohMetPsi} used to estimate the computing cost scaling.
} 
\label{fig:coh_met_det}
\end{figure}

To analytically study the scaling of $\mathcal{N}_{{\rm coh},1}$ as a function 
of $T_{{\rm coh},1}$, 
the function $\Psi(T_{{\rm coh},1})$ can be well approximated by
\begin{equation}
\Psi(T_{{\rm coh},1}) \approx 
	\begin{cases}
	\frac{\Omega_E^3 \, T_{{\rm coh},1}^3}{12\sqrt{15}}, &  T_{{\rm coh},1} < 
	0.572 {\rm yr} \\
	1, &  T_{{\rm coh},1} \geq 0.572 {\rm yr} 
	\end{cases}\,.
	\label{e:CohMetPsi}
\end{equation}
The validity of this approximation is illustrated in \Fref{fig:coh_met_det}. 
%
%--------------------------------------------
%--------------------------------------------
%
Hence, the total number of grid points required in a coherent search is %
\begin{equation}
  \mathcal{N}_{{\rm coh},1} = 
  \frac{\pi^4}{48\sqrt{15}} \,  
  \left(\frac{\Omega_E^3}{12\sqrt{15}}\right)^{(a-3)/3} r_{\rm E}^2\, m^{-2} 
  \, f^2_{\rm 
  max} \, T_{{\rm 
  coh},1}^a\, \mathcal{U} \,, 
  \label{e:Ncohscaling}
\end{equation}
where 
\begin{equation}
a \approx 
	\begin{cases}
	6, &  T_{{\rm coh},1} < 0.572 {\rm yr} \\
	3, &  T_{{\rm coh},1} \geq 0.572 {\rm yr} 
	\end{cases}\,.
	\label{e:a-scaling}
\end{equation}
\Eref{e:Ncohscaling} tells us that
for coherent integration times much shorter than half a year
the sky metric components also still scale with $T_{{\rm coh},1}$,
such that $\mathcal{N}_{{\rm coh},1}$ increases approximately 
as~$T_{{\rm coh},1}^6$. After half a year of coherent integration the
sky metric components quickly approach the resolution saturation 
as the maximum baseline (1 AU) is reached, and thereafter become 
approximately independent of~$T_{{\rm coh},1}$.
Therefore $\mathcal{N}_{{\rm coh},1}$ scales only as~$T_{{\rm coh},1}^3$
in this regime.

\subsection{Coherent Search Sensitivity at Fixed Computing Cost}\label{s:compcost}

For computational efficiency, we use the fast Fourier transform (FFT) 
algorithm \citep{FFTW05} to scan
the $f$-dimension. There are two steps involved in calculating an FFT,
each
with an associated computational cost. Firstly, it is necessary to construct a
discrete time series by interpolating (e.g. by binning) the photon arrival times
into equidistant samples. The cost of this step is proportional to the number
of photon arrival times which must be interpolated. Secondly, the discrete time
series must be transformed into a discretely sampled frequency spectrum, using
the FFT algorithm. For a maximum frequency of $f_{\rm max}$, and a
coherent
integration time of $T_{{\rm coh},1}$ there are $f_{\rm max} T_{{\rm coh},1}$ 
frequency
samples, and the computational cost of calculating the FFT is proportional
to
$f_{\rm max} T_{{\rm coh},1} \log_2(f_{\rm max} T_{{\rm coh},1})$. We assume 
that the 
cost of calculating the FFT is much larger than the cost of creating the 
discrete time series. Compared to the cost 
of computing $\mathcal{P}_1$ explicitly for $N$ photon times at $f_{\rm max} T_{{\rm coh},1}$ 
frequencies, which is proportional to $N f_{\rm max} T_{{\rm coh},1}$, 
it is clear that the FFT method offers more efficiency provided 
$N \gg \log_2(f_{\rm max} T_{{\rm coh},1})$.  

The spacing of frequency samples output by the FFT is $1/T_{{\rm coh},1}$.
According to the metric [see \Eref{e:cohGff}] this implies a worst-case mismatch
due to frequency offsets of  \mbox{$m = G_{ff}/(4 T_{{\rm coh},1}^2) = \pi^2/12 = 0.82$},
which obviously also leads to a high average mismatch.
However, as we will discuss in \Sref{s:fdinterpolation}, it 
is possible to reduce this mismatch at almost no extra computational cost by 
interpolating the frequency spectrum. In the following derivations, we 
therefore separate the total mismatch $m_{\rm tot}$ into two components: a 
constant 
mismatch due to the frequency spacing, $m_f$ determined by the interpolation 
method used, which has a negligible effect on the overall computing cost; and 
the mismatch due to offsets in the remaining parameters, $m$, 
which can 
be freely varied to construct an optimal grid.

For every grid point in $\{\dot{f},\alpha,\delta\}$  an FFT must be computed, 
and hence the overall computation time for the search is simply the cost 
of calculating one FFT multiplied by the number of FFTs that must be
computed. 
The total cost, $C_{{\rm coh},1}$ (measured in units of time), is
\begin{equation}
C_{{\rm coh},1} = K_{\rm FFT} f_{\rm max} T_{{\rm coh},1} \log_2 (f_{\rm max} 
T_{{\rm 
coh},1} ) \frac{\mathcal{N}_{{\rm coh},1}}{\mathcal{N}_f}\,,
\end{equation}
where $K_{\rm FFT}$ is an implementation and computing hardware dependent 
constant, and where $\mathcal{N}_f$ is the number of frequency samples that 
would be 
calculated using a grid with an arbitrary maximum mismatch per dimension of $m$,
\begin{equation}
\mathcal{N}_f = \frac{f_{\rm max}}{2} \sqrt{\frac{G_{ff}}{m}} =  
\frac{\pi}{2\sqrt{3m}}\, f_{\rm max} T_{{\rm coh},1} \,.
\end{equation}
The total computational cost is therefore
\begin{equation}
  C_{{\rm coh},1} = K_{{\rm coh},a} \; m^{-3/2} \; T_{{\rm 
  coh},1}^{a} 
  \, 
  \log_2(T_{{\rm coh},1}f_{\rm max})  \,,
  \label{e:Ccoh1}
\end{equation}
where the constant $K_{{\rm coh},a}$ depends on $a$,
\begin{equation}
  K_{{\rm coh},a} = K_{\rm FFT} \, \frac{ \pi^3\,r_E^2\, f_{\rm 
  max}^2\,\mathcal{U}}{24\sqrt{5}} \left( \frac{\Omega_E^3}{12\sqrt{15}} 
  \right)^{(a-3)/3}  \,.
  \label{e:Kcoha}
\end{equation}
For a search grid constructed with maximum mismatch $m_{\rm tot} = m_f + 3m$, 
the search
sensitivity will scale with the average mismatch~$\langle m_{\rm tot} \rangle = 
\langle m_f \rangle + 3 \xi m$ 
as
$\sqrt{1 - \langle m_{\rm tot}\rangle}$ \citep{PrixShaltev2012}.
Thus, from \Eref{e:searchsens2} it follows that the search sensitivity 
without harmonic summing scales as
\begin{equation}
   p_{{\rm coh},1}^{-1} = \frac{\sqrt{(1-\langle m_{\rm tot} 
   \rangle)\;\mu\;T_{{\rm 
   coh},1}}}{\theta_1^{\ast}} \; |\gamma_1| \,.
   \label{e:pthcoh1a}
\end{equation}
For a computing cost~$C_{{\rm coh},1}$, \Eref{e:Ccoh1} can be used to obtain 
(numerically)
the maximum $T_{{\rm coh},1}$. Substituting this value of $T_{{\rm coh},1}$ in \Eref{e:pthcoh1a} 
finally yields the search sensitivity at the given computational cost.

\subsection{Efficiency of Harmonic Summing at Fixed Computing Cost}\label{s:harmsum}

Based on the results of the previous sections, we now investigate
the efficiency of incoherent harmonic summing under computational
cost constraints. More precisely,
we address the question of whether it is more efficient
in blind searches to sum $M$ harmonics, or to instead 
use a longer coherent integration time without harmonic summing
at the same computing cost.

Thus, we consider the test statistic $Z_M^2$, which
incoherently sums Fourier powers $\mathcal{P}_n$ from $M$
higher harmonics. In Appendix~\ref{s:appcohmetharmsum} we derive
the parameter space metric for the $Z_M^2$ statistic, denoted by $\tilde G$, 
and find that $\sqrt{\det \tilde{G}} = r^4 \,\sqrt{\det G}$, 
where 
$r$ represents a refinement factor due to harmonic summing, and
$G$ is the metric tensor for $\mathcal{P}_1$ of \Eref{e:detGcohText}.
Therefore, to ensure equal sensitivity throughout the original parameter 
space\footnote{This constraint is imposed to eliminate any detection bias 
in favor of pulsars with low frequencies and frequency derivatives, 
allowing for estimates of the true astrophysical pulsar populations.}
the required number of grid points increases by the factor of~$r^4$ compared to using 
$\mathcal{P}_1$ only.
The value of~$r\geq1$ depends on the pulse profile~$\gamma_n$.
For a sinusoidal pulse profile ($|\gamma_{1}|=1/2$ and $|\gamma_{n>1}|=0$), 
obviously $r=1$ (i.e. no refinement), 
and for a Dirac delta function ($|\gamma_{n}|=1$), one finds $r \sim M$, 
as derived in \Eref{e:rsq}. In principle, one could construct a grid with $r^4 
\mathcal{N}_{{\rm coh},1}$ points, and calculate and sum $M$ values of 
$\mathcal{P}_n$ at each point, leading to the cost of a harmonic summing search 
being simply $Mr^4$ times greater than that of a coherent search at the 
fundamental frequency with the same coherent integration time. 

In practice, to 
utilize the efficiency of the FFT, it would be necessary to construct a 
sub-optimal grid in which the range in $f$ and $\dot{f}$ is extended by a 
factor of $M$, and the coherent powers summed appropriately over harmonics. The 
sky-grid in this case may still be constructed using the refinement factor $r$, 
leading to the computing cost being $M^2 r^2$ times $C_{{\rm coh},1}$ at the 
same coherent integration time. While this method may quickly become infeasible due to 
the amount of memory required, we use this only as a theoretically efficient 
method to compare to an equally costly search using only the fundamental 
harmonic power.

We here assume that the small extra cost of actually 
summing the $\mathcal{P}_n$ is negligible.\footnote{Note that this makes
the computing cost estimate generous in favor of the harmonic summing 
approach in this comparison.} The computational expense for incoherent
harmonic summing, $C_{{\rm coh},M}$, using the $Z_M^2$ statistic for a coherent
integration time~$T_{{\rm coh},M}$ becomes
\begin{equation}
  C_{{\rm coh},M} = K_{{\rm coh},a} \; m^{-3/2} \; T_{{\rm coh},M}^{a} 
  \, 
  M^2\,r^2\;\log_2(T_{{\rm coh},M}f_{\rm max}\,M)  \,.
    \label{e:CcohM}
\end{equation}
From \Eref{e:searchsens} above, we found that the search sensitivity 
of incoherent harmonic summing is given by
\begin{equation}
   p_{{\rm coh},M}^{-1} =  \frac{\sqrt{(1- \langle m_{\rm tot} 
   \rangle)\;\mu\;T_{{\rm 
   coh},M}}}{M^{1/4}\,\theta_M^\ast}\;
   \left[\sum_{n=1}^M |\gamma_n|^2 \right]^{1/2}\,.
   \label{e:pthcohM}
\end{equation}

Hence, to compare the search sensitivities $p^{-1}_{{\rm coh},1}$ and $p^{-1}_{{\rm coh},M}$
at fixed computing cost, in principle the following steps are required.
First, for a given computing cost~$C_{{\rm coh},1}$, 
Equations~\eref{e:Ccoh1} and \eref{e:pthcoh1a}
provide the corresponding coherence time~$T_{{\rm coh},1}$ 
and sensitivity~$p^{-1}_{{\rm coh},1}$, respectively. 
Second, by equating $C_{{\rm coh},1} = C_{{\rm coh},M}$, \Eref{e:CcohM} then
can be solved (numerically) for $T_{{\rm coh},M}$, which finally is used to 
obtain the sensitivity~$p^{-1}_{{\rm coh},M}$ from \Eref{e:pthcohM}. 
It should be noted that in comparing $p^{-1}_{{\rm coh},1}$ and $p^{-1}_{{\rm coh},M}$
the same values of $P_{\rm FA}^{\ast}$ and $P_{\rm DET}^{\ast}$ must be
assumed. We here also assume the same mismatch~$m$ in either case, 
because as shown in Appendix~\ref{a:optmismatchcoherent},
the optimal mismatch at fixed computing cost is independent
of coherent integration time, number of harmonics summed, and computing power available.
Notably, a similar result has been found previously by
\citet{PrixShaltev2012} in the context of gravitational-wave pulsar searches.

\begin{figure}[t]
\centering
  \includegraphics[width=0.99\columnwidth]{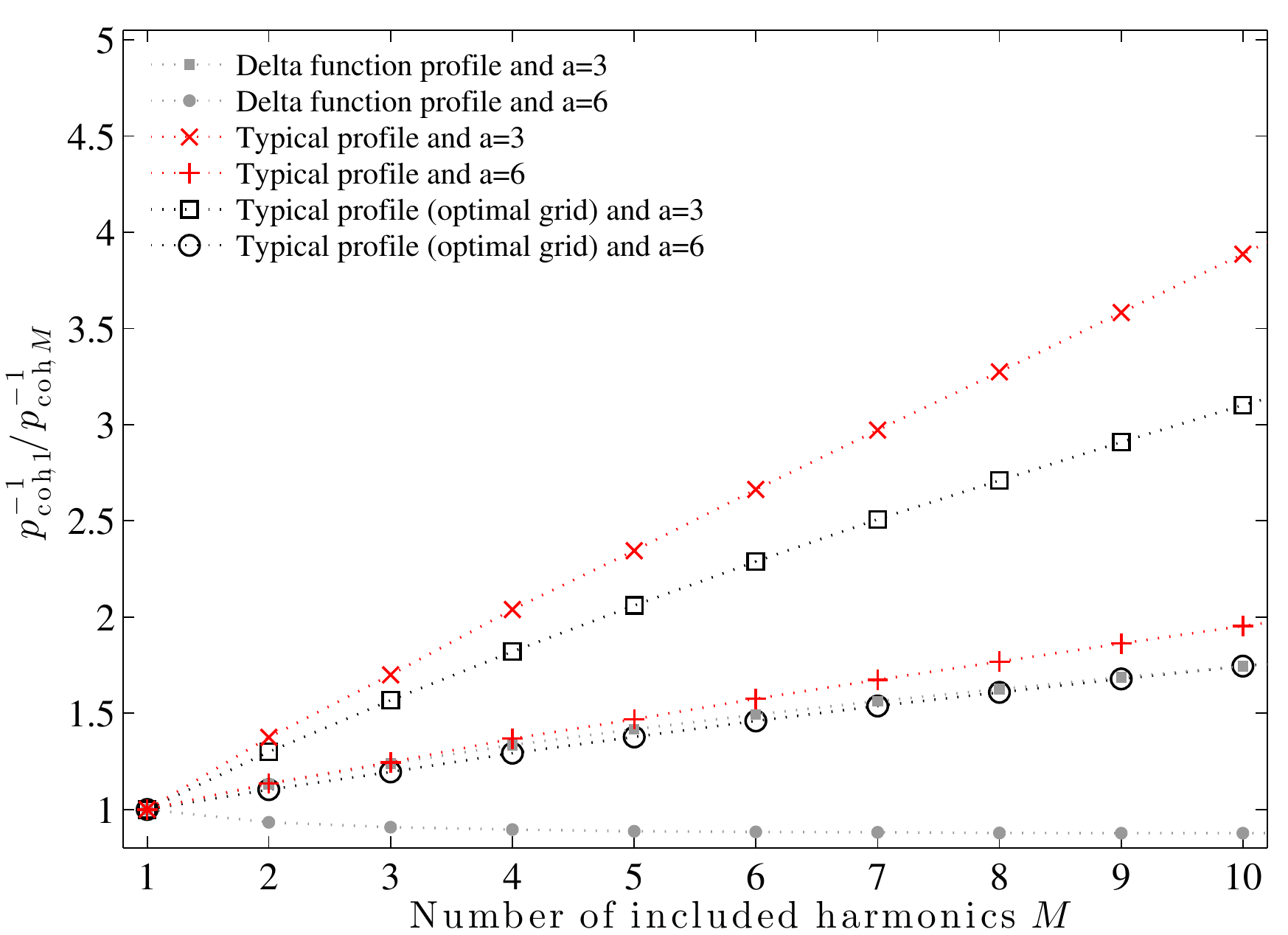}
   \caption{Ratio of search sensitivities $p^{-1}_{{\rm coh},1}$ (without harmonic summing) and
    $p^{-1}_{{\rm coh},M}$ (summing power from $M$ harmonics) at fixed computational cost.  
   The $Z_M^2$ test statistic used here, assumes a delta function pulse profile,
   so optimizing the search grid for this profile leads to the curves shown by
   the filled squares and circles. 
   The red crosses and pluses are for the same statistic and grid, but where
    the signal pulse profile is a more typical one (derived from averaging those of the known gamma-ray pulsars 
    of \Fref{f:coh-sens-scale}).
   The open squares and circles are for the same statistic, but using 
   the same typical signal pulse profile and a grid that
   is also optimized for that same pulse profile.
   For each case, the results of two different scalings of the computing cost with $T_{{\rm coh},M}^a$ 
   are shown, corresponding to $a=6$ and $a=3$ (see text for details). 
   While all points shown are for
   $P_{\rm DET}^{\ast} = 90\%$ and $P_{\rm FA}^{\ast} = 1\%$, they remain qualitatively
   similar for lower $P_{\rm FA}^{\ast}$ values, too.
\label{f:sens-ratio-harmsum}
}
\end{figure}

In the following, we describe an analytical approximation to the numerical approach 
above which we show to be sufficiently accurate for typical search setups.
This approximation is based on ignoring the slowly varying $\log_2$ factors 
in Equations~\eref{e:Ccoh1} and \eref{e:CcohM}, such that
\begin{equation}
  C_{{\rm coh},M} \sim K_{{\rm coh},a} \; m^{-3/2} \; T_{{\rm coh},M}^{a} \, 
  M^2\,r^2  \,.
    \label{e:CcohMapprox}
\end{equation}
Then from 
$C_{{\rm coh},1} = C_{{\rm coh},M}$, it immediately follows that
$T_{{\rm coh},M}$ must be shorter by the factor~$(M^2\,r^2)^{(1/a)}$,
\begin{equation}
  T_{{\rm coh},M} = T_{{\rm coh},1} \left(M^2\,r^2\right)^{-1/a} \,.
  \label{e:TcohM}
\end{equation}
We show in Appendix~\ref{a:AnaApproxHarmSum} that the $T_{{\rm coh},M}$
obtained from this approximation slightly \emph{overestimates} the 
sensitivity~$p^{-1}_{{\rm coh},M}$, while being accurate to within less than about 
$1\%$ for typical search setups.
Using \Eref{e:TcohM} to substitute~$T_{{\rm coh},M}$ in \Eref{e:pthcohM} one 
obtains for the ratio of search sensitivities,
\begin{equation}
  \frac{p^{-1}_{{\rm coh},1}}{p^{-1}_{{\rm coh},M}} =
   \; \frac{M^{1/4+1/a}\; r^{1/a}\; \theta_M^\ast}{\theta_1^\ast}  \; 
   |\gamma_1| \, \left[\sum_{n=1}^M |\gamma_n|^2 \right]^{-1/2} \,,
  \label{e:coh1-cohM-ratio}
\end{equation}
which remarkably is independent of $T_{{\rm coh},1}$ and  $T_{{\rm coh},M}$.
This sensitivity ratio \mbox{$p^{-1}_{{\rm coh},1}/p^{-1}_{{\rm coh},M}$}
of \Eref{e:coh1-cohM-ratio} is shown in \Fref{f:sens-ratio-harmsum} and is
found to be greater than unity for typical gamma-ray pulsars.
Only for unrealistically narrow pulse profiles (i.e. a Dirac delta function),
the sensitivity ratio can remain close to or slightly below unity.
It also should be pointed out that we obtained these results despite the generous
assumptions in favor of the harmonic summing approach. First,
we ignored the extra costs of summing the $M$ power values.
Second, we neglected the possible extra trials when one would maximize 
the test statistics over different~$M$. Third, the analytical approximation
of \Eref{e:TcohM} overestimates the true $T_{{\rm coh},M}$ 
(and hence the sensitivity $p^{-1}_{{\rm coh},M}$) as
we show by numerical evaluation in \Fref{f:Mscale-approx}.

%---------------------------------------------------------------------------------------------------
%---------------------------------------------------------------------------------------------------

Hence the basic moral is clear: 
For blind searches for isolated gamma-ray pulsars, whose sensitivity is limited
by computing power rather than the amount of available data, a more sensitive
search strategy is to employ a longer coherence time instead of using
incoherent harmonic summing at the same computational cost.

\section{Semicoherent test statistics}\label{s:semicoherent}

The key property of the semicoherent test statistics is that only pairs of photon arrival times ($t_j, t_k$)
whose separation \mbox{$\tau_{jk} = t_j - t_k$}, also called \emph{lag}, 
is at most~$T$ (which is much shorter than $T_{\rm obs}$) 
are combined coherently, otherwise incoherently.
Hence, we refer to~$T$ as  the \emph{coherence window} size and denote
by $R$ the ratio of total observational data time span $T_{\rm obs}$ of the 
semicoherent search and~$T$,
\begin{equation}
  R = T_{\rm obs}/T\,.
  \label{e:R}
\end{equation}
Compared to fully coherent methods, this semicoherent approach drastically reduces the computing
cost since fewer search grid points are required (due to the lower parameter-space resolution
as will be described in \Sref{s:semicohmetric}) at the expense of reduced search sensitivity. 
In \Sref{s:sccompcost} we argue that this tradeoff is a profitable one, because at fixed
given computing cost the overall search sensitivity of the semicoherent searches outperform
fully coherent searches restricted to data spans shorter than $T_{\rm obs}$ by the computational
constraints.

To derive a semicoherent test statistic, notice the (unnormalized) coherent Fourier 
power from \Eref{e:Pn} for the fundamental frequency (first harmonic) can also be 
written in the following form,
\begin{equation}
  \mathcal{P}_{1} \propto  \left| \sum_{j=1}^N w_j \; e^{-i \phi(t_j)}  \right|^2 
  = \sum_{j,k=1}^N \; w_j\; w_k  \; e^{-i [\phi(t_j)-\phi(t_k)]}\,.
  \label{e:P1tau}
\end{equation}
Thus, the semicoherent statistic~$S_1$ is formed by multiplying the terms in the above
double sum with a real \emph{lag window}~$\hat{W}_T(\tau_{jk})$, such that
\begin{equation}
  S_1 = \sum_{j,k=1}^N \; w_j\; w_k \; e^{-i [\phi(t_j)-\phi(t_k)]}\; \hat{W}_T(\tau_{jk}) \,,
  \label{e:S1b}
\end{equation}
where the lag window has an effective size~$T$,
\begin{equation}
   \int_{-\infty}^{\infty} \hat{W}_T(\tau) \; d\tau = T \,,
   \label{e:lagwinsize}
\end{equation}
and thus must fall off rapidly outside the interval $[-T/2,T/2]$. 
\citet{BlackmanTukey1958} were the first to consider power 
spectral estimators of the form of $S_1$, which can be seen as the Fourier transform 
of the lag-windowed covariance sequence \citep{Stoica2005}.
The semicoherent statistic~$S_1$ is just a more general version
of the classic Blackman-Tukey method \citep{BlackmanTukey1958}
in spectral analysis, e.g. if the phase model was simply $\phi(t_j)=2\pi f t_j$ only.
Hence, $S_1$ can also be seen as a local spectral average of $\mathcal{P}_1$ values
over neighboring frequencies weighted according to the frequency 
response of~$\hat{W}_T$ \citep{Stoica2005}.

As outlined in \citep{Pletsch+2012-9pulsars}, for special forms of the lag window, 
$S_1$ can also be obtained by summing time-windowed coherent power from 
overlapping subsets of data. This implies a lag window that must be always
positive semidefinite, because it is formed by the convolution of the time window 
with itself in this case \citep{Stoica2005}, whereas the
more general form as of \Eref{e:S1b} in principle can have arbitrary lag windows.

In general, the choice of lag-window function~$\hat{W}_T(\tau)$
has an impact on the sensitivity of the statistic~$S_1$. In tests 
with simulated LAT data, for the purpose of pulsation detection we found 
that the best sensitivity is provided by the simple rectangular lag window,
\begin{equation}
   \hat{W}_T^{{\rm\tiny rect}}(\tau) = 	
      \begin{cases}
	1, & |\tau| \leq T/2 \\
	0, & \text{otherwise}  \,.
 	\end{cases}
	\label{e:rectlagwin}
\end{equation}
which also allows for an efficient implementation as will be described in more
detail in \Sref{s:implementation}.
The usage of the rectangular lag window could also be motivated from the following
viewpoint. Considering the significant sparseness of the LAT data, typically all pairs of photon times
fall at different lags (for any practical sampling time, see \Sref{s:dft_form}). Therefore, one could argue that 
optimally (for minimum variance) all lags (i.e., all photon pairs) should be weighted equally when forming~$S_1$,
which is exactly what $\hat{W}_T^{{\rm\tiny rect}}(\tau)$ implements.
Thus, in the remainder of this manuscript we will keep using the
rectangular lag window $\hat{W}_T^{{\rm\tiny rect}}(\tau)$ to calculate~$S_1$.

\subsection{Statistical Properties }\label{s:semicohstats}

To examine the statistical properties of the semicoherent statistic, $S_1$, it is useful to rewrite \Eref{e:S1b} as
\begin{equation}
S_1 = \sum_{j=1}^N w_j^2 + 2 \sum_{j=1}^N\,\sum_{k=j+1}^N w_j w_k \, \cos[\phi(t_j) - \phi(t_k)]\, \hat{W}_T^{{\rm\tiny rect}}(\tau_{jk})\,.
\label{e:S1c}
\end{equation}
Under the null hypothesis, $p=0$ and assuming $N \gg 1$, 
we show in Appendix~\ref{s:statsemicohpow}
that $S_1$ follows a normal distribution, whose
first two moments of the noise distribution of $S_1$  are:
\begin{align}
   E_0[S_1] &=  \sum_{j=1}^N w_j^2 \,,\\
   Var_0[S_1] &=  2 \sum_{j=1}^{N} \sum_{k=j+1}^{N} \; w_j^2 \; w_k^2  \,\left[\hat{W}_T^{{\rm\tiny rect}}(\tau_{jk})\right]^2
\,,\label{e:Var0S1}
\end{align}
Now consider that the photon data contains a pulsed signal (i.e. $p > 0$)
with a pulse profile defined by Fourier coefficients~$\gamma_n$. 
Then the expectation value of $S_1$ is obtained as
\begin{align}
  E_p[S_1] &\approx E_0[S_1] \nonumber\\
  &\;\;+ 2\, E_p\left[\sum_{j=1}^N \sum_{k=j+1}^N w_j w_k \cos (\phi(t_j)-\phi(t_k)) \,\hat{W}_T^{{\rm\tiny rect}}(\tau_{jk})\,
  \right]\,.
  \label{e:EpS}
\end{align}
Thus, for $S_1$ we can identify the amplitude S/N $\theta_{S_1}$ as
\begin{align}
 \theta_{S_1}^2 &= \frac{E_p[S_1]-E_0[S_1]}{\sqrt{Var_0[S_1]}} \nonumber\\
 & = \frac{ \sqrt{2} E_p\left[\sum_{j=1}^N \sum_{k=j+1}^N w_j w_k \cos (\phi(t_j)-\phi(t_k))\hat{W}_T^{{\rm\tiny rect}}(\tau_{jk})\right]}{ 
  \sqrt{\sum_{j=1}^{N} \sum_{k=j+1}^{N} \; w_j^2 \; w_k^2 \; 
   \left[\hat{W}^{{\rm\tiny rect}}_T(\tau_{jk})\right]^2}}\,.
   \label{e:SNRforS}
\end{align}

To extract the scalings of the semicoherent S/N~$\theta_{S_1}$ in terms of the relevant
search parameters, we assume hard photon-selection cuts, i.e., binary photon weights,
for the remainder of this section. Then \Eref{e:S1c} reduces to
\begin{equation}
   S_1 = N  +  2 \sum_{j=1}^N \sum_{k=j+1}^N \; \cos [\phi(t_j)-\phi(t_k)] \;
    \hat{W}_T^{{\rm\tiny rect}}(\tau_{jk}) \,.
   \label{e:S1d}
\end{equation}
In this case, as derived in Appendix \ref{s:statsemicohpow}, the first two moments of the noise distribution are
\begin{equation}
   E_0[S_1]  =  N \,,\quad   Var_0[S_1] \approx  N^2 \, R^{-1} \,.
   \label{e:S_noise_moments}
\end{equation}
We show in Appendix \ref{s:statsemicohpow}, that for moderately strong signals 
the first two moments of the distribution of $S_1$ are approximately given by
\begin{subequations}
\begin{align}
E_p[S_1] &\approx N + p^2 N^2 \left|\gamma_1\right|^2 R^{-1} \,,\\
Var_p[S_1] &\approx  N^2 R^{-1} \left(1 + 2 p^2 N \left|\gamma_1\right|^2 R^{-1}\right) \,,\label{e:VarpS1ap}
\end{align}
\end{subequations}
and the squared S/N of \Eref{e:SNRforS} becomes
\begin{equation}
   \theta_{S_1}^2 \approx p^2\,N\,R^{-1/2}\, \left|\gamma_1 \right|^2\,.
   \label{e:SNRforSsimple}
\end{equation}

As shown in Appendix \ref{s:statsemicohpow},
the probability density function of~$S_1$ can
be approximated by a normal distribution with the above
expectation values and variances.
The sensitivity of a semicoherent search is the lowest
threshold pulsed fraction~$p$ for a given number of photons~$N$
and at given false alarm probability~$P_{\rm FA}^{\ast}$ 
to achieve a certain detection probability~$P_{\rm DET}^{\ast}$.
For a threshold $S_{1,\rm{th}}$ the false alarm probability is computed as
\begin{align}
  P_{\rm FA}(S_{1,\rm{th}}) &\approx \int_{S_{1,\rm{th}}}^{\infty} \; 
  \mathcal{N}\left\{S_1;E_0[S_1],Var_0[S_1]\right\}\; d S_1 \nonumber\\
  &\approx \frac{1}{2}\erfc\left(\frac{S_{1,\rm th} - E_0[S_1]}{\sqrt{2\,Var_0[S_1]}}\right) \,.
  \label{e:pfaS}
\end{align}
Where, in this context, $\mathcal{N}\left\{X;\mu,\sigma^2\right\}$ denotes a 
normal distribution with mean $\mu$ and variance $\sigma^2$, and should not be 
confused with the number of grid-points, $\mathcal{N}_{{\rm coh},1}$.
We compute the probability of detection using 
\mbox{$Var_p[S_1] \approx Var_0[S_1] (1+2p^2 N \left|\gamma_1\right|^2 R^{-1})$} as
\begin{align}
  &P_{\rm DET}(S_{1,\rm{th}}, \theta_{S_1}^2) \approx \int_{S_{1,\rm{th}}}^{\infty} \; \mathcal{N}\left\{S_1; E_p[S_1],Var_p[S_1]\right\} \; d S_1 \nonumber\\
  &\;\;\approx \frac{1}{2}\erfc\left\{\left( \frac{S_{1,\rm th} - E_0[S_1]}{\sqrt{Var_0[S_1]}} - \theta_{S_1}^2\right)
  \frac{1}{\sqrt{2+4p^2 N \left|\gamma_1\right|^2 R^{-1}}}\right\} \,.
  \label{e:pdetS}
\end{align}
The minimum detectable pulsed fraction is obtained by first inverting \Eref{e:pfaS} to get 
$S_{1,\rm{th}}(P_{\rm FA}^{\ast})$, which in a second step is substituted in \Eref{e:pdetS}
to obtain the threshold S/N~$\theta_{S_1}^{\ast}$ as
\begin{align}
   \theta_{S_1}^{\ast} &=
   \theta_{S_1}(P_{\rm FA}^{\ast},P_{\rm DET}^{\ast}) \nonumber\\
   &\approx\left[ \sqrt{2} \erfc^{-1}(2 P_{\rm FA}^{\ast}) -\sqrt{2+4p^2 N \left|\gamma_1\right|^2 R^{-1}}\, \erfc^{-1}(2 P_{\rm DET}^{\ast})\right]^{1/2} \,.
   \label{e:SNRthSemicoh}
\end{align}
Finally, using \Eref{e:SNRthSemicoh} one can convert \Eref{e:SNRforSsimple} into the 
threshold pulsed fraction~$p^{-1}_{{\rm scoh}, 1}$,
determining the semicoherent sensitivity as
\begin{equation}
  p^{-1}_{{\rm scoh}, 1} 
    = \frac{\sqrt{N}\, R^{-1/4}}{\theta_{S_1}^{\ast}} \,  |\gamma_1 | 
    = \frac{\sqrt{\mu\;T}\, R^{1/4}}{\theta_{S_1}^{\ast}} \,  |\gamma_1 | \,,
  \label{e:scsearchsens}
\end{equation}
where we used $N = \mu T R$. This reveals 
the square-root scaling with the coherence window size~$T$ 
and the expected fourth-root scaling with~$R$ of the 
semicoherent sensitivity. 
Furthermore using $R = T_{\rm obs}/T$, we can rewrite the 
previous equation as
\begin{equation}
      p^{-1}_{{\rm scoh},1} 
      = \frac{\sqrt{\mu}\, \left(T\;T_{\rm obs}\right)^{1/4}}{\theta_{S_1}^{\ast}} \,  |\gamma_1 | \,.
     \label{e:scsearchsens2}
\end{equation}
As a comparison, recall that the coherent
sensitivity as of \Eref{e:pthcoh1a}, $p_{{\rm coh},1}^{-1} \propto \sqrt{T_{{\rm coh},1}}$, increases 
with the square root of the coherent integration time~$T_{{\rm coh},1}$.
Here, \Eref{e:scsearchsens2} shows that the semicoherent sensitivity,
\mbox{$p^{-1}_{{\rm scoh}, 1} \propto \sqrt{(T\;T_{\rm obs})^{1/2}}$},
increases with the square root of the geometric mean of the
coherence window size~$T$ and the total observation time~$T_{\rm obs}$.

It should be noted that while the semicoherent method allows for the use of short lag-windows, 
in order to detect pulsations there is the additional requirement that there is at least 
one pair of pulsed photons which arrive within $T$ of each another. 
This sets a fundamental lower limit on $T$. But for typical pulsed fractions and photon arrival 
rates considered in this work, this lower limit is on the order of only a few hours.

\subsection{Grid-point Counting for Semicoherent Search}\label{s:semicohmetric}

To optimally construct the search grid for the semicoherent statistic~$S_1$,
it is necessary to re-evaluate the appropriate metric on parameter space. 
Analog to \Eref{e:mmcoh}, we define the mismatch for $S_1$ as the fractional loss
in semicoherent S/N squared,
\begin{equation}
  \bar{m} = 1 - \frac{ \theta^2_{S_1}(\vDoppler) \;\;\;}{ \theta^2_{S_1}(\vDoppler_\sig)} 
   = 1-\frac{ \theta^2_{S_1}(\vDoppler_\sig + \Delta\vDoppler)}{ \theta^2_{S_1}(\vDoppler_\sig)} \,.     
   \label{e:mmsemicoh}
\end{equation}
Expanding the mismatch~$\bar{m}$ to second order in the offsets~$\Delta\vDoppler$
as in \Eref{e:mm-metric} yields the semicoherent metric tensor~$\bar{G}$,
\begin{equation}
  \bar{m} = \sum_{k,\ell}\bar{G}_{k\ell}\Delta u^k \Delta u^\ell + \mathcal{O}(\Delta\Doppler^3)\,.
\end{equation}
We derive the components $\bar{G}_{k\ell}$ from the phase model 
in Appendix~\ref{s:appsemicohmet}
analog to the methods described in \citep{Pletsch2010}.
Following the same steps as in \Sref{s:cohmetric}, we find that  $\bar{G}$ is also diagonal 
and the total number of grid points for a semicoherent step can thus be written as
\begin{equation}
   \mathcal{N}_{\rm scoh} = \frac{1}{16} \,\mathcal{U}\, \bar{m}^{-2}\, 
   \sqrt{\det{\bar{G}}}
   \label{e:Nscoh1}
\end{equation}
where $\bar{m}$ here represents the maximum mismatch per dimension used for
grid construction.
As derived in Appendix~\ref{s:appsemicohmet}, the determinant of the semicoherent metric 
is
\begin{equation}
 \sqrt{\det \bar{G}} =  \frac{\pi^4}{12\sqrt{3}} \; T^3 \, f^2 \, r_E^2 \, R \,
 \left[1 - \sinc^2\left(\frac{\Omega_E\,T}{2\pi}\right)\right]  \,.
 \label{e:detGsemicoh2a}
\end{equation}
As in \Sref{s:cohmetric}, for practical purposes 
we construct the grid for the highest
frequency searched~$f_{\rm max}$ in a given frequency band.
Thus, we can rewrite \Eref{e:Nscoh1} as
\begin{equation}
   \mathcal{N}_{\rm scoh} = \bar{m}^{-2} \, \frac{\pi^4}{192\sqrt{3}} \; T^3 \, 
   f_{\rm max}^2 \, r_E^2 \, R \,
 \left[1 - \sinc^2\left(\frac{\Omega_E\,T}{2\pi}\right)\right]\, \mathcal{U} \,,
   \label{e:Nscoh2}
\end{equation}
where the proper search volume $\mathcal{U}$ has been defined previously in 
\Eref{e:parspace}.

\begin{figure}[t]
\centering
\includegraphics[width=1.0\linewidth]{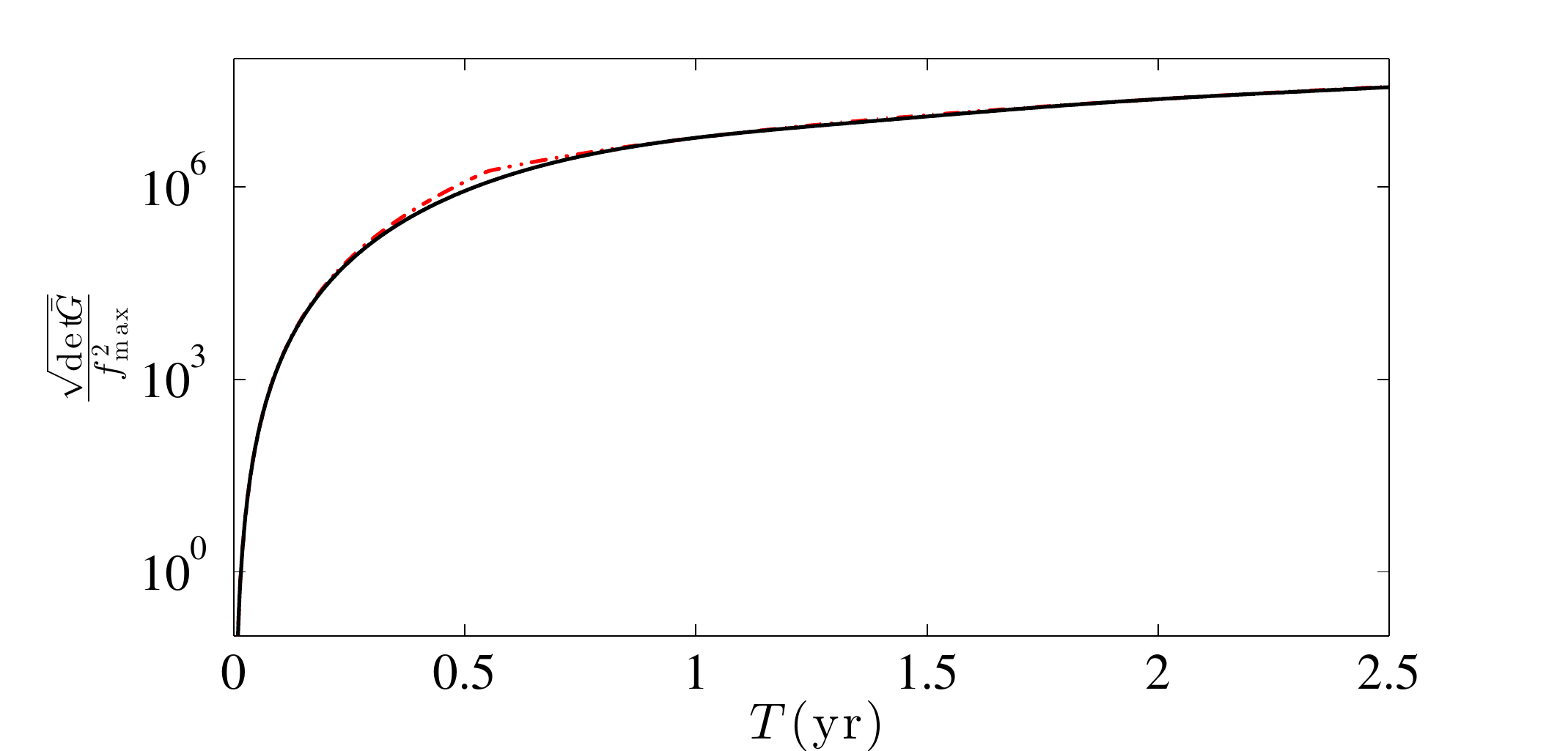}
\caption{
Scaling of the determinant of the semicoherent metric~$\bar G$ as function of 
the coherent window size~$T$ (black solid curve). The red dot-dashed curve shows 
the model for the semicoherent metric determinant from the approximation 
of \Eref{e:SemicohMetSky} used to estimate the computing cost scaling.
}
\label{fig:semicoh_met_det}
\end{figure}

To extract the scaling of $\mathcal{N}_{\rm scoh}$ with $T$,
we use the following approximation,
\begin{equation}
\left[1 - \sinc^2\left(\frac{\Omega_E\,T}{2\pi}\right)\right] \approx 
	\begin{cases}
	\frac{\Omega_E^2\,T^2}{12}, &  T < 0.551 {\rm yr} \\
	1, &  T \geq 0.551 {\rm yr} \,.
	\end{cases}
	\label{e:SemicohMetSky}
\end{equation}
which is illustrated in \Fref{fig:semicoh_met_det}.
%
%------------------------------------------------------------------------------
%------------------------------------------------------------------------------
%
Hence, using $R=T_{\rm obs}/T$ one finds that the total number of grid points 
in the semicoherent search scales as
\begin{equation}
 \mathcal{N}_{\rm scoh} 
 \propto  \bar{m}^{-2} \; \left(\frac{\Omega_E^2}{12}\right)^\frac{(s-3)}{2} \; T^{s-1} \; T_{\rm obs} \; f_{\rm max}^2 \,,
  \label{e:Nsemicohscaling}
\end{equation}
where the exponent $s$ is given by
\begin{equation}
 s \approx
        \begin{cases}
	5, &  T < 0.551 {\rm yr} \\
	3, &  T \geq 0.551 {\rm yr} 
	\end{cases} \,.
\end{equation}

\subsection{Semicoherent Search Sensitivity at Fixed Computing Cost}\label{s:sccompcost}

In analogy to \Sref{s:compcost}, we here adopt a similar model for
the computational cost of a semicoherent search, which is proportional to the number
of search-grid points~$\mathcal{N}_{\rm scoh}$ needed. We again assume that the 
FFT algorithm is used to compute $S_1$ over $f_{\rm max} T$ frequency bins, and 
again split the total mismatch $\bar{m}_{\rm tot}$ into the mismatch due to a 
frequency offset $\bar{m}_f$, and the mismatch due to offsets in the other 
parameters $\bar{m}$.
Hence, using Equations~\eref{e:Nsemicohscaling} and~\eref{e:R}
the semicoherent computing cost model~$ C_{{\rm scoh},1}$ is obtained as
\begin{equation}
   C_{{\rm scoh},1}  = K_{{\rm scoh},s}\,\bar{m}^{-3/2}\;T^{s-1} \, T_{\rm obs} 
   \,   \log_2(T\,f_{\rm max}) \,,
   \label{e:Cscoh1}
\end{equation}
where $K_{{\rm scoh},s}$ denotes a constant of proportionality that depends on $s$, 
\begin{equation}
  K_{{\rm scoh},s} = K_{\rm FFT} \, \frac{\sqrt{2}\,\pi^3\,r_E^2\; f_{\rm 
  max}^2\,\mathcal{U} }{96} \left( \frac{\Omega_E^2}{12} 
  \right)^{(s-3)/2} \,,
  \label{e:Kscohs}
\end{equation}
as well as on the implementation and computing-hardware dependent constant
$K_{\rm FFT}$ as in \Eref{e:Kcoha}.
Analog to \Eref{e:Ccoh1}, we here
also assume that the FFT algorithm is used, hence the $\log_2$ factor in \Eref{e:Cscoh1}.
In \Sref{s:semicohstats} we found the sensitivity of the semicoherent search 
as of \Eref{e:scsearchsens2} can be approximately described by 
\begin{equation}
  p^{-1}_{{\rm scoh}, 1} = \sqrt{(1-\langle\bar{m}_{\rm tot}\rangle) \; \mu} \,
    \frac{\left|\gamma_1\right|}{\theta_{S_1}^{\ast}}\;T^{1/4} \, T_{\rm obs}^{1/4} \,,
    \label{e:pthscoh}
\end{equation}
where $\langle \bar{m}_{\rm tot} \rangle = \langle \bar{m}_f \rangle + 3 \xi 
\bar{m}$ denotes again the total average mismatch of the search grid.

With the sensitivity and computing-cost model at hand, we can
now illustrate the increased efficiency that a semicoherent search offers over
a fully coherent search. We compare the sensitivity $p^{-1}_{{\rm scoh}, 1}$ 
of a semicoherent search with coherence window size~$T$ over a data set which
in total spans the observational time interval $T_{\rm obs}$ to the 
sensitivity $p^{-1}_{{\rm coh}, 1}$ of a fully coherent 
search with coherent integration time $T_{{\rm coh},1}$,
at the same computational cost: \mbox{$C_{{\rm scoh},1} = C_{{\rm coh},1}$}.
For a given computing cost~$C_{{\rm scoh},1}$, and observational data set
spanning~$T_{\rm obs}$, \Eref{e:Cscoh1} determines~$T$. This value of $T$
can then be used to obtain the sensitivity~$p^{-1}_{{\rm scoh}, 1}$ via \Eref{e:pthscoh}.
Similarly, as described in \Sref{s:harmsum}, the given value of $C_{{\rm coh},1}$
determines $T_{{\rm coh},1}$ and thus provides the corresponding $p^{-1}_{{\rm 
coh}, 1}$. 

\begin{figure}[t]
\centering
  \includegraphics[width=0.99\columnwidth]{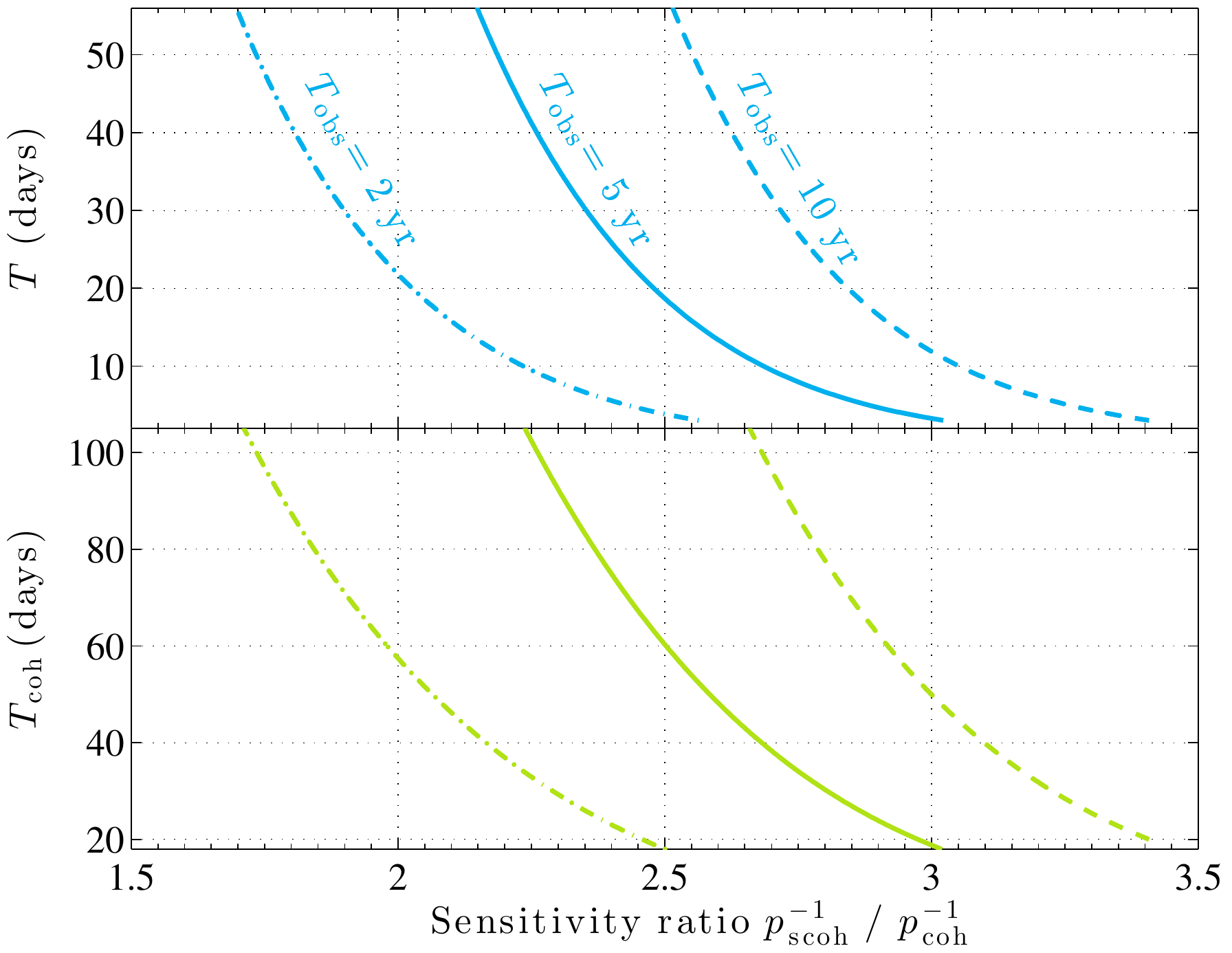}
   \caption{Comparison of a semicoherent and fully coherent search sensitivity 
   at equal computing cost and 
   given observational data time spans of $T_{\rm obs} = 2$\,yr (dotted-dashed curves), 
   $T_{\rm obs} = 5$\,yr (solid curves),  $T_{\rm obs} = 10$\,yr (dashed curves).
   The top panel shows the coherence window size~$T$ of the corresponding
   semicoherent search as a function of the sensitivity ratio.
   The bottom panel shows, for a coherent search, the integration time~$T_{{\rm coh},1}$, 
   i.e. the subset of $T_{\rm obs}$ that could be fully coherently analyzed with the
   same computing cost as the semicoherent search with the corresponding~$T$ shown in the upper panel.
   The sensitivity is for $P_{\rm FA}^{\ast} = 10^{-3}$ and $P_{\rm DET}^{\ast} = 0.9$ in each case and
   a typical pulsed signal with  $p=0.1$ and $|\gamma_1|^2=0.35$ (cf. \Fref{f:2pc-harm}).
   Since the sensitivity ratio is in all practically relevant cases much greater than unity, 
   the semicoherent search approach more efficient.
\label{f:sens-ratio-cohscoh}
}
\end{figure}

The so-obtained ratio of sensitivities $p^{-1}_{{\rm scoh}, 1}/p^{-1}_{{\rm coh}, 1}$ 
is studied numerically in \Fref{f:sens-ratio-cohscoh} for realistic computational power 
available, such as \EAH{}. In both cases the optimal mismatch parameters are assumed,
which are independent of computing cost (see Appendices~\ref{a:optmismatchcoherent} and \ref{a:optmismatchsemicoherent}).
As can be seen in the figure, this sensitivity ratio is always greater than unity and 
increases as $T$ decreases, which is representative 
of the fact that the sensitivity of a semicoherent search decreases more slowly than that of a 
coherent search as the available computing power decreases. Whilst this ratio decreases 
as $T$ (and, therefore, the computing cost) increases, the absolute search sensitivity always 
increases with $T$, and so it is still beneficial to use the largest achievable lag-window 
size $T$ at the available computational power. 

Using a simplified approximation for the semicoherent computing cost
model of \Eref{e:Cscoh1} allows us to obtain some analytical insight
 into the ratio $p^{-1}_{{\rm scoh}, 1}/p^{-1}_{{\rm coh}, 1}$ 
at fixed computing cost, similar to what has been done in \Sref{s:harmsum}.
Ignoring the slowly varying $\log_2$ term gives the approximate semicoherent 
computing cost model as
\begin{equation}
   C_{{\rm scoh},1}  \sim K_{{\rm scoh},s}\,\bar{m}^{-3/2}\;T^{s-1} \, T_{\rm 
   obs} \,.
   \label{e:Cscoh1approx}
\end{equation}
With this simplified model, \mbox{$C_{{\rm scoh},1} = C_{{\rm coh},1}$} can be rewritten
using the approximation of \Eref{e:CcohMapprox} as
\begin{equation}
  \frac{K_{{\rm scoh},s}\; T_{\rm obs} \; T^{s-1}}{\bar{m}^{3/2}} \,  
  = \frac{K_{{\rm coh},a}\;T_{{\rm coh},1}^a }{m^{3/2}} \,.
  \label{e:fixedcompcost}
\end{equation}
Furthermore, using Equations~\eref{e:Kcoha} and \eref{e:Kscohs} to replace $K_{{\rm coh},a}$
and $K_{{\rm scoh},s}$, we can rewrite \Eref{e:fixedcompcost} as
\begin{equation}
  T = \left(\frac{4\,\Omega_E\; \bar{m}^{3/2} \; T_{{\rm coh},1}^6}{5\sqrt{6} \; 
  m^{3/2} \; T_{\rm obs}}\right)^{1/4} \,.
    \label{e:TvsTcoh}
\end{equation}
where we assume $a=6$ and $s=5$, since coherent integration times $T_{{\rm 
coh},1}$ less than half a year will be practically feasible in the near future.
This relation can then be used to substitute $T$ in the ratio $p^{-1}_{{\rm scoh}, 1}/p^{-1}_{{\rm coh}, 1}$ 
using Equations~\eref{e:pthscoh} and \eref{e:pthcoh1a}, yielding 
\begin{equation}
  \frac{p^{-1}_{{\rm scoh}, 1}}{p^{-1}_{{\rm coh}, 1}} \approx 
  2 \frac{\theta^{\ast}_{1}}{\theta^{\ast}_{S_1}} 
  \; \left(\frac{T_{\rm obs}}{1 \rm{yr}}\right)^{1/16}
  \; \left(\frac{T_{\rm obs}}{T_{{\rm coh},1}}\right)^{1/8} \,,
  \label{e:pth-ratio2} 
\end{equation}
where we again assumed the optimal mismatch choices for $m$ and $\bar m$ 
(see Appendices~\ref{a:optmismatchcoherent} and \ref{a:optmismatchsemicoherent}) that are independent of
computational cost. For $a=6$ and $s=5$ these are $m_{\rm opt}=0.172$ 
and \mbox{$\bar{m}_{\rm opt} = 0.146$}.
Hence, as \Fermi-LAT data spans several years (implying \mbox{$T_{\rm obs} \gtrsim 1 \rm{yr}$})
and typically \mbox{$\theta^{\ast}_{1} \gtrsim \theta^{\ast}_{S_1}$},
the sensitivity ratio of \Eref{e:pth-ratio2}  exceeds unity in all practically relevant cases. 
This clearly indicates that at fixed computational cost, a semicoherent blind search is always more sensitive 
than a fully coherent search over the same parameter space.

%---------------------------------------------------------------------------------------------------
%---------------------------------------------------------------------------------------------------

\section{Efficient Implementation of a Multistage Search Scheme} \label{s:implementation}

In Section~\ref{s:coherent}, we argued that under computational cost constraints, 
blind fully coherent searches without harmonic summing are more efficient, i.e. can typically 
achieve higher search sensitivity. In Section~\ref{s:semicoherent}, we showed that at fixed 
computing cost semicoherent searches are more efficient than fully coherent searches
to scan wide parameter space.

These considerations motivate a multistage search strategy, 
in which the first and by far most computationally expensive stage 
uses the most efficient method (i.e. a semicoherent search) to 
explore the entire parameter space. In subsequent stages, the most promising 
candidates are automatically ``followed up'' in further, more sensitive steps, ultimately using fully 
coherent search methods. Since the parameter space relevant for these candidates 
has been previously narrowed down by the first-stage search, the computing cost 
constraints are relaxed (i.e. the computing cost of the follow-ups is negligible compared 
to the overall cost of the first stage of the blind search). 
Hence then the usage of fully coherent methods offering the highest sensitivity 
is made possible.

In this multistage search scheme, before statistically significant candidates from 
the first-stage semicoherent search are followed-up with fully coherent methods,
it is advisable to refine the location of each semicoherent candidate by searching, 
again semicoherently, using a refined grid with a smaller mismatch. 
We then ``zoom in'' on each significant candidate by performing a fully coherent search 
of the local parameter space around the refined location of the semicoherent candidate, 
using the full observational data time span, $T_{\rm obs}$. 
The search-grid construction of each stage is guided by the metric, as described in  
Appendices \ref{s:appcohmet}, \ref{s:appsemicohmet} and~\ref{mstdisr:s:skygrid}.

When searching for weak signals in the presence of noise, this can cause the refined 
semicoherent candidate to occur at a small but unknown offset from the true signal 
parameters. This offset depends on the candidate S/N; candidates with higher S/N have 
a smaller uncertainty region. In order not to miss weak signals, the coherent follow-up has to 
cover a conservative region in each dimension around the semicoherent candidate location. 
Since the parameter space which must be searched coherently has been greatly reduced, 
this step represents a very small fraction of the overall cost of the search. If the ratio
of the coherence window size $T$ used in the first stage and $T_{\rm obs}$ is very
large, it is more efficient to insert another intermediate zooming stage that
does another semicoherent
search with a coherence window size between $T$ and $T_{\rm obs}$. This would further
reduce the parameter space to be searched in the fully coherent step, ensuring that the 
follow-up remains a negligible fraction compared to overall search.
Finally, candidates from this coherent follow-up step are then ranked for further investigation 
(e.g. by taking into account higher harmonics, or a more complex phase model) according 
to their false alarm probability.

Since this multistage scheme is designed such that the largest computational burden is 
associated with the first stage, it is important to optimize this method 
of calculating the semicoherent test statistic~$S_1$ as much as possible. 
In the following, we describe various complementary methods which 
improve the efficiency and sensitivity of a computationally limited semicoherent search.

\subsection{Efficient Computation of Semicoherent Test Statistic}
\label{s:dft_form}

For each sky-position grid point of the search region the barycentric corrections
are applied directly to the LAT-registered arrival times $t_{\tiny\rm LAT}$,
to obtain the corresponding photon arrival times $t$ at the SSB.
The semicoherent detection statistic~$S_1$ as of \Eref{e:S1b} is then 
computed over the $f$- and $\dot f$-ranges.
However, directly computing $S_1$ from \Eref{e:S1b} is computationally inefficient.
Therefore, we here discuss more efficient ways of how to do this.
 
Making the dependence of $S_1$  on the search parameters $f$ and $\dot f$ explicit
for clarity, we rewrite \Eref{e:S1c} as
\begin{equation}
   S_1(f,\dot f) =  \sum_{j,k=1}^N \; w_j\; w_k \; e^{-i [\phi(t_j;f,\dot f)-\phi(t_k;f,\dot f)]}\; \hat{W}_T^{{\rm\tiny rect}}(\tau_{jk}) \,,
   \label{e:hatS1-1}
\end{equation}
where the phase differences in terms of $f$ and $\dot f$ are given by
\begin{align}
   \phi(t_j;f,\dot f)-\phi(t_k;f,\dot f) &= 2\pi f \tau_{jk} + \pi \dot{f} \left[(t_j-t_0)^2 - (t_k-t_0)^2\right] \nonumber\\
   &=2\pi f \tau_{jk} + \pi \dot{f} \left[ t_j^2-t_k^2 -2 t_0 \tau_{jk}  \right]   \,.
   \label{e:phase_diff}
\end{align}
Thus, $S_1$ of \Eref{e:hatS1-1} takes the following form,
\begin{equation}
  S_1(f,\dot f) = \sum^{N}_{j,k = 1} w_j w_k \,
    e^{-\pi i \dot{f} \left[ t_j^2-t_k^2 -2 t_0 \tau_{jk}  \right]} \;
     \hat{W}_T^{{\rm\tiny rect}}(\tau_{jk}) \;  e^{-2\pi i f \tau_{jk}} \,,
  \label{e:hatS1-2} 
\end{equation}
which allows us to utilize the efficiency of the FFT 
to scan along the $f$-direction.
In the following we describe how to achieve this.
First, we construct an
equidistant lag series whose separation is the sampling 
interval~$\delta_{\tau} = 1/ (2 f_{\rm max})$, where $f_{\rm max}$
is equal to the Nyquist frequency $f_{\rm Ny}$.
Then for each pair of times $(t_j,t_k)$ having a lag $\tau_{jk}$ smaller than the 
lag window (i.e. for which \mbox{$\hat{W}_T^{{\rm\tiny rect}}(\tau_{jk}) = 1$}), 
we determine the corresponding bin index~$b$ of the equidistant lag series
via interpolation. While we study the efficiency of different lag-domain-interpolation 
schemes below, let us assume here nearest-neighbor interpolation for simplicity.
Thus, we just round to the nearest lag-bin index $b$, 
\begin{equation}
 \label{e:lagbins}
 b = \textrm{round}\left[\tau_{jk} / \delta_{\tau}\right] \,.
\end{equation}
The FFT performance is generally best for input sizes
that are a power of $2$ (radix-$2$ FFTs). Therefore, we choose $T$ 
and $f_{\rm max}$, such that the total number of lag bins 
\mbox{$B_T = T/\delta_{\tau} = 2 T f_{\rm max}$} is a power of $2$.
We denote the lag-interpolated version of $S_1$ from \Eref{e:hatS1-2} 
by $\hat{S}$, which can be written using the lag-bin index~$b$ as
\begin{equation}
   \hat{S}(f,\dot f)= \sum_{b=-B_T/2}^{B_T/2} \; Y_b(\dot f)\; e^{-2\pi i \, f \, \delta_{\tau} b} \,,
   \label{e:SgDFT1}
\end{equation}
where terms depending on $\dot f$ and the photon weights have been absorbed 
into the complex numbers~$Y_b(\dot f)$. More precisely, each $Y_b(\dot f)$
is the sum of pairwise weight and $\dot f$ phase factors, falling into the same
lag bin~$b$,
\begin{equation}
  Y_b(\dot f) = \sum_{j=1}^N\; y_j(b;\dot f) \,,
\end{equation}
where
\begin{equation}
   y_j(b,\dot f) = 
   \begin{cases}
   w_j \; w_k \; e^{-\pi i \dot{f} \left[ t_j^2-t_k^2 -2 t_0 \tau_{jk}  \right]} \,, & \textrm{round}\left[\tau_{jk} / \delta_{\tau}\right] = b\,,\\
   0, & \textrm{else}\,.
   \end{cases}
\end{equation}
Note that the so-constructed lag series $Y_b$ has Hermitian symmetry, i.e. $Y_b = Y_{-b}^{\ast}$,
and therefore $\hat{S}$ remains entirely real-valued.
The above expression for $\hat{S}$ in \Eref{e:SgDFT1} can be seen as a Fourier transform
of the complex lag series $Y_b$, and so $\hat{S}$ can be computed efficiently at many discrete frequencies by exploiting the FFT algorithm, i.e. by calculating
\begin{equation}
  \hat{S}_g(\dot{f}) = \sum_{b=-B_T/2}^{B_T/2} Y_b(\dot{f}) \;  e^{-2\pi i\, g\, b \, / \,B_T}\,.
  \label{e:SgDFT3}
\end{equation}
where the frequency at the $g$th bin is $f=g/T$.
There exist efficient FFT algorithms  \citep{FFTW05} which can be used to evaluate this
complex-to-real (c2r) transform of \Eref{e:SgDFT3}, and which only require the positive lag portion of $Y_b$ to be calculated as an input.

The above formulation of the semicoherent detection statistic, $\hat{S}_g$, is very similar to the $D_{\ell}$ statistic, described in A06 as the DFT of the discrete autocorrelation function of the (binned) photon arrival times. However, there are some key differences. While further differences are discussed in the following subsections as we encounter them, we here note a first difference between the methods related to the correction of the frequency derivative 
$\dot{f}$. When calculating $D_{\ell}$, the frequency derivative is corrected by constructing a new time series in which the photon arrival times are stretched out 
according to \mbox{$t_j = \tilde{t}_j + \frac{1}{2}\frac{\dot{f}}{f}\tilde{t}^2_j$}.
In order to search the $\{f,\dot{f}\}$ parameter space, the ratio $\dot{f}/f$ is increased by small increments. According to this scheme, the search points in the $\{f,\dot{f}\}$ plane lie along straight lines with increasing gradient, intersecting at the origin. 
As a result, the search grid point density is highly non-uniform in the $\{f,\dot{f}\}$ plane,
decreasing from low to high search frequencies. The result is that the search parameter space is highly oversampled in the $\dot f$ dimension at low frequencies. 
This sub-optimal grid-point density implies that far more grid points are needed to cover the parameter space. Decreasing the lag-window size to account for this extra computational cost causes a reduction in sensitivity which more than accounts for the decrease in the average 
mismatch\footnote{This is because despite the reduced mismatch in the $\dot f$ dimension,
the contributions of the other three dimensions still remain and dominate the total mismatch 
that is relevant for the search sensitivity.}.
Calculating $\hat{S}_g$ in the manner described above, where the effect of the frequency derivative is accounted for by the complex lag-series, $Y_b(\dot{f})$, allows us to uniformly sample the 
$\{f,\dot{f}\}$ plane with the optimal average mismatch.

\subsection{Frequency Domain Interpolation}\label{s:fdinterpolation}

\begin{figure}[t]
\centering
\includegraphics[width=0.9\columnwidth]{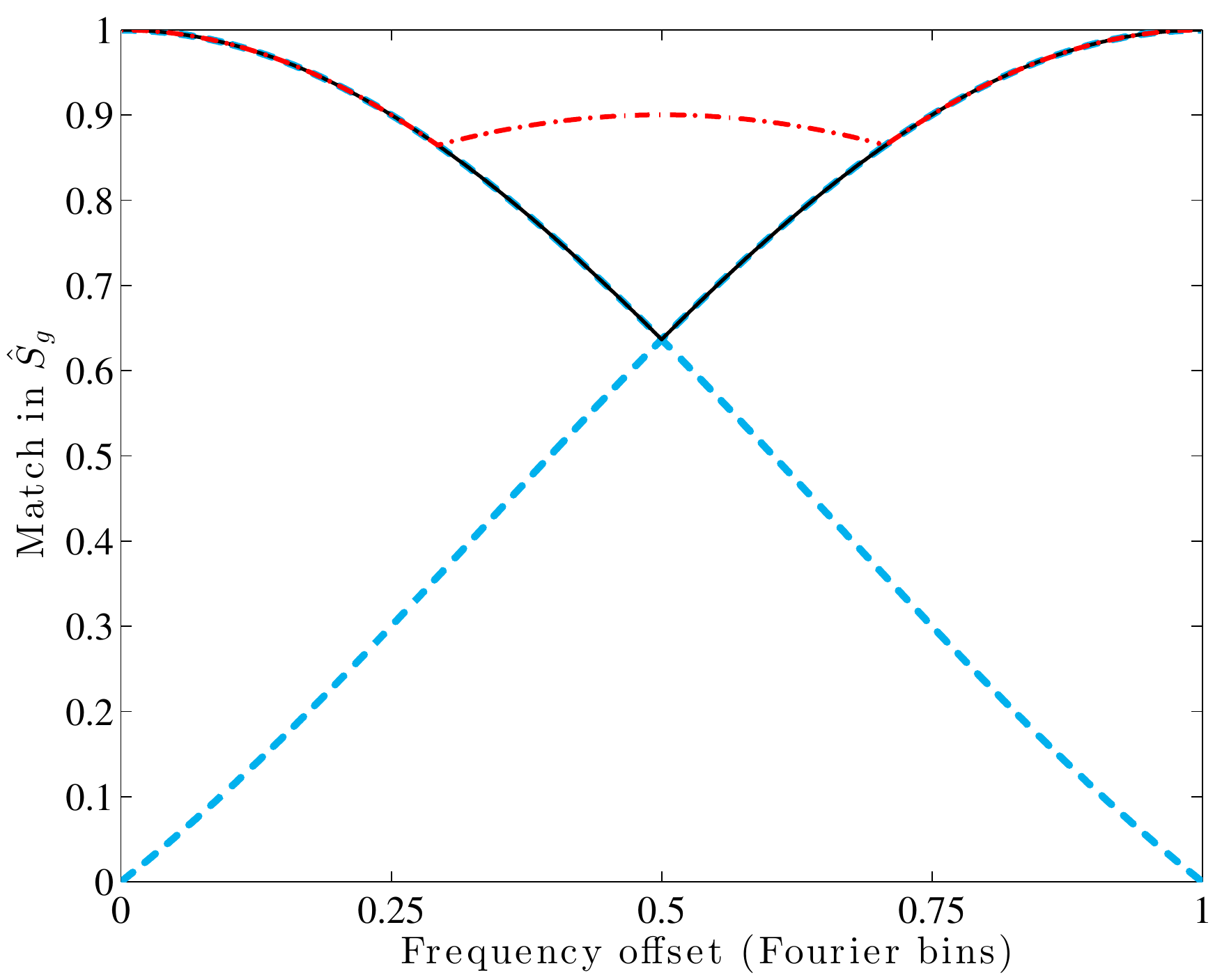}
\caption{Illustration of frequency domain interpolation.
The dashed blue curve shows the relative response (match) of $\hat{S}_g$  at neighboring 
Fourier bins as a function of the signal frequency offset. The solid black curve represents the overall DFT response. 
The overlaid dotted-dashed red curve is the overall DFT response match obtained via the 
frequency domain interpolation as described in the text.}
\label{f:interbinning}
\end{figure}

When performing a semicoherent search using $\hat{S}_g$, computed via the FFT as in \Eref{e:SgDFT3}, for a pulsar signal frequency that does not lie exactly at a Fourier frequency 
(i.e. not at an integer multiple of $1/T$) a loss in signal power (mismatch) will result. 
To evaluate the response of $\hat{S}_g$ to signals at a non-Fourier frequency, we consider the case when the lag-series contains a pure sinusoid, with amplitude $\hat{S}_0$, at a frequency $h/T$. Including an appropriate normalization factor of $1/B_T$ for the Fourier transform, so that
\begin{equation}
Y_b(0) = \frac{\hat{S}_0}{B_T}\, e^{2\pi i h b / B_T}\,.
\label{e:sine_wave}
\end{equation} 
This represents the (unlikely) case of a strong signal, in the absence of noise, where the frequency derivative and sky location have been perfectly matched. Using \Eref{e:SgDFT3} 
the response at the $g$th frequency bin is therefore:
\begin{align}
\hat{S}_g &= \sum_{b=-B_T/2}^{B_T/2} Y_b(0) \;  e^{-2\pi i\, g\, b \, / \,B_T} \nonumber\\
&=\frac{\hat{S}_0}{B_T}\, \sum_{b=-B_T/2}^{B_T/2} e^{-2\pi i\, b\,(g-h)\ / \,B_T}\,.
\label{e:Sg_response}
\end{align}
The above summation over $b$ can be explicitly calculated and is also called the
\emph{Dirichlet} kernel, which is given by
\begin{equation}
 \mathcal{D}_N(x) = \sum_{b=-N}^{N} e^{-i\,b\,x} = \frac{\sin\left( (N+1/2) x\right) }{\sin(x/2)} \,.
 \label{e:DiriKernel}
\end{equation}
Using this identity gives rise to rewrite \Eref{e:Sg_response},
\begin{align}
\hat{S}_g &= \frac{\hat{S}_0}{B_T}  \mathcal{D}_{B_T/2-1}\left( 2\pi(g-h)/B_T \right) \nonumber \\
 &= \frac{\hat{S}_0}{B_T} \; \frac{ \sin\left(\pi\, (g-h) \, (1-1/B_T) \right) }{ \sin\left( \pi \,(g-h)/B_T\right)} \nonumber \\
 &\approx  \frac{\hat{S}_0}{B_T} \; \frac{ \sin\left(\pi\, (g-h) \right) }{ \sin\left( \pi \,(g-h)/B_T\right)} \,\nonumber \\
 &\approx \hat{S}_0 \;\sinc(g-h) \,,
 \label{e:Sg_response2}
\end{align}
where in the approximation made in the third step we assumed that \mbox{$1/2 \gg 1/B_T$},
and in the fourth step we used in addition the following approximation
\mbox{$\sin(\pi(g-h)/B_T) \approx \pi(g-h)/B_T$}, 
since typically for nearby frequency bins $B_T \gg (g-h)$.
Therefore, the match is well described by a sinc function for signals at non-Fourier frequencies and is smallest 
(i.e. greatest mismatch) if the signal lies exactly halfway between two Fourier frequencies. This
is shown in \Fref{f:interbinning}, which displays the approximated response of \Eref{e:Sg_response2}.

%-----------------------------------------
%-----------------------------------------

This loss can be reduced by interpolating the Fourier response halfway between two Fourier 
frequency bins. 
One method of interpolating the Fourier transform output, known as zero-padding, is to extend the original lag series (or time series) to twice its original length by adding zeros onto the end. However, this requires calculating a Fourier transform which is twice as long, and therefore more than twice as costly. To avoid increasing the computational cost, we use a more efficient interpolation technique in the frequency domain, also known as ``interbinning'' \citep{VanderKlis1989,Ransom2002}. 
Note that \cite{Ransom2002} gives a formulation for calculating interbin amplitudes 
for real- or complex-to-complex Fourier transforms.
However, in our case, where $\hat{S}_g$ is entirely real-valued, it is sufficient to calculate interbins by summing the amplitude of neighboring frequency bins, 
\begin{equation}
  \hat{S}_{g+ 1/2} = \frac{1}{\sqrt{2}}\left(\hat{S}_g + \hat{S}_{g+1}\right)\,.
  \label{e:real_interbinning}
\end{equation}
It is also important to emphasize that our chosen normalization differs from that used by \cite{VanderKlis1989,Ransom2002}, where the interbins are normalized to ensure that all of the signal power is recovered in an interbin if the signal lies exactly halfway between two Fourier bins. 
Instead, we here use a normalization factor of $1/\sqrt{2}$ ensuring that interbins have the same noise variance as the standard Fourier bins \citep[as was first done by][]{AstoneInterbinning}. Whilst the method used in \Eref{e:real_interbinning} results in a mismatch even for signals at the center of an interbin, ensuring that the noise variance is consistent between bins and interbins facilitates semicoherent candidate ranking for follow-up procedures.

The overall response for signals at non-Fourier frequencies before and after interbinning is shown 
in \Fref{f:interbinning}. Using the interbinning method, the average mismatch due to a frequency offset is reduced from $\sim 0.13$ to $\sim 0.075$, whilst the maximum mismatch is reduced from $\sim 0.36$ to $\sim 0.14$. Thanks to their simplicity, interbins can be calculated very quickly, and so this performance gain comes at negligible extra computing cost (when compared to the dominant FFT computing cost).

\subsection{Complex Heterodyning}\label{s:heterodyning}

Searching a wide range of frequencies (i.e., large $f_{\rm max}$) using the test statistic $\hat{S}$ would require computing a single FFT of large size, $B_T$. 
The length of an FFT which can be computed is limited by the amount of memory accessible. In particular, extending the frequency search band to the millisecond pulsar regime (i.e. near $1$kHz frequencies) would require a large increase in the sampling rate, and would potentially require decreasing the lag-window size (and hence the sensitivity of the search) to make the FFT short enough to fit into memory. 

To address this problem, we divide the total frequency range into smaller bands of 
size~$\Delta f$ (that can be efficiently searched in parallel) using complex heterodyning, 
without sacrificing sensitivity.
Using this method, the center frequency, $f_{\rm H}$, of a given subband is shifted to DC, which
in the lag domain corresponds to multiplying each lag bin by $e^{-2\pi i \, f_{\rm H} \,\delta_{\tau}b}$. The heterodyned lag series is therefore defined as
\begin{equation}
   Y^\prime_b(\dot{f},f_{\rm H}) = Y_b(\dot{f}) \;  e^{-2\pi i\, f_{\rm H}\delta_{\tau}b} \,,
\end{equation}
and the frequency at the $g$th bin becomes 
\begin{equation}
f = g /T + f_{\rm H}\,.
\label{e:het_f}
\end{equation}
One can therefore compute $\hat{S}_g(\dot{f})$ over the 
subband \mbox{$[f_{\rm H}-\Delta f/2; f_{\rm H}+\Delta f/2]$} via
\begin{equation}
  \hat{S}_g (\dot f) = \sum_{b=-B_T/2}^{B_T/2} \; Y^\prime_b(\dot{f},f_{\rm H}) \; e^{-2\pi i\, g \, b / B_T} \,,
 \label{e:heterodyning}
\end{equation}
in the same way as described in \Eref{e:SgDFT3}, but using a sampling interval 
of only~$\delta_\tau = 1/(\Delta f)$. Hence, we can search subbands 
in the millisecond-pulsar regime, while the FFT size
remains at \mbox{$B_T = T \Delta f$}.

\subsection{Lag Domain Interpolation}\label{s:tdi}

As outlined above, before the FFT can be performed the lags~$\tau_{jk}$ 
have to be binned into an equidistant lag series. Because the lags~$\tau_{jk}$
will in general not coincide with the lag-bin centers, the nearest-neighbor interpolation
of \Eref{e:lagbins} introduces an additional, frequency-dependent loss 
(mismatch) of signal power across the frequency band 
analyzed \citep[e.g.,][]{VanderKlis1989,Ransom2002}.

\begin{figure}[t]
   \centering
	        \subfigure
		{\includegraphics[width=1.0\linewidth]{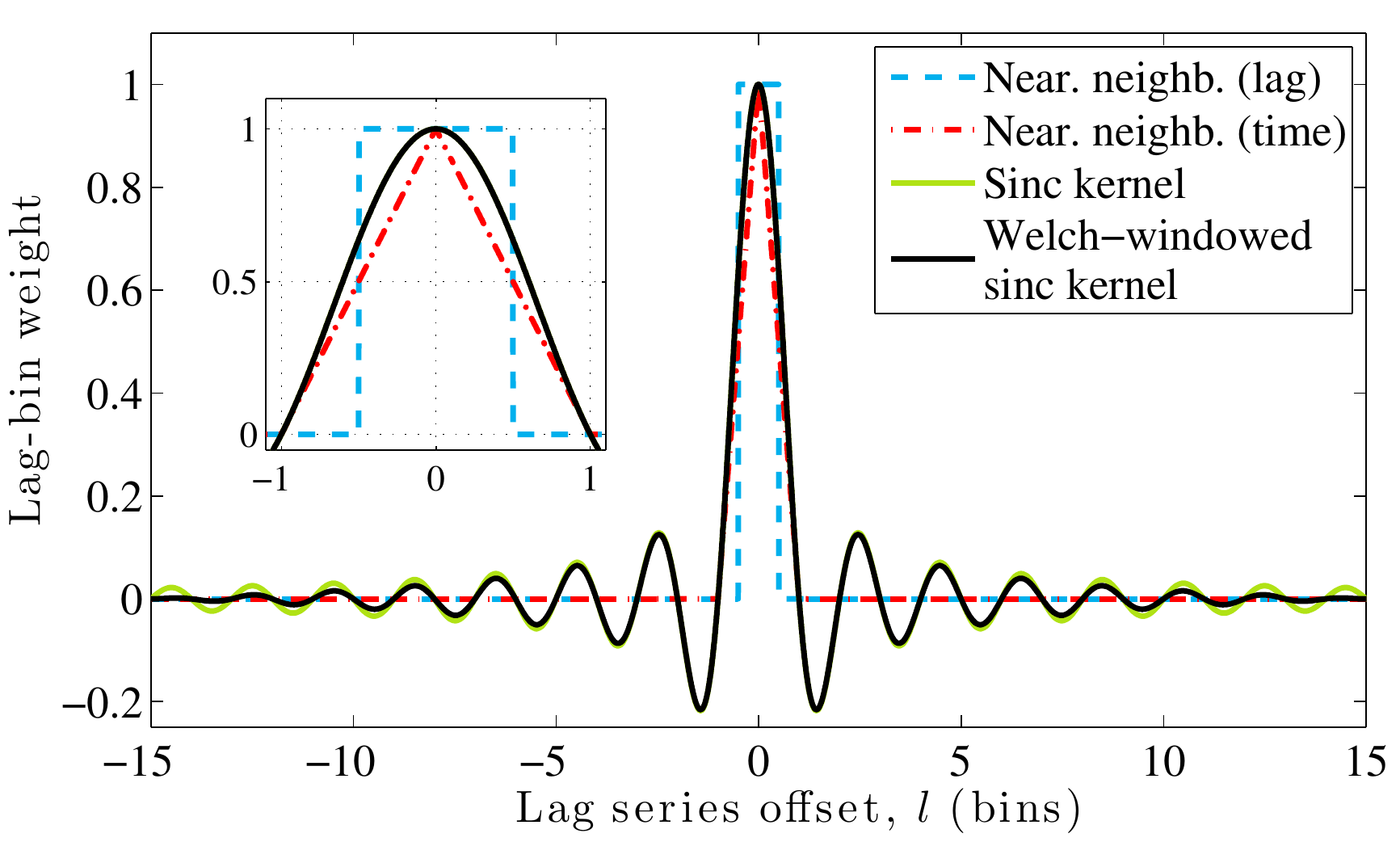}}\\
		\subfigure
		{\includegraphics[width=1.0\linewidth]{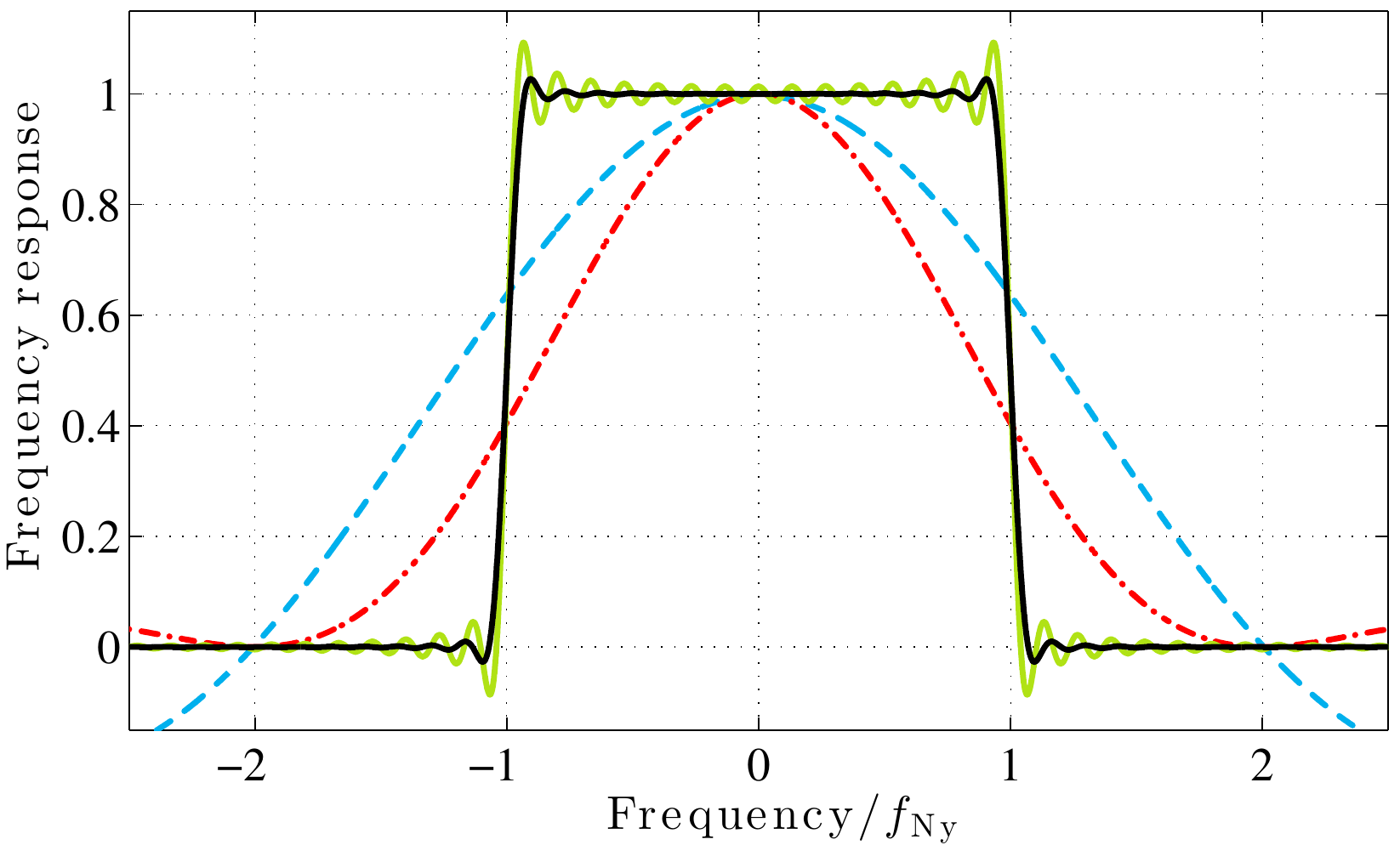}}
\caption{
Upper panel:
comparison of different lag domain interpolation functions, with  
the interpolating bin weights calculated 
over the range of the nearest 15 lag bins on either side of the center. 
For clarity the inset shows a zoom of the central region.
Lower panel:
overall frequency response of each interpolation function from the upper panel, 
that $\hat S$ is multiplied with in the frequency domain. 
The ideal response would be unity within the search band, i.e.
for $-f_{\rm Ny} < (f-f_{H}) < f_{\rm Ny}$, and zero otherwise. 
For this specific choice of using the nearest 15 bins for the 
interpolation, the average mismatch (loss in signal power) across the search band from is 
$\sim23\%$ for the rectangular binning function in time,
$\sim13\%$ for the rectangular binning function in lag domain, but only 
$\sim1\%$ for the sinc kernel and also for the Welch-windowed sinc kernel
that shows reduced Gibbs oscillations. 
}
\label{f:interpolation}
\end{figure}

The process of binning in lag can be thought of as convolving the lag series with a binning function. 
By the convolution theorem, the resulting response across the frequency band is the Fourier
transform of this convolving function. For $\hat{S}_g$ as derived above, the binning function 
(for nearest-neighbor interpolation) is a simple rectangular function of width $\delta_{\tau}$, 
leading to the sinc response in the frequency domain.
As a consequence, this results in an average loss (mismatch) in signal power of $\sim 13\%$ 
across the entire search band, illustrated in \Fref{f:interpolation}. 

Improved lag domain interpolation can reduce these losses. 
A given frequency response can be achieved by weighting the lag series bins 
around each $\tau_{jk}$ with an appropriate interpolation function. 
Ideal (i.e. lossless) interpolation would lead to a frequency response that is a rectangular function: 
unity within the search band to remove all bias in the spectrum, 
and zero outwith to prevent any noise from being aliased into the band.
Therefore, this ideal case of a rectangular frequency response requires a lag 
interpolation function that is the sinc function.
However, this interpolation function has infinite extent in the lag domain
and is therefore impossible to realize in practice. 

A practical solution is to truncate the sinc function in the lag domain 
around each~$\tau_{jk}$, such that the computational cost of this interpolation 
remains a negligible fraction of the overall computation time. 
In fact, one can show that using lag domain interpolation with the sinc function 
truncated to only the $d$ nearest lag bins for each~$\tau_{jk}$
is the best $d$th order approximation in the least squares sense to the
ideal (rectangular) response function \cite[e.g.,][]{Percival}.
As a result, the average loss (mismatch) across the frequency search band is 
drastically reduced. In the example shown in Figure~\ref{f:interpolation}, 
with a truncated sinc kernel using the $d=15$ nearest lag bins are on either side 
reduces this average mismatch  to only $\sim1\%$, as compared to the 
nearest-neighbor interpolation. Generally, it is often practical to use even more 
neighboring bins without significantly affecting the computational cost, but 
reducing the average mismatch even further.

%--------------------------------------------------------------
%--------------------------------------------------------------

However, as can also be seen in \Fref{f:interpolation}, an inconvenient property 
of the truncated sinc kernel is the Gibbs oscillation throughout the frequency band.
These oscillations mean that the false alarm probabilities of candidates can vary significantly
across the frequency band, making it difficult to rank candidate pulsars 
for follow-up.
This problem can be mitigated by multiplying the sinc kernel by another 
windowing function \citep[p. 176]{LyonsDSP}. This windowing function is 
required to be simple 
(and therefore efficient) to compute, and must still have a reasonably sharp 
fall-off in frequency near the edges of the bands. We find that the Welch 
window (an inverted parabola) provides a useful compromise between these  
requirements. 
The interpolated lag series, $\tilde{Y}_b$, is constructed by spreading the 
original lag-series $Y_b^\prime$ amongst the first $d$ bins on either side of 
the nearest bin to a single photon pair with lag~$\tau_{jk}$, 
\begin{align}
    \tilde{Y}_{b + l} (\dot{f},f_{\rm H}) \;=\;  Y_b^\prime(\dot{f},f_{\rm H}) \;&\; \sinc\left(b + l - \frac{ \tau_{jk}}{\delta_{\tau}}\right) \nonumber \\
    &\;\;\;\times\;  \left[1 - \left(b + l - \frac{\tau_{jk}}{\delta_{\tau}}\right)^2 \frac{1}{d^2}  \right] \,,
\end{align}
for $l = 0,\pm1,...,\pm d$.
The frequency response of the Welch-windowed sinc kernel is displayed \Fref{f:interpolation}. 
Whilst the average mismatch with the Welch-windowed sinc 
kernel is comparable to the truncated sinc kernel, the reduced 
Gibbs oscillation means that the false alarm probabilities of candidates are 
much more consistent across the frequency band, allowing candidate pulsars to 
be more easily ranked, albeit with almost no increase in the cost of 
interpolating the lag-series. 
Fortunately, the interpolation functions can be efficiently computed using 
trigonometric look-up tables and 
recurrence relations. When this efficiency is combined with the typical 
sparseness of the lag-series, the interpolation step
remains a negligible fraction of the overall computation time.

Within this framework of lag domain interpolation, another key difference to the A06 method
is worth pointing out. In A06, the SSB photon arrival times~$t_j$ are binned directly \emph{prior} 
to calculating the lags~$\tau_{jk}$ and the DFT (the $D_{\ell}$ in their notation). 
This implies a rectangular window function in time, which then is convolved with itself 
leading to a triangular window shape in the lag domain. Hence, the resulting frequency response 
is effectively the sinc function squared  (also shown in Figure~\ref{f:interpolation}). 
This causes significant loss in signal power, especially at the edges of the frequency band,
and amounts to a loss of $\sim23$\% averaged across the entire frequency band.
For comparison, by using the lag domain interpolation technique with the 
Welch-windowed sinc kernel as presented above, this average loss can be reduced by 
more than an order of magnitude, 
from $\sim23$\% to $\sim1$\%, at about the same computational expense.

\section{Performance Demonstration} \label{s:perfdemo}

In order to validate the expected sensitivity gain  from the improved methods 
presented in this paper, we perform extensive Monte-Carlo simulations. 
The false alarm probabilities are obtained using 
simulated data sets with different realizations of $8000$ photon arrival times (with unit weights), 
spanning a realistic observation time of $T_{\rm obs} = 5$\,yr.
To find the detection probabilities (for a given false alarm probability)
simulated pulsar signals are added, which have the same pulse profile
of Gaussian shape whose Fourier coefficient at the fundamental 
frequency is $|\gamma_1|=0.82$, and varying pulsed fractions~$p$.

While for computational reasons, the actual parameter space searched in each simulation
was chosen smaller than in a real search, the main conclusions from these results
are unaffected by this.
In each simulation, the search covered a frequency bandwidth of $1$\,Hz and a frequency derivative range of $10^{-13}$\,Hz\,s$^{-1}$. Each simulation searched the nearest nine sky positions around the signal location, at a uniformly random location on the sky.
In the semicoherent search stage we used a coherence window size of $T =
2^{20}\,\textrm{s}\approx 12\, \textrm{d}$.

For further comparison, we also apply the A06 method to the simulated data sets.
However, here we obtain a generous sensitivity estimation.
This is because the non-uniform sampling of the $\{f,\dot{f}\}$ parameter space (discussed in more detail in \Sref{s:dft_form}) was not accounted for. While this is justifiable for a search for isolated millisecond pulsars, at lower frequencies and larger frequency derivatives (i.e. where most young pulsars are found) this non-optimal sampling requires reducing the lag-window size (and therefore reducing the sensitivity) to achieve the same computational cost. 

\begin{figure}[t]
\centering
\includegraphics[width=\columnwidth]{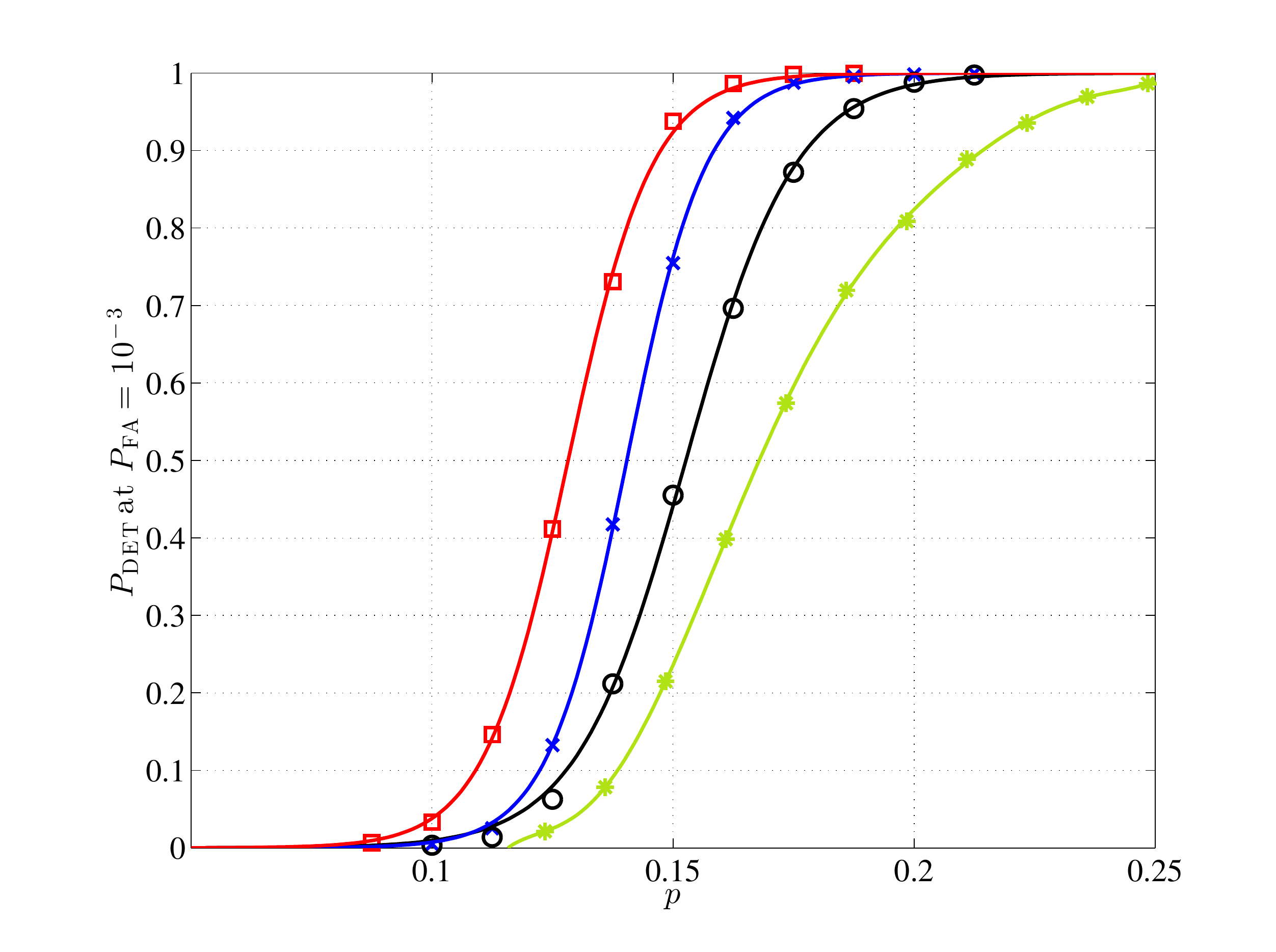}
\caption{Comparison of search efficiency of different search methods at fixed computational cost.
Shown is the detection probability $P_{\rm DET}$ at increasing pulsed fractions~$p$ for 
the simulated gamma-ray pulsar signals at $P_{\rm FA} = 10^{-3}$. 
The solid curves represents fits to each set of data points. 
Green stars: estimated sensitivity using the A06 method for the same computing cost.
Black circles: semicoherent search method, using only nearest-neighbor 
lag-domain interpolation. 
Blue crosses: semicoherent search method, using interbinning
frequency-domain interpolation, and lag-domain interpolation with a
Welch-windowed sinc kernel.
Red squares: multistage search method
(Semicoherent search method using lag-domain and interbinning frequency-domain
interpolation, plus fully coherent follow-up).
In all cases, the number of simulations was chosen
large enough so that the uncertainties of the data points 
become smaller than the size of the data markers.
\label{f:sens_curve}
}
\end{figure}

The results from all simulations are summarized in \Fref{f:sens_curve}, which shows the detection probability as a function of pulsed fraction for each of the search methods discussed in this paper. 
From best-fit curves (of typical sigmoid shape) shown in \Fref{f:sens_curve}, we compare the pulsed fraction required to give a detection probability of $95\%$ at a false alarm probability of $0.1\%$.
We find that this pulsed fraction is around $48\%$ lower for the full multistage method presented here than for the A06 method with approximately the same computational cost. This sensitivity increase is due to several improvements described in previous sections, in particular: use of the parameter space metric to allow optimally spaced grid-points; lag- and frequency-domain interpolation to reduce mismatch; and an automated coherent follow-up step to increase sensitivity to weak gamma-ray pulsar signals.

%------------------------------------------------------------------------
%------------------------------------------------------------------------

\section{Conclusions}\label{s:concl}

We have presented optimized strategies to improve the efficiency of blind searches 
for isolated gamma-ray pulsars, whose search sensitivity is computationally limited.
Under these conditions, our results confirm that fully coherent searches are generally 
less efficient than semicoherent searches, as well as that 
harmonic summing is typically less efficient than searching only for the strongest 
individual harmonic.
We also derived the parameters for most efficient search grids.
As motivated by these results, we presented and studied the implementation of
a multistage search strategy. 
We have also presented efficient computation and interpolation techniques for 
the semicoherent test statistic, offering further important sensitivity gains. 
Finally, we have conducted realistic simulations which demonstrate the improved 
performance from our combined advances, providing in a substantial increase 
in sensitivity (i.e. lowering the minimum detectable pulsed fraction by almost $50\%$) 
over previous methods at the same computational cost.

The methods presented here are being implemented with the 
\EAH{} volunteer computing project to increase the chances
of detecting new gamma-ray pulsars among the unidentified LAT sources.
While here we have focused on searches for isolated pulsars, 
the methods also apply to searches for pulsars 
in binaries, where partial knowledge of the orbit is available from 
observations at other wavelengths~\citep{Pletsch+2012-J1311}. 

Furthermore, the framework derived in this work in order to obtain 
an improved understanding of the pulsation search sensitivities 
underlying the different methods should also be useful for population studies. 
Specifically, these estimates can facilitate 
identifying the selection biases in the known gamma-ray pulsar sample,
for example due to the difference in pulse profile shape. In future work, 
we shall also explore using this framework to improve the efficiency of 
harmonic summing employing one or more realistic pulse profile templates 
built from the existing population of known gamma-ray pulsars.

\acknowledgements

This work was supported by the Max-Planck-Gesell\-schaft~(MPG), 
as well as by the Deutsche Forschungsgemeinschaft~(DFG) through 
an Emmy Noether research grant PL~710/1-1 (PI: Holger~J.~Pletsch). 
We also thank the anonymous referee for suggestions that helped improve the
manuscript.

\appendix

\section{Derivation of statistical properties of coherent test statistic}
\label{s:statcohpow}

From \Eref{e:Pnbweights} in \Sref{s:Statcoherent} the coherent power $\mathcal{P}_n$ 
can be rewritten as
\mbox{ $ \mathcal{P}_n = c_n^2 + s_n^2$},
where
\begin{align}
  c_n = \sqrt{\frac{2}{N}} \sum_{j=1}^N \cos[n\,\phi(t_j)] \,\\
  s_n = \sqrt{\frac{2}{N}} \sum_{j=1}^N \sin[n\,\phi(t_j)] \,.
\end{align}
Under the null hypothesis $p=0$, 
the phases $\phi(t_j)$ are uniformly distributed on $[0,2\pi]$ and
it is straightforward to show that
\begin{subequations}
\label{e:cnsnE0}
\begin{align}
 & E_0\left[\cos(n\phi(t_j))\right] =  E_0\left[\sin(n\phi(t_j))\right]  = 0\,,\\
 & Var_0\left[\cos(n\phi(t_j))\right] =  Var_0\left[\sin(n\phi(t_j))\right]  = 1/2 \,.
\end{align}
\end{subequations}
Since we have typically $N\gg1$, by appealing to the Central Limit Theorem, the random variables
$c_n$ and $s_n$ are normally distributed with zero mean and unit variance,
\begin{subequations}
\begin{align}
 &E_0[c_n] = E_0[s_n] = 0 \,,\\
 &Var_0[c_n] = Var_0[s_n] = 1 \,.
\end{align}
\end{subequations}
Hence, $\mathcal{P}_n$ follows a central $\chi^2$-distribution 
with $2$ degrees of freedom \citep[e.g., ][]{BlackmanTukey1958}.
Therefore the first two moments are  $E_0 \left[ \mathcal{P}_n \right] = 2$
and  $Var_0 \left[ \mathcal{P}_n \right] = 4$, as given in \Eref{e:PnE0Var0}.

Suppose a pulsed signal is present, $p>0$, with a pulse profile having 
the complex Fourier coefficients $\gamma_n$ as defined by \Eref{e:FourierCoeff}. 
While in this case for the $(1-p)N$ ``non-pulsed'' photons (i.e. background) 
Equations~\eref{e:cnsnE0} still hold, however for the $pN$ ``pulsed'' photons 
(i.e. not background) one obtains
\begin{subequations}
\begin{align}
  &E_p\left[\cos(n\phi(t_j))\right] = \Re(\gamma_n)  \,,\\ 
  &E_p\left[\sin(n\phi(t_j))\right]  = - \Im(\gamma_n)\,,\\
  &Var_p\left[\cos(n\phi(t_j))\right] = \frac{1}{2} + \frac{\Re(\gamma_{2n})}{2} - \Re(\gamma_{n})^2  \,,\\ 
  &Var_p\left[\sin(n\phi(t_j))\right]  = \frac{1}{2} - \frac{\Re(\gamma_{2n})}{2} - \Im(\gamma_{n})^2 \,.
\end{align}
\end{subequations}
Therefore, the random variables
$c_n$ and $s_n$ are normally distributed (since $N\gg1$) 
with the following mean values and variances,
\begin{subequations}
\begin{align}
 &E_p[c_n] = p\,\sqrt{2N}\, \Re(\gamma_n) \,,\\
 &E_p[s_n] = - p\,\sqrt{2N}\, \Im(\gamma_n) \,,\\
 &Var_p[c_n] = 1 + p\,\Re(\gamma_{2n}) + 2p\, \Re(\gamma_{n})^2 \,,\\
 &Var_p[s_n] = 1 - p\,\Re(\gamma_{2n}) - 2p\, \Im(\gamma_{n})^2 \,.
\end{align}
\end{subequations}
For weak signals (i.e. small pulsed fractions) 
and typical gamma-ray pulse profiles (see \Fref{f:2pc-harm}), 
we can approximate these variances as
\begin{equation}
 Var_p[c_n] \approx Var_p[s_n] \approx 1 \,.
\end{equation}
With this approximation, the distribution of $\mathcal{P}_n$ follows a 
noncentral $\chi^2$-distribution \citep[][]{Groth1975,Guidorzi2011}
with $2$~degrees of freedom, whose the first two moments are 
\begin{subequations}
\begin{align}
   &E_p\left[\mathcal{P}_n\right] \approx 2 + 2 p^2 N  \left|\gamma_n\right|^2\,,   \label{e:EpPnA}\\
   &Var_p\left[\mathcal{P}_n\right] \approx 4 + 8 p^2 N \left|\gamma_n\right|^2\,,
   \label{e:VarpPnA}
\end{align}
\end{subequations}
recovering Equations~\eref{e:EpPn} and \eref{e:VarpPn}.
The noncentrality parameter of that distribution is
the second summand in \Eref{e:EpPnA}, \mbox{$2 p^2 N  \left|\gamma_n\right|^2$}.

\section{Coherent metric}
\label{s:appcohmet}

%---------------------------------------------------------------
\begin{figure*}[t]
 \centering
 		\subfigure
 		{\includegraphics[width=0.33\textwidth]{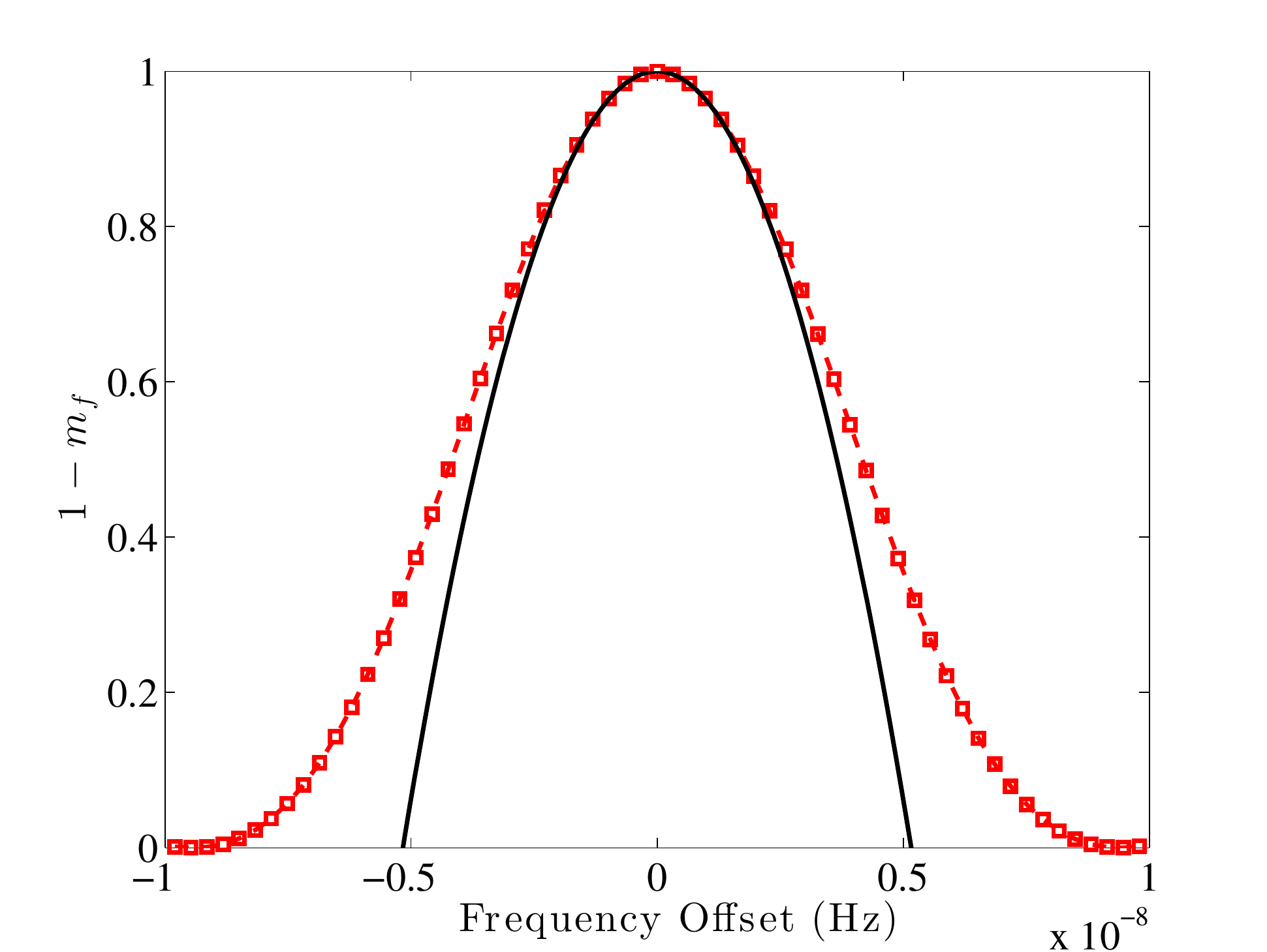}}
 		\subfigure
 		{\includegraphics[width=0.33\textwidth]{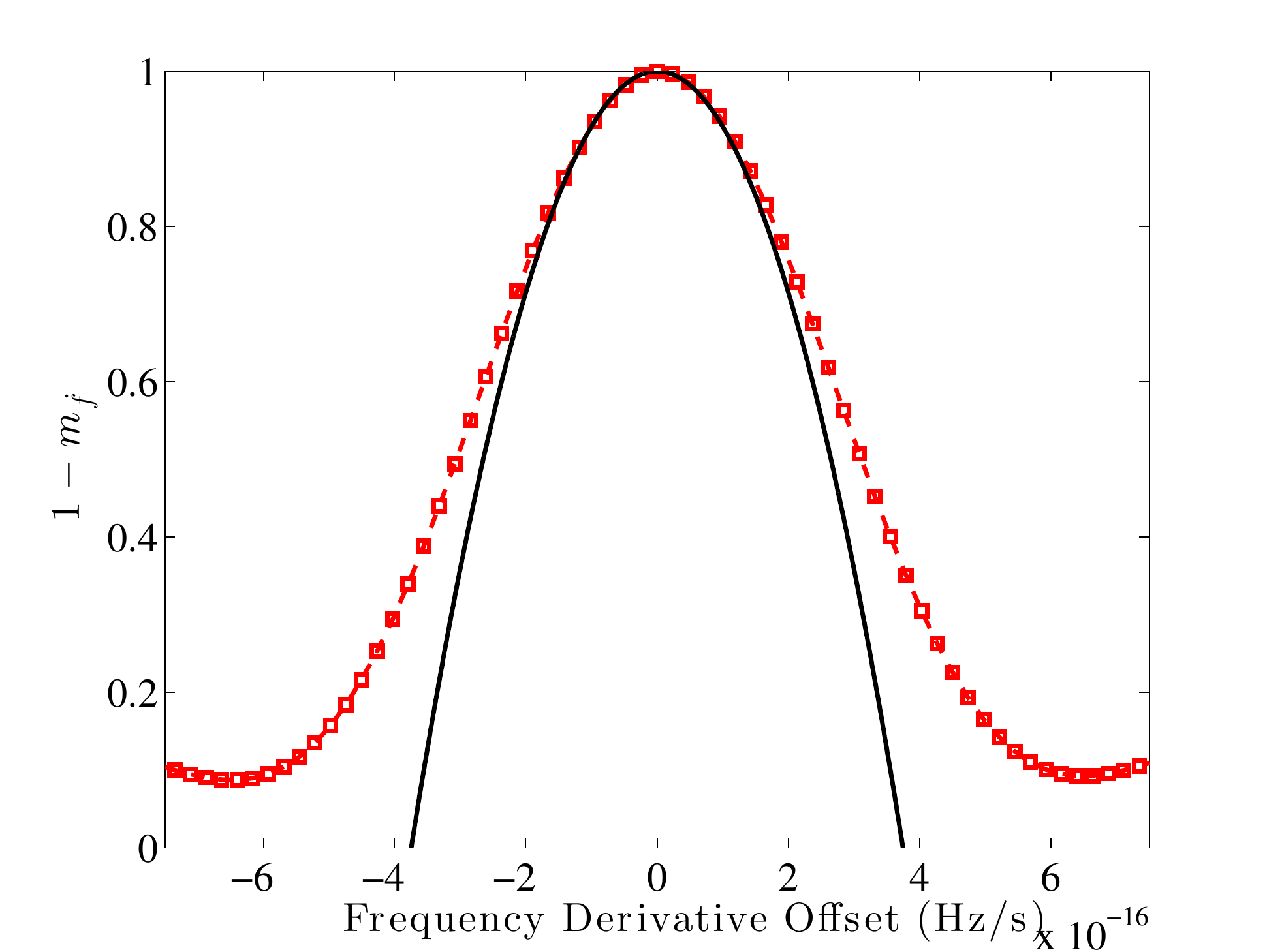}}
 		\subfigure
 		{\includegraphics[width=0.33\textwidth]{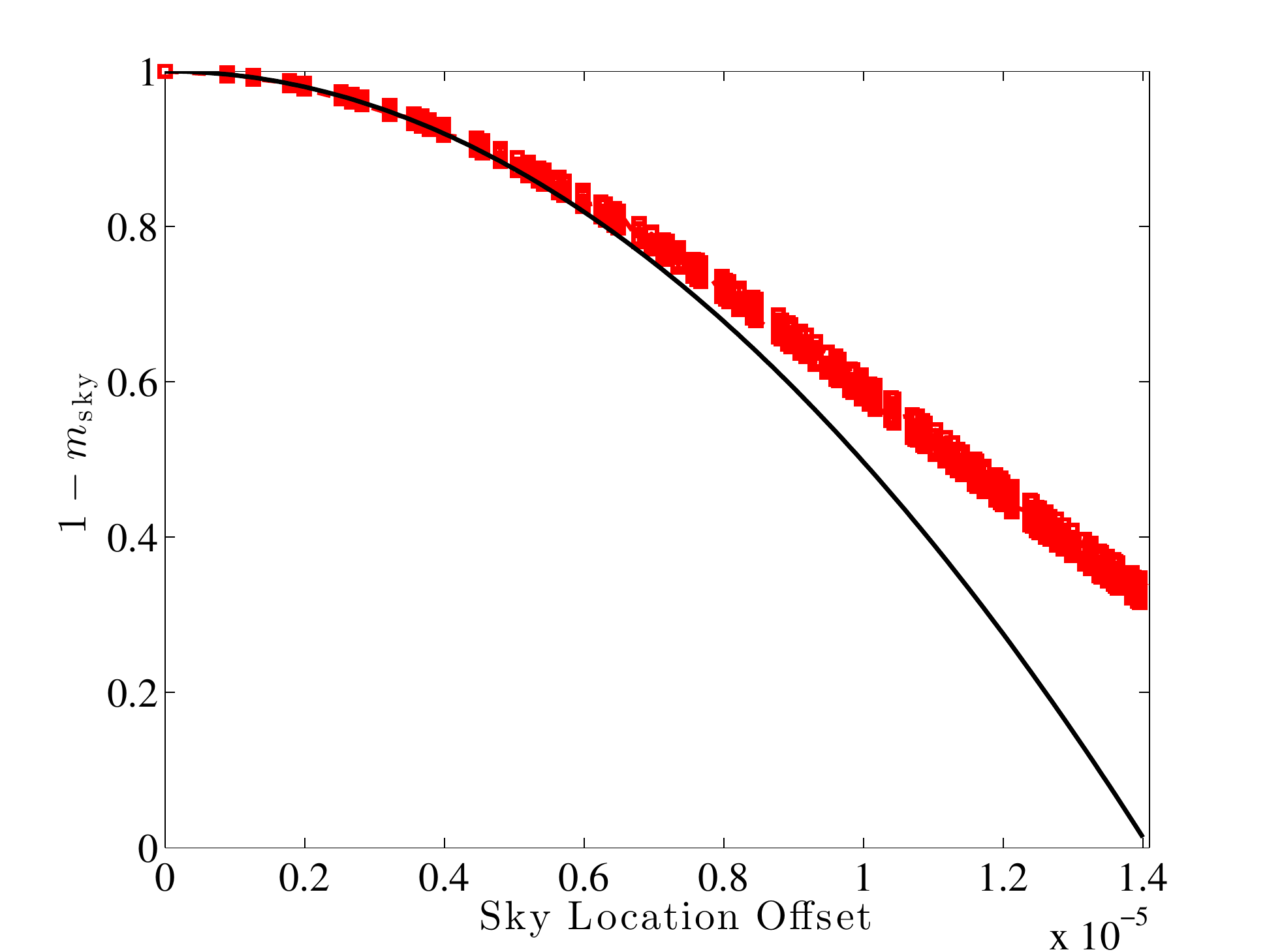}}
\caption{Comparison of mismatch in $\mathcal{P}_1$ (dashed curves) with coherent 
metric prediction (solid curves). 
In each panel the horizontal axis shows the offset from the signal parameters in
$f$ (left), $\dot f$ (middle), and sky position (right). The sky-location offset is
$\sqrt{\Delta n_x^2 + \Delta n_y^2}$, which measures
the offset in coordinates $(n_x,n_y)$ in the ecliptic plane.
The underlying pulsar signal has been simulated with spin parameters
$f = 32$\,Hz, $\dot{f} = -10^{-12}$\,Hz\,s$^{-1}$ for a total coherent observation time of $T_{\rm coh} = 3.4$\,yr. 
\label{f:P_mismatch}}
\end{figure*}
%---------------------------------------------------------------

For the purpose of efficient search-grid construction
we exploit a simplified phase model which captures
the most dominant effects. It is to be emphasized that we do \emph{not}
use this phase model in the actual search when computing the
phases at the photon arrival times.
Thus, we here assume that the LAT data set spans at least
one year, such that the Doppler modulation is dominated
by the Earth motion around the SSB. 

For very short coherent integration times, the orbital motion of the \Fermi{} satellite around the Earth could also introduce further Doppler effects. Comparing this effect to the much larger effect of the Earth's orbital motion around the sun, which is responsible for the behavior of the metric visible in, e.g., \Fref{fig:coh_met_det}, it is clear that this effect would saturate after a small number of orbits. Hence for coherent integration times of more than a few hours, here it is safe to neglect the rapidly oscillating components
of the motion of the \Fermi{} satellite around the Earth. 
Doing so yields the following phase model,
\begin{align}
   \phi(t,\vDoppler) &= 2\pi\,f(t-t_0) + \pi \,\dot f (t-t_0)^2 + 2\pi \, f \frac{\vec{n} \cdot \vec{r}_E(t)}{c}  \nonumber\\
             &= 2\pi\,f(t-t_0) + \pi \,\dot f (t-t_0)^2  \nonumber\\
             &\;\;\;+ 2\pi \, f \, r_{E}\left[ n_x\, \cos(\Omega_E t) + n_y \sin(\Omega_E t) \right]  \,,
             \label{e:phasesimple1}
\end{align}
where $n_x$ and $n_y$ are the components of $\vec n$, the unit vector 
pointing from the SSB to the sky location $(\alpha,\delta)$,
projected into the ecliptic plane (using the obliquity of the ecliptic, $\epsilon$), 
\begin{align}
 n_x &= \cos(\alpha)\,\cos(\delta) \,,\\
 n_y &= \cos(\epsilon) \,\sin(\alpha)\, \cos(\delta) + \sin(\epsilon)\,\sin(\delta) \,,
\label{e:eq2ec-nx-ny}
\end{align}
and $\Omega_E = 2\pi/1$yr,
and $r_{E}=1\textrm{AU}/c\sim 500$s.

In the presence of a small offset~$\Delta\vDoppler$ from a signal's location in parameter 
space~$\vDoppler_\sig $, 
we can write the mismatch, $m^{[t_j]}$, in the coherent power in a window of length $T$, centered on the $j$th photon as 
\begin{align}
m^{[t_j]}  &= 1-\frac{ \left(\theta^2_{\mathcal{P}_1}(\vDoppler_\sig + \Delta\vDoppler)\right)^{[t_j]}}{ \theta^2_{\mathcal{P}_1}(\vDoppler_\sig)} \, \\
&= 1 - \left|\av{e^{-i\phi(t,\Delta\vDoppler)}}^{[t_j]}\right|^2,
\end{align}
where we have replaced the discrete sum of \Eref{e:Pn} for simplicity by a continuous integral over the coherent integration time~$T$, i.e., 
\begin{equation}
   \av{x}^{[t_j]} \equiv \frac{1}{T} \int_{t_j-T/2}^{t_j+T/2}\, x(t)\, \rmd t \,.
\end{equation}
Following the derivation in \cite{Pletsch2010}, the mismatch can be Taylor expanded up to second order in terms of the parameter offsets, $\Delta\Doppler^k$ to give
\begin{equation}
   \mm^{[t_j]} = \sum_{k,\ell} G_{k\ell}^{[t_j]}\,\Delta\Doppler^k \,\Delta\Doppler^\ell +\;\mathcal{O}(\Delta\vDoppler^3) \,.
   \label{e:mm-metric-app}
\end{equation}
The coherent metric components are defined as
\begin{equation}
   G_{k\ell}^{[t_j]} = \av{\partial_k \phi\; \partial_\ell \phi}^{[t_j]} 
   - \av{\partial_k \phi}^{[t_j]}\,\av{\partial_\ell \phi}^{[t_j]} \,,
 \label{e:metric-tensor1}
\end{equation}
where $\partial_k\phi$ is the partial derivative of the phase at the signal location with respect to the $k$th component of the parameter offset:
\begin{equation}
   \partial_k \phi \equiv \left. \frac{\partial \; \phi(t;\vDoppler_\sig+\Delta\vDoppler)}{\partial(\Delta\Doppler^k)} \;
   \right|_{\Delta\vDoppler=\Nullvec} \,.
\end{equation}
Using the simplified phase model of \Eref{e:phasesimple1}, the metric components 
for a coherent window, centered on $t_j$ are given by
\begin{subequations}
\begin{align}
G_{ff}^{[t_j]} &= \frac{\pi^2 T^2}{3} \,,\\
G_{\dot{f}\dot{f}}^{[t_j]} &= \frac{\pi^2 T^4}{180} + \frac{\pi^2(t_j - t_0)^2 T^2}{3} \,,\\
G_{n_x n_x}^{[t_j]} &= 2\pi^2 f^2 r_E^2 \;\bigl[1 + \sinc\left(\Omega_E T/\pi\right)\cos\left(2\Omega_Et_j\right) \nonumber\\
&\;\;\; - 2 \, \sinc^2 \, \left(\Omega_E T/2\pi\right)\cos^2\left(\Omega_E t_j\right)\bigr] \,,\\
G_{n_y n_y}^{[t_j]} &= 2\pi^2 f^2 r_E^2 \; \bigl[1 - \sinc\left(\Omega_E T/\pi\right)\cos\left(2\Omega_Et_j\right) \nonumber\\
&\;\;\; - 2 \, \sinc^2 \, \left(\Omega_E T/2\pi\right)\sin^2\left(\Omega_Et_j\right)\bigr] \,.
\end{align}
\label{e:metricj}
\end{subequations}
For the specific case of the general expressions above, where $t_j = t_0 = 0$,
the metric components for the coherent detection statistic simplify to the following
form,
\begin{subequations}
\label{e:CohMetricComps}
  \begin{align}
  G_{ff} &= \frac{\pi^2 T^2}{3} \,,\label{e:cohGff} \\
  G_{\dot{f}\dot{f}} &= \frac{\pi^2 T^4}{180}  \,, \\
  G_{n_x n_x} &= 2\pi^2 f^2 r_E^2\left[1 +  \sinc\left(\Omega_E\ T/\pi\right) - 2\,  \sinc^2\left(\Omega_E\,T/2\pi\right) \right] \,, \\
  G_{n_y n_y} &= 2\pi^2 f^2 r_E^2\left[1 - \sinc\left(\Omega_E\,T/\pi\right)\right] \,.
\end{align}
\end{subequations}
The mismatches predicted by these derived metric components are compared to the measured mismatches in $\mathcal{P}_1$ for a simulated pulsar signal in Figure \ref{f:P_mismatch}.

Therefore, the determinant of the coherent metric is found as
\begin{align}
 \sqrt{\det G} &= \frac{\pi^4}{\sqrt{135}} \; T^3 \, f^2 \, r_E^2 \, \nonumber\\
 &\;\;\times \left[1 +  \sinc\left(\Omega_E\,T_{{\rm coh},1}/\pi \right) - 2\,  
 \sinc^2\left(\Omega_E\,T_{{\rm coh},1}/2\pi \right)\right] \nonumber\\
 &\;\;\times \left[1 - \sinc\left(\Omega_E\,T_{{\rm coh},1}/\pi \right)\right] 
 \,.
 \label{e:detGcoh}
\end{align}

\section{Coherent Metric with Incoherent Harmonic Summing}
\label{s:appcohmetharmsum}

If a search is performed using the $Z_M^2$ statistic, i.e., incoherently summing the coherent power~$\mathcal{P}_n$ in the first $M$ harmonics, the mismatch, $\tilde{m}$, becomes
\begin{align}
\tilde{m}  &= 1 - \frac{\sum_{n=1}^M\theta_{\mathcal{P}_n}^2(\vDoppler_{\rm sig} + \Delta\vDoppler)}{\sum_{n=1}^M\theta_{\mathcal{P}_n}^2(\vDoppler_{\rm sig})} \nonumber\\
& = 1 - \frac{\sum_{n=1}^M \left|\gamma_n\right|^2 \left| \av{e^{-in\phi(t,\Delta\vDoppler)}}^{[t_j]}\right|^2}{\sum_{n=1}^M  \left|\gamma_n\right|^2} \,.
\label{e:mmZm}
\end{align}
Taylor expanding this mismatch to second order gives the metric components,
\begin{equation}
   \tilde{m} = \sum_{k,\ell} \tilde{G}_{k\ell}^{[t_j]}\,\Delta\Doppler^k \,\Delta\Doppler^\ell +\;\mathcal{O}(\Delta\vDoppler^3) \,,
   \label{e:mm-metric-app}
\end{equation}
which can be expressed using \Eref{e:metric-tensor1} as
\begin{equation}
  \tilde{G}_{k\ell}^{[t_j]} = r^2\; G_{k\ell}^{[t_j]} \,,
  \label{e:harmsummetric}
\end{equation}
where we defined the harmonic refinement factor~$r$ from
\begin{equation}
 r^2 = \frac{\sum_{n=1}^{M} \left|\gamma_n\right|^2 n^2}{\sum_{n=1}^{M} \left| \gamma_n \right|^2} \,.
 \label{e:r-refinement}
\end{equation}
Thus, \Eref{e:harmsummetric} indicates that the parameter space must be sampled 
$r$ times more finely in each dimension when summing the power from $M$ harmonics, 
\begin{equation}
    \sqrt{\det \tilde{G}}  = r^4 \;\sqrt{\det G}   \,.
\end{equation}
The value of this refinement factor~$r$ also depends on the signal pulse profile~$\gamma_n$,
which of course is unknown in advance. However, we can consider the two limiting cases.
First, for the narrowest possible pulse profile, a Delta function, all coefficients are equal,
$|\gamma_n|=1$, such that
\begin{equation}
   r^2 = \frac{1}{M} \sum_{n=1}^{M} n^2 
   = \frac{M^2}{3} + \frac{M}{2} + \frac{1}{6} \,.
   \label{e:rsq}
\end{equation}
Therefore, for $M>1$ the parameter space must be sampled more finely in each dimension
by a factor of approximately~$M^2/3$ (to leading order).
On the other limiting case, for a sinusoidal pulse profile, where $|\gamma_{n>1}|=0$, $r=1$ and thus $\tilde{G}_{k\ell}^{[t_j]} = G_{k\ell}^{[t_j]}$, requiring no refinement.
Therefore, the range of the harmonic-summing refinement factor is approximately limited to~$r\in[1,M]$.

Finally, we would like to point out a further generalization.
Suppose a search is performed using the $Q_M$ statistic and a template pulse
profile~$\alpha_n$, which is not equal to the Dirac delta function (in this case
$Q_M$ would reduce again to $Z_M^2$). Then by a straightforward repetition of  arguments 
from the beginning of this section one obtains the resulting metric
tensor~$\hat{G}_{k\ell}^{[t_j]}$ for the $Q_M$ test statistic as
\begin{equation}
  \hat{G}_{k\ell}^{[t_j]} = \hat{r}^2\; G_{k\ell}^{[t_j]} \,,
  \label{e:harmsummetricQ}
\end{equation}
where the harmonic refinement factor~$\hat{r}$ in this case would be different from
\Eref{e:r-refinement}, namely
\begin{equation}
   \hat{r}^2 = \frac{\sum_{n=1}^{M} 
   \left|\gamma_n\right|^4 n^2}{\sum_{n=1}^{M} \left| \gamma_n \right|^4} \,.
   \label{e:r-refinementQ}
\end{equation}

\section{Approximate harmonic-summing computing cost}
\label{a:AnaApproxHarmSum}

\begin{figure}[t]
\centering
  \includegraphics[width=0.47\textwidth]{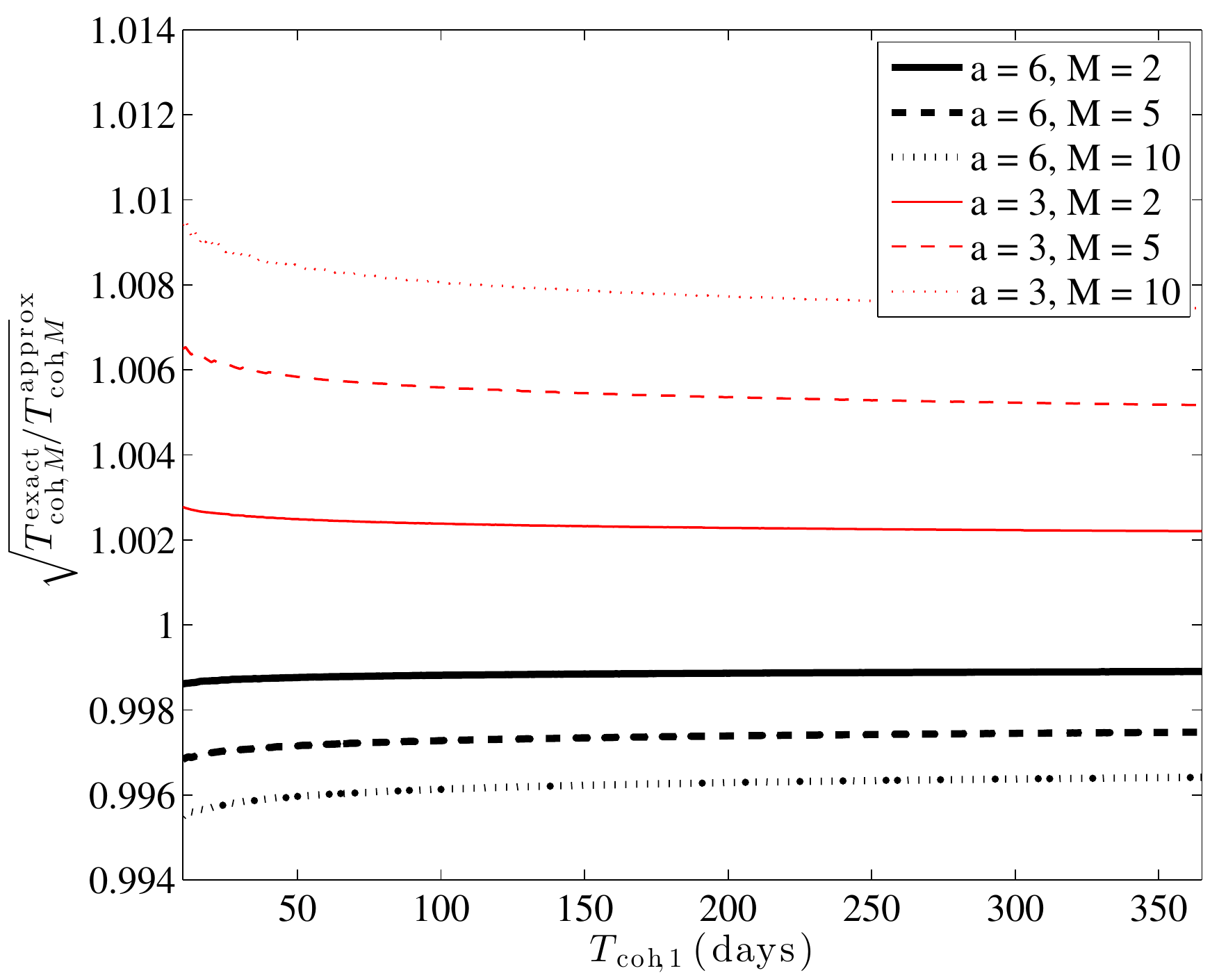}
   \caption{Comparison of the analytical approximation for the harmonic-summing
   computing cost model (leading to $T_{{\rm coh},M}^{\rm approx}$) to the results 
   obtained from fully numerical evaluation (leading to $T_{{\rm coh},M}^{\rm exact}$),
   as a function of $T_{{\rm coh},1}$ corresponding to the same 
   computing cost $C_{{\rm coh},1} = C_{{\rm coh},M}$.
   Since we are interested in the impact on search sensitivity 
   $p_{{\rm coh}, M}^{-1} \propto \sqrt{T_{{\rm coh},M}}$, the vertical axis shows 
   the square root of the ratio. 
   As indicated by the legend the different curves are
   for different values of scaling exponent~$a$ of \Eref{e:a-scaling} and number 
   of harmonics summed~$M$.
   \label{f:Mscale-approx}
}
\end{figure}

In \Sref{s:harmsum}, we describe an analytical approximation for the computing
cost model of incoherent harmonic summing. This approximation is based on 
ignoring the slowly varying $\log_2$ factors in Equations~\eref{e:Ccoh1} and \eref{e:CcohM}. 
If then one equates $C_{{\rm coh},1} = C_{{\rm coh},M}$, it follows that
$T_{{\rm coh},M}$ must be shorter by the factor~$(M^2\,r^2)^{(1/a)}$, as given in \Eref{e:TcohM}.
Here, we study the accuracy of the analytical approximation in terms of the search
sensitivity $p_{{\rm coh}, M}^{-1} \propto \sqrt{T_{{\rm coh},M}}$, 
by comparison to the exact value for $T_{{\rm coh},M}$ obtained from numerical evaluation.
For a given value of $T_{{\rm coh},1}$, we find numerically the exact value of $T_{{\rm coh},M}$ 
such that $C_{{\rm coh},1} = C_{{\rm coh},M}$. We here assume a wide search
frequency range, $f_{\rm max} = 1000$\,Hz.
 The results are displayed in \Fref{f:Mscale-approx}, showing that the approximation is
accurate to within less than $1\%$ for typical search setups. As can also be seen,
for the realistic case of $a=6$ the approximation is generous in favor of the harmonic summing 
approach, because  $T_{{\rm coh},M}^{\rm approx} \gtrsim T_{{\rm coh},M}^{\rm exact}$, 
the approximation overestimates the true search sensitivity.

%---------------------------------------------------------------------------------------------------
%---------------------------------------------------------------------------------------------------

\section{Optimal mismatch in coherent search}
\label{a:optmismatchcoherent}

In this section, we use the method of Lagrange multipliers as in 
\citet{PrixShaltev2012} to obtain 
the optimal average mismatch for a fully coherent search. 
We use the scalings of the sensitivity~$p_{{\rm coh}, M}^{-1}$ and computing 
cost $C_{\rm coh,M}$, ignoring the $\log_2$ FFT scaling factor,
from Equations~\eref{e:searchsens2} and \eref{e:CcohMapprox}, respectively.
In order to find the optimal mismatch at a fixed computing cost $C_0$, 
we search for stationary points of the Lagrange function,
\begin{align}
  L(T_{{\rm coh},M},m,M,\lambda) 
  & = p_{{\rm coh}, M}^{-1} - \lambda (C_{\rm coh,M}  - C_0) \nonumber\\
  &= (1- \langle m_{\rm tot} \rangle)^{1/2}\; T_{{\rm coh},M}^{1/2} \;h^\ast(M) 
  \nonumber\\
  &\;\;\; + \lambda \left( K'_{{\rm coh},a}m^{-3/2}T_{{\rm coh},M}^aM^2r^2(M) - 
  C_0 
  \right),
\end{align}
where $\lambda$ is a Lagrange multiplier, 
and we defined $K'_{{\rm coh},a} = K_{{\rm coh},a}f_{\rm max}^2$, as well as
the function $h^\ast(M)$ as,
\begin{equation}
  h^\ast(M) 
  = \frac{1}{M^{1/4} \; \theta_M^{\ast}} \,  
  \left[\sum_{n=1}^M |\gamma_n|^2 \right]^{1/2} \,,
  \label{e:hofM}
\end{equation}
using $^\ast$ to indicate the implicit dependence on 
$P_{\rm FA}^{\ast}$ and $P_{\rm DET}^{\ast}$ through $\theta_M^{\ast}$.
Taking partial derivatives with respect to $T_{{\rm coh},M}$, $m$ and $M$ 
respectively yields:
\begin{align}
  \frac{\partial L}{\partial T_{{\rm coh},M}} &= \frac{1}{2}(1- \langle m_{\rm 
  tot} \rangle)^{1/2}T_{\rm 
  coh}^{-1/2}h^\ast(M) + \frac{a \lambda C_{\rm coh,M}}{T_{{\rm coh},M}} = 0 
  \,,\\
  \frac{\partial L}{\partial m} &= \frac{1}{2}(1-\langle m_{\rm 
  tot}\rangle)^{-1/2} 
  3\xi\,T_{\rm 
  coh}^{1/2}h^\ast(M) + \frac{3 \lambda C_{\rm coh,M}}{2 \,m} = 0 \,,\\
  \frac{\partial L}{\partial M} &= (1-\langle m_{\rm tot} \rangle)^{1/2}T_{{\rm 
  coh},M}^{1/2} 
  \frac{\partial h^\ast(M) }{\partial M} \nonumber \\&+ \lambda C_{\rm 
  coh,M}\left(\frac{2}{M}+ \frac{2}{r(M)}\frac{\partial 
  r}{\partial M}\right)= 
  0 \,.
\end{align}
Equating these and rearranging for $\xi m$, we find that the optimal average 
mismatch for a fully coherent search is
\begin{equation}
  3 \xi\,m_{\rm opt} = \frac{1 - \langle m_f \rangle}{\frac{2a}{3}+1} \,.
\end{equation} 
As we argue in \Sref{s:compcost}, practical fully coherent searches are 
computationally
limited to integration times~$T_{{\rm coh},M}$ less than half a year, implying 
$a = 
6$. If the frequency dimension is interpolated using interbinning, $\langle m_f 
\rangle \approx 0.14$, giving $m_{\rm opt} = 0.172$ for a total average 
mismatch 
of $\langle m_{\rm tot} \rangle = 0.312$. 
It is noteworthy that this result is independent of the computational cost, 
the coherent integration time, and the number of harmonics summed.

In principle, one can also rearrange for $M$ to find the optimal number of 
harmonics,
which then requires solving a complicated differential equation.
However, the derivatives of the functions $h^\ast(M)$ 
defined in \Eref{e:hofM}, and $r(M)$ defined in \Eref{e:r-refinement} are 
difficult to obtain for most pulse profiles. Therefore, we followed the 
approach presented in \Sref{s:harmsum}
to find the optimal~$M$ at fixed computing cost, which 
does not require calculating these derivatives.

\section{Derivation of statistical properties of semicoherent test statistic}
\label{s:statsemicohpow}

From \Eref{e:S1b}, the expectation value of $S_1$ can be written as
\begin{equation}
E_0\left[S_1\right] = E_0\left[\sum_{j,k}^N\, w_j w_k e^{-i(\phi(t_j) - \phi(t_k)} \,\hat{W}_T^{{\rm\tiny rect}}(\tau_{jk})\right]\,.
\label{e:E_S}
\end{equation}
In order to evaluate this expectation value, we must take into account terms in the double sum where the photon indexes $(j,k)$ are equal, giving
\begin{equation}
E_0\left[S_1\right] = \sum_{j=1}^N w_j^2 \hat{W}_T(0) + \sum_{j\neq k }^N w_j w_k\, E_0\left[e^{-i(\phi(t_j) - \phi(t_k))}\right] \,\hat{W}_T(\tau_{jk})\,,
\end{equation}
where $\sum_{j \neq k}^N$ denotes a double sum over all photons, excluding terms where $j=k$. Under the null hypothesis, $p=0$, it holds
\begin{equation}
E_0\left[e^{-i\phi(t_j)}\right] = E_0\left[e^{i\phi(t_k)}\right] = 0\,,
\label{e:E0_phase_terms}
\end{equation}
and hence we find that the expectation value of $S_1$ is simply
\begin{equation}
E_0[S_1] = \sum_{j=1}^N w_j^2 \, \hat{W}_T(0)\,.
\end{equation}
To find the variance of $S_1$, we must evaluate
\begin{equation}
E_0\left[S_1^2\right] = E_0\left[\sum_{j,k,l,m}^N \, e^{-i(\phi(t_j) - \phi(t_k) + \phi(t_l) - \phi(t_m)} \, \hat{W}_T(\tau_{jk}) \, \hat{W}_T(\tau_{lm})\right] \,.
\label{e:E_S_2}
\end{equation}
Again, taking into account terms where photon indexes are equal, and using \Eref{e:E0_phase_terms}, 
we find that
\begin{equation}
E_0\left[S_1^2\right] = \sum_{j=1}^N \, w_j^4 \hat{W}_T(0)^2 \, + \sum_{j\neq k}^N \,w_j^2\, w_k^2\, \hat{W}_T(0)^2 + \sum_{j \neq k}^N \,w_j^2\, w_k^2\, {\hat{W}_T(\tau_{jk})}^2\,, 
\end{equation}
and hence the variance of $S_1$ under the null hypothesis is
\begin{align}
Var_0\left[S_1\right] &= E_0\left[S_1^2\right] - E_0\left[S_1\right]^2 \nonumber\\
&= \sum_{j \neq k}^N \,w_j^2\, w_k^2\, {\hat{W}_T(\tau_{jk})}^2 \,.
\end{align}
From now on in this section, we will use the rectangular lag-window $\hat{W}_T^{{\rm\tiny rect}}(\tau_{jk})$
of \Eref{e:rectlagwin}.
In addition, we assume binary photon weights for simplicity. 
In this case one obtains
\begin{equation}
   E_0[S_1]  =  N \,,\quad   Var_0[S_1] \approx  N^2 \, R^{-1} \,.
   \label{e:S_noise_moments_app}
\end{equation}
%

%---------------------------------------------------------
\begin{figure}[t]
   \centering
   \includegraphics[width=0.47\textwidth]{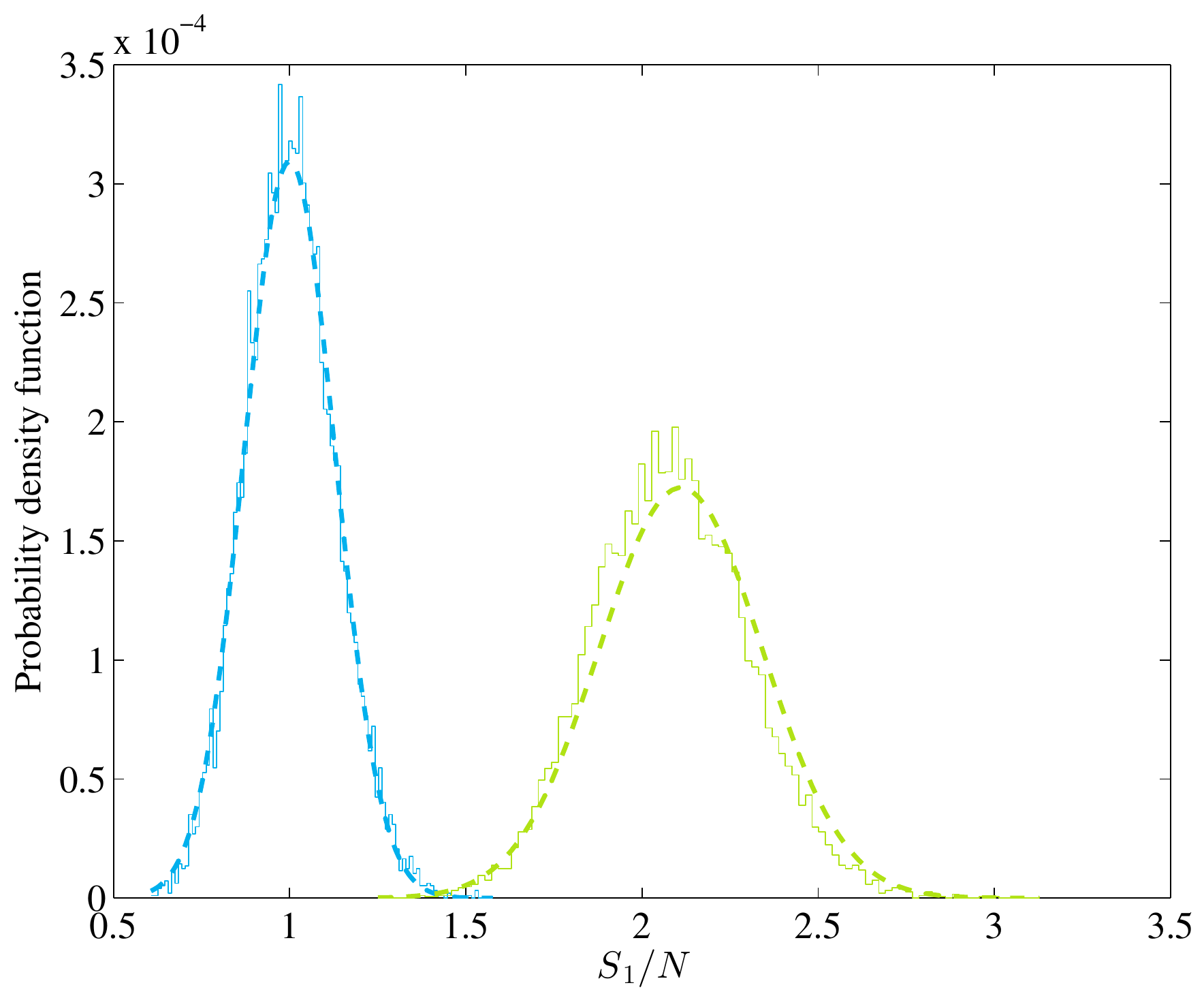}
    \caption{Comparison of empirical and analytically predicted 
    probability density function (PDF) of the semicoherent test statistic~$S_1$.
    The blue curves (left) refer to the noise-only case, where $S_1$ has been 
    calculated from many simulated data sets of $N=10^4$ unit-weight photons
    with $R=60$ to obtain the empirical PDF (solid curve) which is compared 
    to the analytical PDF (dashed).
    The green curves represent the PDF of $S_1$ for simulated data sets containing 
    signals with a pulsed fraction of $p = 0.1$ and a profile with $|\gamma_1|^2 = 0.668$,
    where again the empirical PDF (solid curve) is compared to the analytical PDF (dashed).
 }
\label{f:s_dist}
\end{figure}
%---------------------------------------------------------

To derive the moments of the distribution of $S_1$ in the presence of a perfectly-matched signal, we need 
to distinguish times $t_j$ of non-pulsed photons (i.e., background) from pulsed photons 
by denoting the latter times as $t_j^\prime$. We then use the definitions of the Fourier coefficients of the pulse profile to evaluate the expectation values
\begin{equation}
  E_p\left[e^{-in\phi(t^\prime_j)}\right] = \gamma_n \,,\qquad
  E_p\left[e^{in\phi(t^\prime_j)}\right] = \gamma_n^\ast \,.
\label{e:Ep_phase_terms}
\end{equation}
Evaluating Equations \eref{e:E_S} and \eref{e:E_S_2}, using the expectation values from Equations \eref{e:E0_phase_terms} and \eref{e:Ep_phase_terms}, with a  pulsed fraction, $p \sim \mathcal{O}(10^{-1})$ and a typical pulse profile~$\gamma_n$ (cf. \Fref{f:2pc-harm}), gives the first two moments of the distribution of $S_1$ in the presence of a weak signal as
\begin{align}
  E_p\left[S_1\right] &\approx N + p^2 N^2 \left|\gamma_1\right|^2 R^{-1} \,,\\
  Var_p\left[S_1\right] &\approx \frac{N^2}{R}\left(1 + 2p^2 N \left|\gamma_1\right|^2 R^{-1}\right)\,, 
\end{align} 
where we have assumed a large number of photons $N \gg 1$, and that $R$ is large enough such that edge effects (e.g., effectively shorter windows near the end of the observational data time span) become negligible.

Again, appealing to the central limit theorem (i.e., assuming that there are many photon pairs within the double sums of \Eref{e:E_S}), we can approximate the distribution of $S_1$ by a normal distribution with the same mean and variance. By comparison with numerical simulations \Fref{f:s_dist} validates this approximation for the
purpose of the sensitivity estimation as presented in \Sref{s:semicohstats}.

\section{Semicoherent Metric}
\label{s:appsemicohmet}

\begin{figure*}[t]
 \centering
 		\subfigure
 		{\includegraphics[width=0.33\textwidth]{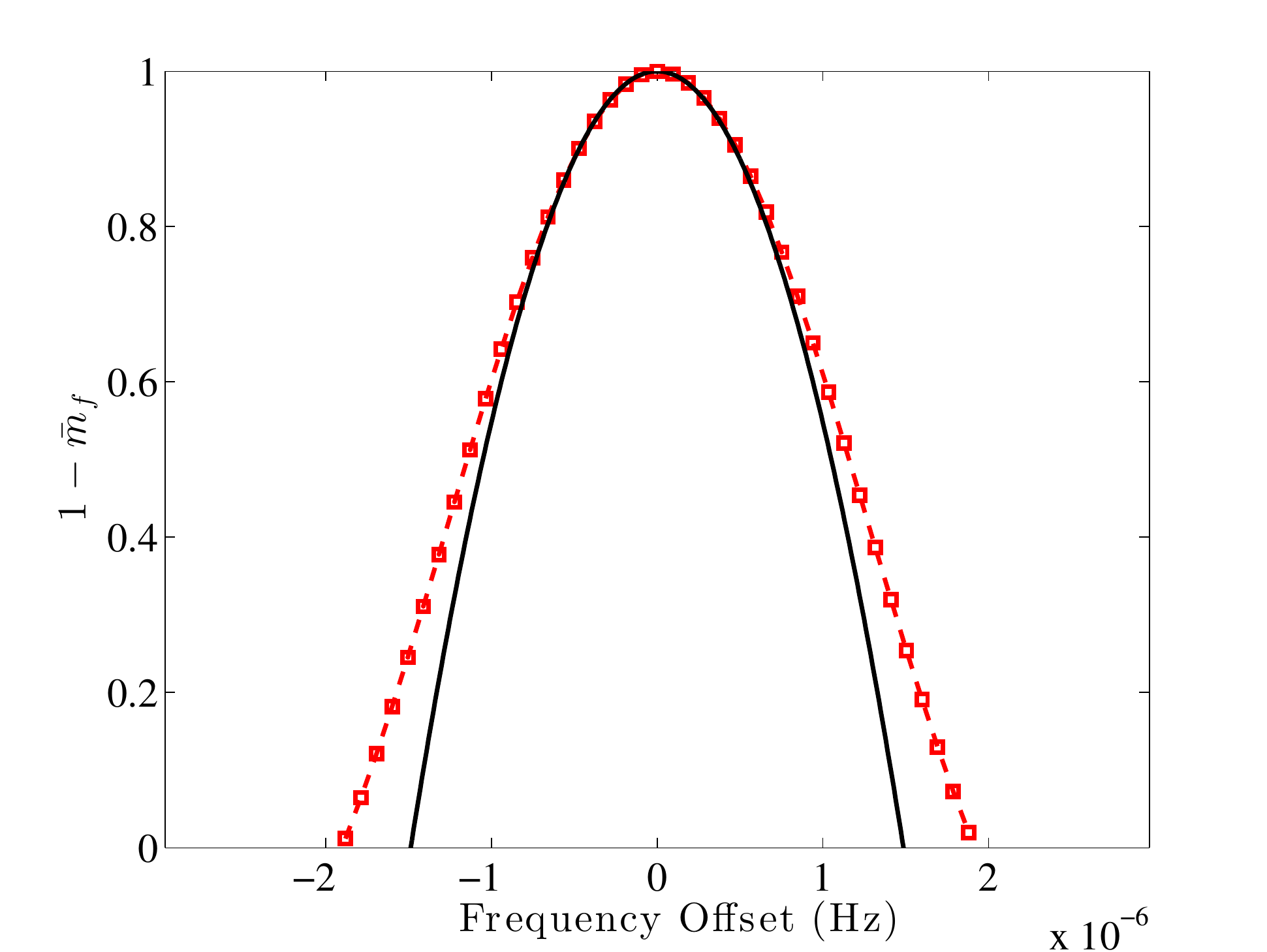}}
 		\subfigure
 		{\includegraphics[width=0.33\textwidth]{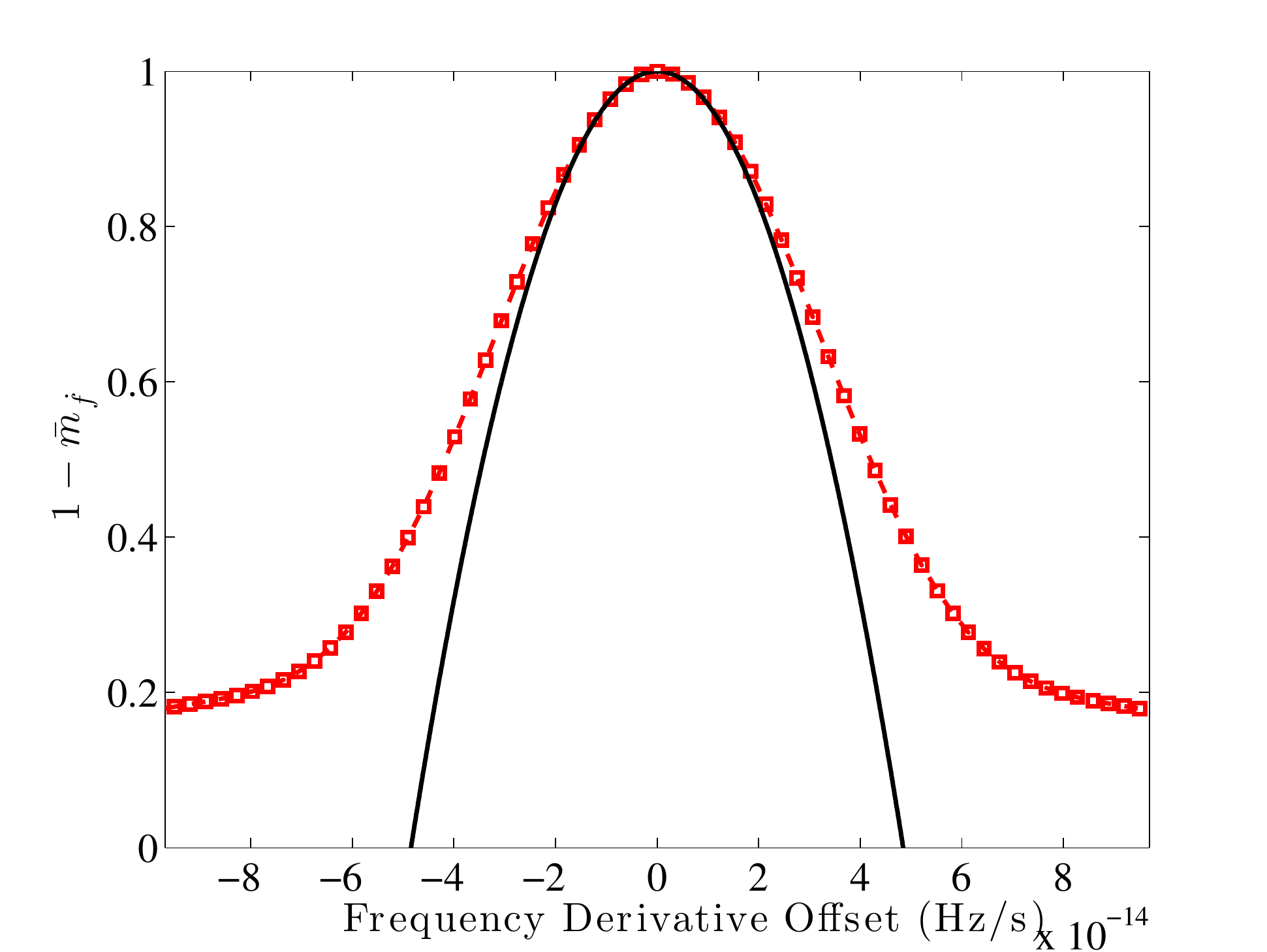}}
 		\subfigure
 		{\includegraphics[width=0.33\textwidth]{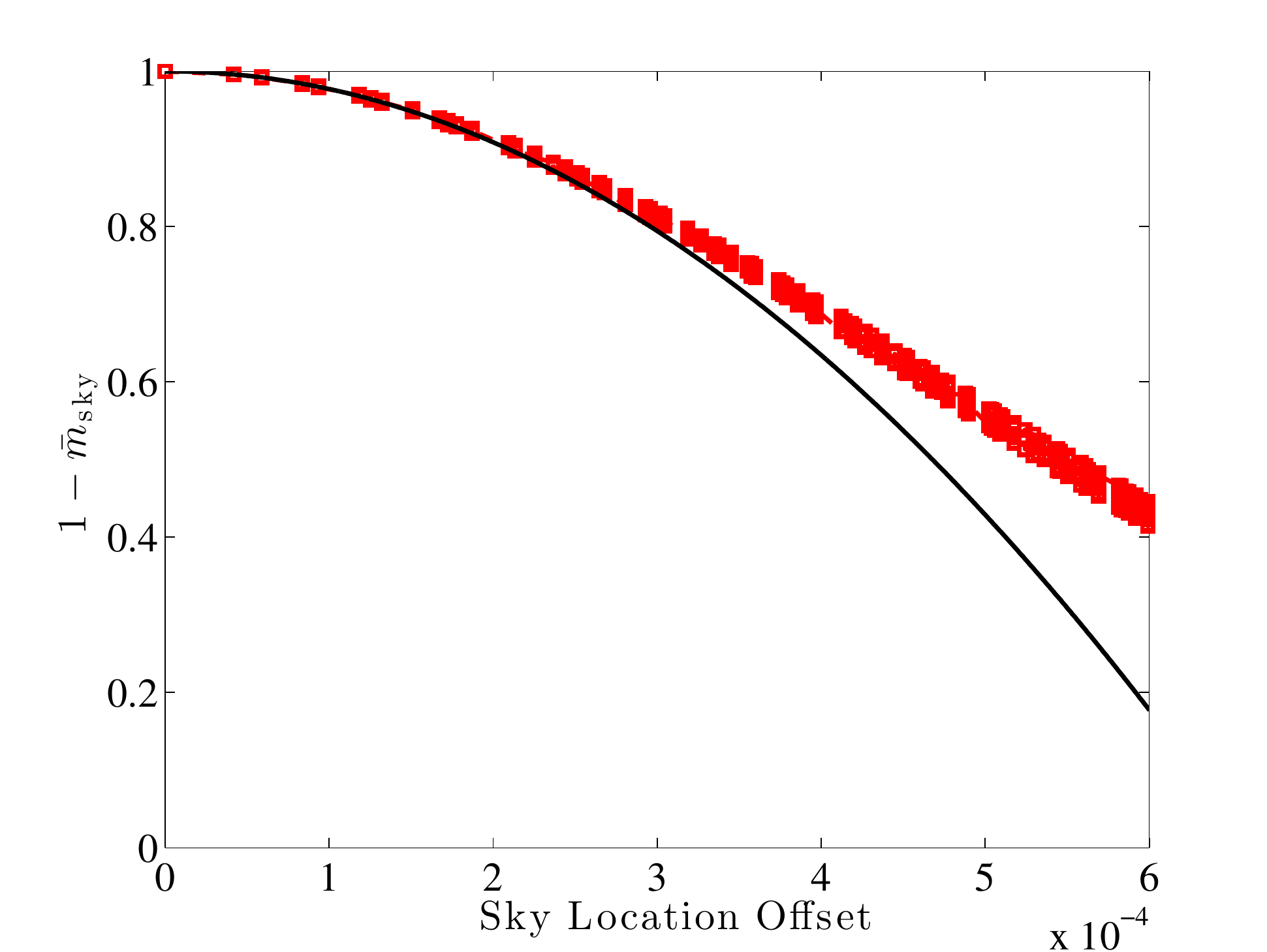}}
 \caption{Comparison of mismatch in $S_1$ (dashed curves) with semicoherent 
 metric prediction (solid curves). 
In each panel the horizontal axis shows the offset from the signal parameters in
$f$ (left), $\dot f$ (middle), and sky position (right).
The sky-location offset is
$\sqrt{\Delta n_x^2 + \Delta n_y^2}$, which measures
the offset in coordinates $(n_x,n_y)$ in the ecliptic plane.
The underlying pulsar signal has been simulated with parameters
$f = 32$\,Hz, $\dot{f} = -10^{-12}$\,Hz\,s$^{-1}$ for a total observational data time span of $T_{\rm obs} = 3.4$\,yr
and a coherent window size of $T = 524288$\,s. 
\label{f:S_mismatch}}
\end{figure*}

To derive the semicoherent metric, we investigate the mismatch in the semicoherent detection statistic in the presence of a strong signal. Starting from \Eref{e:S1b}, using binary photon weights and the rectangular lag window,
\begin{align}
  S_1 &= \sum_{j=1}^N \; \sum_{k=1}^N \; e^{-i [\phi(t_j)-\phi(t_k)]}\; \hat{W}_T^{{\rm\tiny rect}}(\tau_{jk}) \,\nonumber \\
  &= \sum_{j=1}^N e^{-i\phi(t_j)} \; \sum_{k=1}^N \; e^{i\phi(t_k)} \; \hat{W}_T^{{\rm\tiny rect}}(\tau_{jk}) \,.
  \label{e:S1b_app}
\end{align}
Again, replacing the sum over $k$ with a continuous integral allows us to write the mismatch as:
\begin{equation}
\bar{m} = 1 - \frac{\sum_{j=1}^N \, e^{-i\phi(t_j,\vDoppler_{\rm sig} + \Delta\vDoppler)} \, \av{e^{i\phi(t,\vDoppler_{\rm sig} + \Delta\vDoppler)}}^{[t_j]}}{\sum_{j=1}^N \, e^{-i\phi(t_j,\vDoppler_{\rm sig})} \,\av{e^{i\phi(t,\vDoppler_{\rm sig})}}^{[t_j]}}\,.
\label{e:scoh_mm}
\end{equation}
Assuming that each coherent window contains the same power (and hence has the same S/N at $\vDoppler_\sig$), this can be simplified to:
\begin{equation}
\bar{m} = 1 - \frac{1}{N}\sum_{j=1}^N \, e^{-i\phi(t_j,\Delta\vDoppler)} \, \av{e^{i\phi(t,\Delta\vDoppler)}}^{[t_j]} \,.
\label{e:scoh_mm2}
\end{equation}
Taylor expanding this mismatch around $\Delta\vDoppler = 0$ to second order in $\Delta\vDoppler$ gives:
\begin{eqnarray}
\bar{m} &=& \frac{i}{N}\sum_{j=1}^{N} \left(\left. \partial_k \phi \right|_{t=t_j} - \av{\partial_k\phi}^{[t_j]}\right)\Delta\Doppler^{k} \notag\\
&&+ \;\frac{1}{2N}\sum_{j=1}^{N} \left(\left.\partial_k\phi\right|_{t=t_j}\left.\partial_{\ell}\phi\right|_{t=t_j} + \av{\partial_k \phi \partial_{\ell} \phi}^{[t_j]}\right)\Delta\Doppler^k\Delta\Doppler^{\ell} \notag\\
&&- \;\frac{1}{N} \sum_{j=1}^N \left(\left.\partial_k\phi\right|_{t=t_j} \av{\partial_{\ell}\phi}^{[t_j]}\right) \Delta\Doppler^k \Delta\Doppler^\ell \notag\\
&&+ \;\frac{i}{2N} \sum_{j=1}^N \left(\left.\partial_{k}\partial_{\ell}\phi\right|_{t=t_j} - \av{\partial_{k}\partial_{\ell}\phi}^{[t_j]}\right) \Delta\Doppler^k \Delta\Doppler^{\ell}\notag\\
&&+\;\mathcal{O}(\Delta\vDoppler^3)\,,
\label{e:scoh_mm_taylor}
\end{eqnarray}
where there are implicit sums over repeated indices.
Evaluating the partial derivatives at $t_j$, under the assumption that $T \ll T_{\rm obs}$, gives:
\begin{align}
\left.\partial_{k}\phi\right|_{t=t_j} &\approx \av{\partial_{k}\phi}^{[t_j]}
\label{e:partial_av} \\
\left.\partial_{k}\partial_{\ell}\phi\right|_{t=t_j} &\approx  \av{\partial_{k}\partial_{\ell}\phi}^{[t_j]}
\end{align}
Thus, the mismatch of \Eref{e:scoh_mm_taylor} becomes,
\begin{align}
\bar{m} &\approx \frac{1}{2N} \sum_{j=1}^N \left(\av{\partial_k \phi \partial_{\ell} \phi}^{[t_j]} - \av{\partial_k \phi}^{[t_j]}\av{\partial_{\ell}\phi}^{[t_j]} \right) \Delta\Doppler^k \Delta\Doppler^{\ell} \nonumber\\
& = \frac{1}{2N} \sum_{j=1}^{N} G_{k\ell}^{[t_j]} \Delta\Doppler^k \Delta\Doppler^{\ell} \,.
\label{e:scoh_mm_final}
\end{align}
Hence, the semicoherent metric components can be found by taking half the average of the coherent metric components of Equations \eref{e:metricj} over all photons in the observation time. Using the approximations given in \citep{Pletsch2010}, which are valid under the assumption that the data set spans many years, we find
\begin{subequations}
\begin{align}
  \bar{G}_{ff} &= \frac{\pi^2 T^2}{6} \,,\\
  \bar{G}_{\dot{f}\dot{f}} &= \frac{\pi^2 T^4}{360}\,\gamma^2 \,,\\
 \bar{G}_{n_x n_x} = \bar{G}_{n_y n_y} &=  \pi^2 f^2 r_E^2 \left[1 - \sinc^2(\Omega_E T/2 \pi)\right] \,,
 \label{e:scohmet_comps}
\end{align}
\end{subequations}
where $\gamma$ is the semicoherent refinement factor~\citep{PletschAllen2009,Pletsch2010} defined as
\begin{equation}
  \gamma^2 = 1 + \frac{60}{N}\sum_{j=1}^N\frac{(t_j-t_0)^2} {T^2} \,.
\end{equation} 
The mismatches predicted by these derived metric components are compared to the measured mismatches in $S_1$ for a simulated pulsar signal in Figure \ref{f:S_mismatch}.

For the purpose of the analytic study of the computing cost scaling in this paper,
we employ the approximation \mbox{$\gamma \approx \sqrt{5} T_{\rm obs}/T = \sqrt{5} R$}.
Hence the determinant of the semicoherent metric is obtained as,
\begin{equation}
 \sqrt{\det \bar{G}} \approx  \frac{\pi^4}{4\sqrt{27}} \; T^3 \, f^2 \, r_E^2 \, R \,
 \left[1 - \sinc^2\left(\frac{\Omega_E\,T}{2\pi}\right)\right]  \,.
 \label{e:detGsemicoh2aAppendix}
\end{equation}
%

%---------------------------------------------------------------
%---------------------------------------------------------------

\section{Optimal mismatch in semicoherent search}
\label{a:optmismatchsemicoherent}

Following the same steps as in Appendix~\ref{a:optmismatchcoherent}, we can find the 
optimal average mismatch 
for a semicoherent search with sensitivity $p^{-1}_{{\rm scoh},1}$ 
at a fixed computing cost $C_0$ by consideration of 
the following Lagrange function:
\begin{align}
  L(T,\bar{m},\lambda) &= p^{-1}_{{\rm scoh},1} + \lambda (C_{\rm scoh} -C_0) 
  \nonumber\\
  &= (1-\langle \bar{m}_{\rm tot}\rangle)^{1/2}T^{1/4} + \lambda(K'_{\rm 
  scoh}\bar{m}^{-3/2}T^{(s-1)} - C_0) \,.
\end{align}
Applying the method of Lagrange multipliers as above, we find that
\begin{equation}
   3 \xi\bar{m}_{\rm opt} = \frac{1 - \xi \bar{m}_f}{\frac{4(s-1)}{3} + 1} \,.
\end{equation}
As argued in \Sref{s:sccompcost}, an efficient strategy uses coherence window 
sizes~$T$
much less than half a year. In this regime of interest, $s = 5$. Using 
interbinning to interpolate the frequency spectrum gives $\langle m_f \rangle 
\approx 
0.075$, giving the optimal maximum mismatch in the remaining three parameters 
as $\bar{m}_{\rm opt} = 0.146$.

\section{Sky-grid Construction}\label{mstdisr:s:skygrid}

From the metrics derived above, in Appendices~\ref{s:appcohmet} and \ref{s:appsemicohmet}, 
we know when searching over a grid of sky locations that these grid points 
should be defined by a uniform grid in the ecliptic plane.

To construct the sky search grid for a source within an angular radius of $\theta$ from $(\alpha_0,\delta_0)$, this central point is rotated from equatorial to ecliptic coordinates according to the Earth's axial tilt (using the obliquity of the ecliptic, $\epsilon$) and projected into the ecliptic plane, with Cartesian coordinates $(x_0,y_0)$,
\begin{align}
x_0 &= \cos(\alpha_0)\,\cos(\delta_0) \,,\\
y_0 &= \cos(\epsilon) \,\sin(\alpha_0)\, \cos(\delta_0) + \sin(\epsilon)\,\sin(\delta_0) \,.
\label{e:eq2ec}
\end{align}
A square of side length $\theta$ on the unit circle is calculated around this point, and sampled (using the semicoherent or coherent metric components as appropriate) with spacings
\begin{align}
  \Delta n_x = \Delta n_y &= 2\sqrt{m/G_{n_x n_x}} \,.
  \label{e:grid_spacing}
\end{align}
These grid points are then projected back onto the unit sphere, and rotated into equatorial coordinates for barycentering.

Since a square region is sampled in the ecliptic plane, many of the resulting sky-points lie outwith the radius defining the search region on the sky. These points are simply discarded, resulting in the original circular search region on the sky in equatorial coordinates, sampled by a uniform grid defined in the ecliptic plane.

A possible problem arises when the search region crosses the ecliptic equator, since when the square is constructed in the ecliptic plane, some points lie outwith the unit circle, and therefore cannot be projected onto a unit sphere. This can be overcome by reflecting points, $(x,y)$, which lie outside the unit circle back into the sphere around the ecliptic longitude, $l$, of the center of the search region:
\begin{subequations}
\begin{align}
\label{e:ecliptic_longitude}
l &= \tan^{-1}\left(y_0/x_0\right) \,,\\
\label{e:reflect_point_x}
x' &= \cos(l) - \left[x - \cos(l)\right] \,,\\
\label{e:reflect_point_y}
y' &= \sin(l) - \left[y - \sin(l)\right] \,.
\end{align}
\end{subequations}
The new points $(x',y')$ are then projected into the \emph{opposite} hemisphere from the central point of the search region, resulting in a grid which covers an area of the sky which wraps around the ecliptic equator.

\bibliographystyle{apj}

\bibliography{msfgrp} 

\end{document}